\documentclass[a4paper,fleqn,usenatbib]{mnras}
\usepackage{newtxtext,newtxmath}
\usepackage[T1]{fontenc}
\usepackage{ae,aecompl}

\usepackage{graphicx}	
\usepackage{amsmath}	
\usepackage{multirow}
\usepackage{caption} 
\usepackage{subcaption} 
\title[]{Sub-millimetre compactness as a critical dimension to understand the Main Sequence of star-forming galaxies }

\author[A. Puglisi et al.]{Annagrazia Puglisi$^{1}$\thanks{E-mail: annagrazia.puglisi@durham.ac.uk (AP)},
Emanuele Daddi$^{2}$,
Francesco Valentino$^{3,4}$,
Georgios Magdis$^{3,4,5,6}$,
\newauthor 
Daizhong Liu$^{7}$,
Vasily Kokorev$^{3,4}$,
Chiara Circosta$^{8}$,
David Elbaz$^{2}$,
Frederic Bournaud$^{2}$,
\newauthor
Carlos Gomez-Guijarro$^{2}$,
Shuowen Jin$^{9,10}$,
Suzanne Madden$^{2}$,
Mark T. Sargent$^{11,12}$ and
\newauthor
Mark Swinbank$^{1}$
\\
\\
$^{1}$ Centre for Extragalactic Astronomy, Department of Physics, Durham University, Durham, UK \\
$^{2}$ CEA, Irfu, DAp, AIM, Universit\`e Paris-Saclay, Universit\`e de Paris, CNRS, F-91191 Gif-sur-Yvette, France \\
$^{3}$Cosmic Dawn Center (DAWN), Copenhagen, Denmark \\
$^{4}$Niels Bohr Institute, University of Copenhagen, Jagtvej 128, DK-2200, Copenhagen, Denmark \\
$^{5}$ DTU-Space, Technical University of Denmark, Elektrovej 327, DK-2800 Kgs. Lyngby, Denmark \\
$^{6}$ Institute for Astronomy, Astrophysics, Space Applications and Remote Sensing, National Observatory of Athens, GR-15236 Athens, Greece \\
$^{7}$ Max-Planck-Institut f{\"u}r Extraterrestrische Physik (MPE), Giessenbachstr. 1, D-85748 Garching, Germany\\
$^{8}$Department of Physics \& Astronomy, University College London, Gower Street, London WC1E 6BT, UK \\
$^{9}$Instituto de Astrofísica de Canarias, La Laguna, Spain \\
$^{10}$Departamento de Astrofísica, Universidad de La Laguna, La Laguna, Spain \\
$^{11}$Astronomy Centre, Department of Physics and Astronomy, University of Sussex, Brighton BN1 9QH, UK \\
$^{12}$International Space Science Institute (ISSI), Hallerstrasse 6, CH-3012 Bern, Switzerland}

\date{Accepted XXX. Received YYY; in original form ZZZ}

\pubyear{2021}

\begin{document}
\label{firstpage}
\pagerange{\pageref{firstpage}--\pageref{lastpage}}

\maketitle

\begin{abstract}
{
We study the interstellar medium (ISM) properties as a function of the molecular gas size for 77 infrared-selected galaxies at $z \sim 1.3$, having stellar masses $10^{9.4} \lesssim M_{\star} \lesssim 10^{12.0}$ M$_{\odot}$ and star formation rates $12 \lesssim {\rm SFR}_{\rm FIR} \lesssim1000$ M$_{\odot}$ yr$^{-1}$.}
Molecular gas sizes are measured on ALMA images that combine CO(2-1), CO(5-4) and underlying continuum observations, and include CO(4-3), CO(7-6)+[CI]($^3 P_2-^3P_1$), [CI]($^3 P_1-^3P_0$) observations for a subset of the sample.
The $\gtrsim 46 \%$ of our galaxies have a compact molecular gas reservoir, and lie below the optical disks mass-size relation.
Compact galaxies on and above the main sequence have higher CO excitation and star formation efficiency than galaxies with extended molecular gas reservoirs, as traced by CO(5-4)/CO(2-1) and CO(2-1)/$L_{\rm IR, SF}$ ratios. 
Average CO+[CI] spectral line energy distributions indicate higher excitation in compacts relative to extended sources.
{Using CO(2-1) and dust masses as molecular gas mass tracers}, and conversion factors tailored to their ISM conditions, we measure lower gas fractions in compact main-sequence galaxies compared to extended sources.
We suggest that the sub-millimetre compactness, defined as the ratio between the molecular gas and the stellar size, is an unavoidable information to be used with the main sequence offset to describe the ISM properties of galaxies, at least above $M_{\star} \geqslant 10^{10.6}$ M$_{\odot}$, where our observations fully probe the main sequence scatter.
Our results are consistent with mergers driving the gas in the nuclear regions, enhancing the CO excitation and star formation efficiency. 
Compact main-sequence galaxies are consistent with being an early post-starburst population following a merger-driven starburst episode, stressing the important role of mergers in the evolution of massive galaxies. 
\end{abstract}

\begin{keywords}
galaxies: evolution -- galaxies: star formation -- galaxies: ISM
\end{keywords}


\section{Introduction}

The majority of star-forming galaxies (SFGs) are observed to follow a correlation in the stellar mass ($M_{\star}$) versus star formation rate (SFR) plane. 
In the mainstream scenario, the existence of this so-called Main Sequence \citep[MS, ][to mention a few]{Noeske07, Elbaz07, Daddi07, Wuyts11, Kashino13, Rodighiero14, Sargent14, Speagle14, Whitaker, Pannella14, RenziniPeng, Schreiber15} and its tight scatter \citep[$\sim 0.3 \ dex$, e.g.][]{Rodighiero11, Sargent12, Schreiber15} are interpreted as evidence that star formation in most galaxies is a fairly ordered process. In particular, galaxies on the main sequence appear to be secularly evolving \citep{Daddi10, Genzel15} clumpy disks \citep{FS09} growing inside out \citep{Nelson12, Nelson16b}. 
Galaxies above the main sequence are believed to undergo a starburst mode of star formation associated with stochastic processes like major mergers \citep{Kartaltepe12, Hung13, Silverman15a, Silverman18, Cibinel19}. These so-called starbursts only represent a small percentage of the star-forming galaxy population and seem to have a minor impact on the cosmic star formation history \citep{Rodighiero11, Schreiber15}.

On the other hand, recent studies at far-infrared (FIR) /sub-millimetre (sub-mm) wavelengths of galaxies at $z \textgreater 1$ are revealing a somewhat different picture.
In particular, the properties of galaxies at long wavelengths seem to be poorly correlated with their main sequence position, in seeming contrast with the current interpretation of the main sequence.
For example, the spatial extent of the molecular gas reservoir does not show correlations with the offset from the main sequence ($\Delta$MS = SFR/SFR$_{\rm MS}$) or, almost equivalently, with the specific SFR \citep[sSFR, ][]{Puglisi19}. 
The star formation efficiency (SFE) and molecular gas excitation are also somewhat weakly correlated with the main sequence offset \citep[e.g.][and references therein]{Tacconi20} and little difference is observed when comparing average CO spectral line energy distributions (SLEDs) of galaxies on and above the main sequence \citep{Valentino20}.
This echoes studies of the infrared spectral energy distribution (SED) showing that there is only a little difference between the far-infrared SED shape of main-sequence and starburst galaxies at $z \sim 2$ \citep[][see also \citealt{Burnham21}]{Bethermin15}. 

In addition to an overall weak correlation between the interstellar medium (ISM) properties and the main sequence offset, it has been recently discovered that a large fraction of massive ($M_{\star} \gtrsim 10^{10.5}$ M$_{\odot}$) star-forming galaxies within the main sequence have compact molecular gas reservoirs embedded in a more extended stellar structure \citep{Tadaki17, Elbaz18, JimenezAndrade19, JimenezAndrade21, Puglisi19, Franco20, Tadaki20, GomezGuijarro21}. 
These ``sub-millimetre compact'' main-sequence galaxies have short depletion time-scales, as derived from dust-continuum measurements \citep{Elbaz18, Franco20}, and single-object studies show that these objects have highly excited ISM \citep{Popping17}. These properties are consistent with the properties expected from merger-driven starbursts \citep{MihosHernquist96, Papadopoulos12, Hodge16}.
However, these sources display "main sequence" levels of star-forming activity. 
Furthermore, ALMA observations reveal that some of these compact, star-forming cores are rotating and thus possibly disks \citep{Talia18, Tadaki17, Tadaki17b, Kaasinen20}.
Understanding the formation mechanism of such objects is important since compact star-forming galaxies might represent a key phase in the passivization mechanisms of galaxies \citep{Barro13, Barro14, Elbaz18, Puglisi19, GomezGuijarro19}.

Are galaxies within the main sequence uniquely associated with a secular mode of star-formation? 
Do we really observe two star-forming galaxy populations with different properties (extended disks in secular evolution and merger-driven starbursts) or do we rather observe a broad variety of properties in massive star-forming galaxies at $z \geqslant 1$? 
What are the mechanisms responsible for the formation of massive galaxies with a compact molecular gas reservoir within the main sequence?

In this paper we aim to address the above questions by taking advantage of the unique coverage provided by our Atacama Large Millimeter Array (ALMA) survey presented in \cite{Valentino20}.
This multi-cycle campaign has allowed us to obtain several carbon monoxide (CO) and neutral atomic carbon ([CI]) line detections enabling a characterization of the molecular gas excitation conditions in a statistical sample of far-infrared selected star-forming galaxies at $z \sim 1.3$.
While CO scaling relations as a function of the main sequence position have been extensively discussed in \cite{Valentino20}, here we investigate the dependence of ISM properties on the spatial extent of the molecular gas reservoir.
We also take advantage of the availability of different observables to estimate the gas content of our sample from multiple tracers (namely the CO(2-1) luminosity, $L'_{\rm CO(2-1)}$, and the dust mass, $M_{\rm dust}$), building on the study of the ISM conditions to physically motivate the choice of the conversion factors required to convert the observables into gas masses.

The paper is organised as follows. 
In Section \ref{sec:sample} we report information on the measurements that are essential for the analysis presented in this paper. These include details on the galaxy size measurements, galaxy integrated properties 
and our sources classification criterion. 
In Sect. \ref{sec:Results}, we present the results on the ISM properties of the sample.
We discuss these results in Sect. \ref{sec:discussion}. Finally, we summarize the main findings of this paper in Sec. \ref{sec:summary}.
Throughout this paper we use a \cite{Chabrier} initial mass function and a standard $\Lambda$CDM cosmology ($H_{0} = 70$ km s$^{-1}$ Mpc$^{-1}$, $\Omega_{\rm m} = 0.3$, $\Omega_{\Lambda} = 0.7$).

\section{The sample}
\label{sec:sample}

In this work we study the molecular gas properties of a statistical sample of {123 galaxies at $1.1 \leqslant z \leqslant 1.7$ in the COSMOS field \citep{scoville07}} selected in the far-infrared to have {a total infrared luminosity} $L_{\rm IR} \gtrsim 10^{12} \ L_{\odot}$.
These galaxies were observed with ALMA in Band 6 with an average circularized beam of $\sim 0.7''$ to detect the CO(5-4) emission (Program-ID 2015.1.00260.S, PI E. Daddi). For 75 of the 123 galaxies detected at high significance we obtained follow-up observations of the CO(2-1) transition in ALMA Band 3 with an average circularized beam of $\sim 1.5''$ (Program-ID 2016.1.00171.S, PI E. Daddi). 
We also acquired CO(7-6)+[CI]($^3 P_2 - ^3P_1$) observations from a follow-up program targeting 15 of the 123 galaxies in the main program (Program-ID 2019.1.01702.S, PI F. Valentino). 
Finally, we complemented these observations with available CO(4-3) and [CI]($^3 P_1 - ^3P_0$) observations from an independent campaign (Program-IDs 2016.1.01040.S and 2018.1.00635.S, PI: F. Valentino) for 15 out of 123 galaxies in the primary sample.
{For a detailed description of the selection criteria, the full ALMA data-set, the data reduction, and measurements of emission line fluxes and flux upper limits we refer the reader to \cite{Valentino20}.
Here we report details of the ALMA size measurements that are used in this paper (Sect. 2.1). We also provide details of the measurements of stellar masses, near-infrared sizes, star formation rates, dust masses and the intensity of the radiation field for these galaxies (Sect. 2.2). Finally, we present the source classification criterion adopted in our analysis, which is based on the size of the molecular gas component relative to the position of the source with respect to the optical mass-size relation (Sect. 2.3).} 

\begin{table}
\begin{center}
\caption{ {Summary of ALMA size measurements and CO emission lines statistics for our sample.}}
\label{tab:Summary}
\begin{tabular}{lccc} 
\hline
\hline
		Quantity & $N_{\rm obj}$ \\
		\hline
		ALMA size & 39/123 \\
		ALMA size upper limit & 50/123\\
		ALMA size or size upper limit \& M$_{\star}$ \& SFR$_{\rm FIR}^{\rm (a)}$ & 82/123 \\
		ALMA size or size upper limit \& M$_{\star}$ \& SFR$_{\rm FIR}$ \& $f_{\rm AGN} \textless 80 \%$ & 77/123 \\
		This work with CO(5-4)$^{\rm (b)}$ & 46/77 \\
		This work with CO(2-1)$^{\rm (c)}$ & 33/77 \\
		This work with CO(5-4) and CO(2-1)$^{\rm (d)}$ & 31/77 \\
\hline
\end{tabular}
\end{center}
{\bf Notes.} 
 {
\\$^{\rm (a)}$We exclude 7 out of the 89 galaxies with a ALMA size or size upper limit from our analysis. Four of these galaxies have no far-infrared counterpart in the \citet{Jin18} catalogue. Three objects are AGN and have no measurements of the stellar mass. 
\\$^{\rm (b)}$Galaxies with CO(5-4) detections or robust upper limits (corresponding to Flag$_{\rm CO(5-4)} \geqslant 0.5$ in the \citealt{Valentino20} catalogue). 
\\$^{\rm (c)}$Galaxies with CO(2-1) detection or robust upper limits (Flag$_{\rm CO(2-1)} \geqslant 0.5$). 
\\$^{\rm (d)}$Galaxies with CO(5-4) and CO(2-1) detection or robust upper limits.}
\end{table}

\subsection{ALMA size measurements}
\label{subsect:ALMA_data}

In this work we measure sizes from ALMA images following the methodology detailed in \cite{Puglisi19}, \cite{Valentino20} and references therein. 
As already discussed in these papers, the source sizes are measured by combining all the available ALMA observations in the {\it uv} plane, allowing for an arbitrary re-normalization of the signal for all tracers. 
These include the CO(2-1), CO(4-3), CO(5-4), CO(7-6), [CI] and underlying continua. 
This is to maximize the accuracy of the size measurements and the sample statistics. 
To determine the best-fit size and its 1$\sigma$ uncertainty we compare the {\it uv} distance vs. amplitude distribution to circular Gaussian models. 
{Our ALMA size measurement method is exemplified in Figure 1 of \citet{Puglisi19}.}
The best-fit size of each galaxy $R_{\rm eff}$ is defined as half of the full width half maximum (FWHM) of the best-fit circular Gaussian model ($R_{\rm eff} = {\rm FWHM}/2$). This corresponds to the half-light radius for a 2D Gaussian profile.
We quantify the probability $P_{\rm unres}$ of each galaxy to be unresolved by comparing the best-fit circular Gaussian $\chi^2$ to the $\chi^2$ for a point source. We consider a galaxy to be resolved when  $P_{\rm unres} \leqslant 10\%$ with this threshold corresponding to 0.5 galaxies expected to be spuriously resolved in in our total sample of 123 sources.
Recent results have suggested that high-redshift galaxies have far-infrared profiles consistent with exponential disks  \citep[][and references therein]{HodgeDaCunha20}. 
However, most of these studies rely on higher resolution ALMA observations \citep[$\lesssim 0.25 ''$, e.g.,][]{Hodge16}.
A circular Gaussian profile provides a good fit to galaxies in our sample given the signal-to-noise ratio (SNR) and beam of our observations. 
In addition, fitting an exponential profile to our data gives consistent size measurements within the errors. 

We measure sizes or size upper limits for 89 out of the 123 sources observed in our ALMA program.
Of these, 39 galaxies are resolved { with an average size/size-error ratio of $\sim$ 10. We measure 1$\sigma$ size upper limits  for 50 galaxies in the sample.
Finally, we cannot measure sizes for 34 sources in the sample. 
Of these, 18 objects lack a line detection and have uncertain emission line upper limits as a result of a poor optical redshift. 
In addition, 15 sources have uncertain upper limits on the CO(5-4) transition based on high-quality optical redshifts only and no continuum detection. 
Finally, we discard one object because of an unreliable size measurement ($P_{\rm unres} = 29 \%$) due to low SNR detections in CO and continuum.
We include in Table \ref{tab:Summary} a summary of the ALMA size measurements statistics.
We note that size measurements are updated with respect to those presented in \cite{Puglisi19}. 
This is because in this work we have included additional CO(7-6)+[CI]($^3 P_2 - ^3P_1$) and CO(4-3) and [CI]($^3 P_1 - ^3P_0$) observations in our average size estimate. This results in 6 additional sources with a measured size. Furthermore, this results in an average size/size-error that is $\sim 1.9 ~ \times$ higher than in our previous analysis \citep[average size/size-error of $\sim$ 5.3 in][]{Puglisi19}.}

{
We extensively tested the reliability of our procedure to measure sizes using Monte-Carlo simulations, as discussed in \citet{Puglisi19} and \citet{Coogan18}. 
In brief, we created 1000 mock realizations of our data-sets by perturbing the best-fit models within the measured uncertainties in the {\it uv}-amplitudes plane, assuming a Gaussian distribution for the noise. We applied our procedure to measure sizes for each of these 1000 synthetic realizations. 
We performed this test on galaxies covering a broad range in sizes and flux SNR corresponding to that covered by our measurements ($R_{\rm eff} = 0.1'' - 1''$, flux SNR $= 5 - 10$).
We find no residual systematic uncertainties in the recovered sizes within 5\%. 
For galaxies that are significantly larger than our average beam (i.e. above $\sim 1.2''$), we find a small systematic uncertainties with sizes from simulations that are $\sim$10\% smaller than input sizes. This systematics does not depend on the SNR of the source but it seems to be rather related to the intrinsic size of the object. 
However, we find that the largest discrepancy is $\sim$10\% for the most extended sources, thus not affecting our results.
We also find that the average size errorbars are consistent with the 1$\sigma$ dispersion of simulated size measurement, validating the robustness of our 1$\sigma$ size error-bars. 
The robustness of the 1$\sigma$ size errorbars is also indicated by the $\chi^2$ distribution and the corresponding probability of the individual fits in the {\it uv} plane, which is consistent with pure noise.}

\subsubsection{{Potential biases in our ALMA size estimates}}
\label{subsub:ALMA_biases}

{
Our method that combines several molecular gas tracers to measure sizes might be sensitive to different components of the ISM depending on the excitation properties of each individual galaxy. 
However, we verified that in all cases (but for three single notable exceptions) where multiple tracers were combined to a single size measurement, the tracers’ signal versus amplitude trends agree among themselves. This suggests that for our sample there are no significant size variations between low-to-high J CO tracers and underlying dust continuum within the beam of our observations \citep[$\sim 0.7'' - 1.5''$, see Figure 1 in][]{Puglisi19}. 
Furthermore,we find a 1:1 correlation between the ALMA sizes and the sizes of CO(5-4) or CO(4-3) and/or the underlying dust continuum ($\lambda_{\rm rest} \sim 520 - 650 \ \mu$m), suggesting that ALMA sizes are mostly driven by the high-J CO transitions and/or the underlying dust continuum.
When comparing CO(5-4) or CO(4-3) and 1 mm dust continuum sizes, we find that the scatter among the two independent size measurements is consistent with pure noise, leaving no space for additional systematics. 
Finally, for both extended and compact galaxies with robust CO(5-4) sizes (i.e. size/size error $\geqslant$ 3), we have verified that the CO(5-4) size is consistent within the errors with the average size obtained as the weighted average of the individual size tracers and excluding the CO(5-4).
These tests justify our method to combine different molecular gas size tracers and suggests that our ALMA sizes are representative of the typical extension of the molecular gas and cold dust in our galaxies.
Future ALMA observations with higher spatial resolution in multiple tracers will help us to investigate the presence of size variations as a function of the molecular gas tracer for this sample.}

{Another potential source of bias in our size measurements might be associated to the presence of an active galactic nucleus (AGN), since we detect AGN features in the $\sim 40 \%$ of our sample \citep[][see also Sect. \ref{nir}]{Valentino20}. However, our ALMA sizes are mostly derived from the CO(5-4) emission and we do not expect a significant contribution of X-ray dominated regions (XDR) at these intermediate CO transitions (see also discussion in Sect. \ref{subsec:Excitation}). Even when size measurements are driven by the continuum underlying the CO(5-4) emission, we expect the AGN to provide a marginal contribution to the size. In fact, our decomposition of the far-infrared SED (see below) shows a marginal AGN contribution at $\lambda_{\rm rest} \sim 500 \mu$m. 
This is in line with other studies showing that there is currently no evidence of the AGN driving the FIR continuum sizes of galaxies \citep[e.g.][]{Chen20}. 
}

\subsection{Galaxy integrated properties and near-infrared sizes}
\label{nir}

We obtain stellar mass measurements for the galaxies in our sample from the \cite{Laigle16} catalogue. 
For the sources with a significant contribution from an AGN (see below), we repeat the fitting procedure of the UV-to-NIR photometry, following the approach detailed in \cite{Circosta18}. 

{As in \cite{Puglisi19}, for the 82 galaxies that have an ALMA size, a stellar mass and a star formation rate measurement (see Table \ref{tab:Summary}), we measure near-infrared rest-frame sizes ($\lambda_{\rm obs} \sim 1 \ \mu$m) from UltraVISTA K$_{\rm s}$-band images  \citep{McCracken12}. The UltraVISTA images allow us to estimate sizes with a $\sim 20 \%$ accuracy \citep[e.g.][]{Faisst17} and have an average seeing of $\sim 0.7''$, which is comparable to our best ALMA beam. 
For consistency with the approach adopted to measure sizes on the ALMA images, we measure near-infrared sizes by fitting circular Gaussian profiles with \textsc{galfit} \citep{Peng10_GALFIT} 
We note that in principle, Sersic models would be more appropriate for fitting the optical/near-infrared emission of galaxies. 
However, our sources are only marginally resolved in the UltraVISTA observations and circular Gaussian profiles provide an adequate fit to the K$_{\rm s}$-band images. Using an exponential disk or a Sersic profile with a free Sersic index does not significantly affect our conclusions.
As already discussed in \citet{Puglisi19}, 10 out of the 82 objects considered for the analysis show a prominent point-like emission in the K$_{\rm s}$-band imaging, likely associated with an AGN. These galaxies are not shown in the left panel of Figure \ref{fig:mass_size_selection}. This does not affect our results, since K$_{\rm s}$-band sizes are not considered for the quantitative analysis presented in this paper.

{The other physical properties that are relevant for this work are the total IR luminosity $L_{\rm IR}$ in the rest-frame wavelength range $\lambda = 8 - 1000 \ \mu$m, the dust mass and the intensity of the radiation field $\langle U \rangle$. These quantities are obtained by modelling the IR photometry from the \cite{Jin18} ``super-deblended'' catalogue, and by adding the dust continuum emission observed with ALMA in the $\lambda_{\rm obs} \in [0.8 - 3.2]$ mm wavelength range. 
The FIR modelling is performed by using the customized $\chi^2$ minimisation tool \textsc{Stardust} \citep{Kokorev21}\footnote{\url{https://github.com/VasilyKokorev/stardust}} and it is based on the \citet{Magdis12} approach which uses \citet{DraineLi07} dust models and a mid-infrared AGN torus component from \citet{Mullaney11}. The total infrared luminosity obtained from the fitting thus consists of a component associated with star formation ($L_{\rm IR, SF}$), and one component arising from the dusty torus ($L_{\rm IR, AGN}$). 
This allows us to estimate the fraction of the total IR luminosity associated with the dusty torus as $f_{\rm AGN} = L_{\rm IR, AGN}/L_{\rm IR}$. Following \citet{Valentino20}, we consider that an AGN component is reliably detected when $f_{\rm AGN} + 1\sigma_{f_{\rm AGN}} \geqslant 20 \%$, and we classify galaxies as AGN-dominated when $f_{\rm AGN}  \geqslant 80 \%$.
We finally measure star formation rates SFR$_{\rm FIR}$ from $L_{\rm IR, SF}$ by using the \cite{Kenni98} calibration rescaled to a \cite{Chabrier} IMF by a factor 1.7.
For more details about the modelling procedure we refer the reader to Sect. 3.3 of \cite{Valentino20} and references therein.}

\subsection{Sub-millimetre compactness}
\label{subsec:compactness}

In Figure \ref{fig:mass_size_selection} we show an updated version of the $M_{\star}$-Size plane presented in \cite{Puglisi19}.
In the left panel we show the $K_{\rm s}$-band sizes, roughly tracing the size of the stellar mass distribution at $z \sim 1.25$, corresponding to the average redshift of our sample. This plot shows that nearly all the galaxies in our sample are located within the scatter of the $M_{\star}$-Size relation for $z \sim 1.25$ star-forming disks \citep[or late type galaxies, LTG, ][blue line in Figure \ref{fig:mass_size_selection}]{vanDerWel14}. 
In the right panel of Figure \ref{fig:mass_size_selection} we show the ALMA sizes of our sources. From this plot we see that the ALMA measurements  are skewed towards small size values.

{
We define the sub-millimetre compactness (or simply compactness) as:
\begin{equation}
 C_{\rm gas} = R_{\rm eff, LTG} / R_{\rm eff, ALMA}, 
\label{compactness}
\end{equation}
where $R_{\rm eff, LTG}$ is the size measured from the \citet{vanDerWel14} LTG relation at the stellar mass of the galaxy, and $R_{\rm eff, ALMA}$ is the ALMA size of the source. 
We then compute the $\chi^2$ of the compactness distribution for the 82 galaxies with an ALMA size (or size upper limit), a measurement of the stellar mass and of the star formation rate  \citep[see Eqn. 1 in ][]{Puglisi19}.}
We find that 34\% of the sample has an ALMA size/size upper limit consistent with the \cite{vanDerWel14} mass-size relation for star-forming disks. 
These galaxies have similar $K_{\rm s}$-band and ALMA sizes. {We identify these galaxies as ``extended''.}
Instead, 46 \% of the galaxies have molecular gas sizes that are $\geqslant 1 \sigma$ more compact relative to the stars as expected from the disks mass-size relation. These galaxies are highlighted in red in both panels of Figure \ref{fig:mass_size_selection}. Through the paper we will refer to these galaxies as ``compacts''.  These compact sources are on average $3.3 \times$ smaller in ALMA than in the $K_{\rm s}$-band, on average, consistent with the measurements from our previous study \citep{Puglisi19}.
{Finally, we find that 20\% of the sources have ALMA size upper limits within the LTG relation (black upper limits in the right panel of Figure \ref{fig:mass_size_selection}). We dub those objects as ``ambiguous'' since we cannot place robust constraints on their compactness. We will consider these objects as a separate category through the paper.} 

While the stellar sizes of our sample are by and large consistent with those of typical $z \sim 1.25$ star-forming galaxies, the left panel of Figure \ref{fig:mass_size_selection} suggests that compact galaxies have slightly smaller $K_{\rm s}$-band sizes than extended sources. 
To quantify this effect, we perform a two-sided Kolmogorov-Smirnov (KS) test after dividing the size inferred from the \cite{vanDerWel14} LTG relation at the stellar mass of the source by the $K_{\rm s}$-band size. We consider only extended and compact galaxies for this test.
We find  median size ratios of $R_{\rm eff, LTG, Extended} / R_{\rm eff, K_{\rm s}-band} = 1.03 \pm 0.38$ and R$_{\rm eff, LTG, Compacts} / R_{\rm eff, K_{\rm s}-band} = 1.23 \pm 0.43$ for extended and compact galaxies, respectively. 
That is, compact galaxies have slightly smaller $K_{\rm s}$-band sizes than extended sources.
However, we cannot reject the null hypothesis at the $\sim 10\%$ level ($p-value = 0.13$). 
This suggests that the $K_{\rm s}$-band size distribution of extended and compact sources in our sample is slightly but not substantially different, as expected if compact galaxies are caught in the process of building up a dense stellar core (see also discussion in Sect. \ref{subsec:compacts}).
Recently, \citet{Popping21} has challenged the idea that the dust-continuum size is more compact than the stellar half-mass radius in $z \geqslant 1$ galaxies, due to strong dust attenuation gradients affecting H-band observations. This however corresponds to an observed-frame wavelength of $\sim 0.5 \mu$m at $z \sim 2$. Here we use $K_{\rm s}$-band observations (corresponding to $\sim 1 \mu$m at $z \sim 1.25$) to sample the stellar half-mass radius and we expect negligible dust attenuation effects in this case. In fact, the emission at these wavelengths has been observed to align with the CO and radio emission even for the most obscured starbursts such as GN20 \citep[e.g.][]{Tan14}.

We highlight with crosses in Figure \ref{fig:mass_size_selection} the galaxies with a reliably-detected AGN contribution to the far-infrared SED. Five of the sources with an ALMA size/size upper limit have $f_{\rm AGN} \geqslant 80 \%$ (red crosses in Figure \ref{fig:mass_size_selection}).  
{We exclude these galaxies from the analysis presented in Sect. \ref{sec:Results} since we cannot derive robust SFR constraints and XDRs might contribute significantly to the CO line ratios.}
We find AGN signatures in $25 \pm 11 \%$ of the extended galaxies.
Instead, we find AGN signatures in $45 \pm 13 \%$ of the compact sources. 
This suggests that compact galaxies have an enhanced AGN fraction.
To test whether AGN are associated to more compact galaxies, we apply a log-rank test to the compactness distribution of galaxies with and without an AGN contribution to the far-infrared SED, accounting for the presence of compactness lower limits. 
We find a 94\% probability that the two distributions are different ($p-value = 0.06$). We note however that the AGN host population contains a larger fraction of compactness lower limits (62\% versus 47\% in galaxies without an AGN) and such lower limits are mostly associated with large compactness values. 
This implies that the AGN hosts' compactness distribution is not well constrained at high compactness values and the log-rank test results might be unreliable. On the other hand, this may suggest that AGN are associated with more compact galaxies for which we have stringent size upper limits. 
We thus conclude that our observations provide marginal evidence that AGN are more likely associated to galaxies with a compact molecular gas reservoir, similarly to that suggested by previous analyses (\citealt{Elbaz18}, \citealt{Puglisi19}, \citealt{Lamperti21}, see also \citealt{Barro14} for optically-compact star-forming galaxies). This might indicate that the mechanisms responsible for fuelling the molecular gas to the nuclear regions is also efficient in feeding the AGN. 

\begin{figure}
\includegraphics[width=\columnwidth]{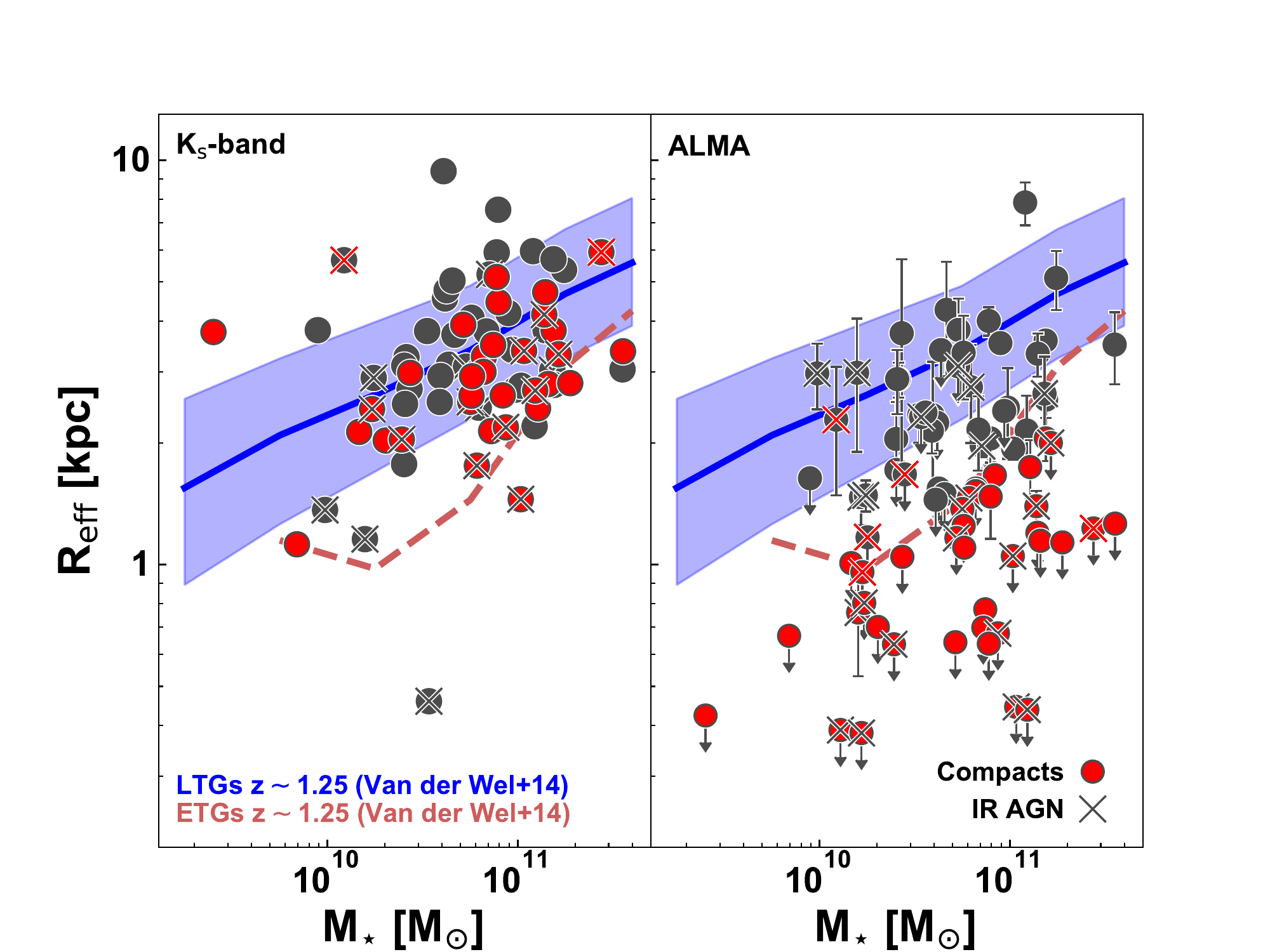}
\caption{
The left panel shows the correlation between the stellar mass and the K$_{\rm s}$-band size of our sample, roughly tracing the stellar mass size at $z \sim 1.25$.
The right panel shows the correlation between the stellar mass and the ALMA size. 
The blue line and blue shaded area indicate the \protect\cite{vanDerWel14} relation and scatter for star-forming disks at $z \sim 1.25$.
The dark-red dashed line represents the \protect\cite{vanDerWel14} relation for early type galaxies and it is shown here as a reference. 
In both panels, red filled circles highlight the compact galaxies in our sample. We highlight with crosses the AGN. Sources with a strong AGN contamination to the far-infrared SED ($f_{\rm AGN} \geqslant 0.8$) are highlighted with a red cross and are excluded from the rest of the analysis.
{In the right panel, errorbars indicate the 1$\sigma$ uncertainty associated with our ALMA size measurements. Down-facing arrows highlight 1$\sigma$ ALMA size upper limits.}
}
\label{fig:mass_size_selection}
\end{figure}

{In Figure \ref{fig:main_sequence} we report the star formation rate as a function of the stellar mass for our sample.
We normalise the star formation rate of each galaxy to the SFR of the main sequence at the average redshift of the sample, to account for the redshift evolution of the main sequence normalisation \citep{Sargent12}.
Similarly to our previous analysis, we identify as ``main sequence galaxies'' sources with $\Delta$MS$\textless 3.5$, whereas we classify as ``off main sequence'' or starbursts galaxies with $\Delta$MS$\geqslant 3.5$.
The color code in Figure \ref{fig:main_sequence} provides information on the sources' compactness. In particular, blue circles represent galaxies with an ALMA size consistent with the LTG relation, red filled circles indicate compact sources and grey filled circles highlight ambiguous galaxies. Finally, small black dots display galaxies for which we cannot measure a size from our ALMA observations (see Tab. \ref{tab:Summary} and Sect. \ref{subsect:ALMA_data}).}
We do not find a clear correlation between the compactness and the main sequence position, similarly to what reported in \citet{Puglisi19}.
Galaxies with a compact molecular gas reservoir make up a significant fraction of the main sequence population above $\sim 5 \times 10^{10}$ M$_{\odot}$, in line with previous studies \citep[e.g.,][]{Tadaki17b, Tadaki20, Elbaz18, Puglisi19, Franco20}.
{Figure \ref{fig:main_sequence} also shows that, while spanning a broad range of stellar masses and star formation rate in the main sequence plane, our observations consist of 
``upper main sequence'' galaxies, that is, strongly star-forming galaxies above the main sequence and galaxies that probe the high specific SFR stripe of the main sequence scatter.}
This is a result of the far-infrared selection, which corresponds to a horizontal cut in the main sequence plane \citep{Rodighiero14}.

\begin{figure}
\centering
\includegraphics[width=\columnwidth]{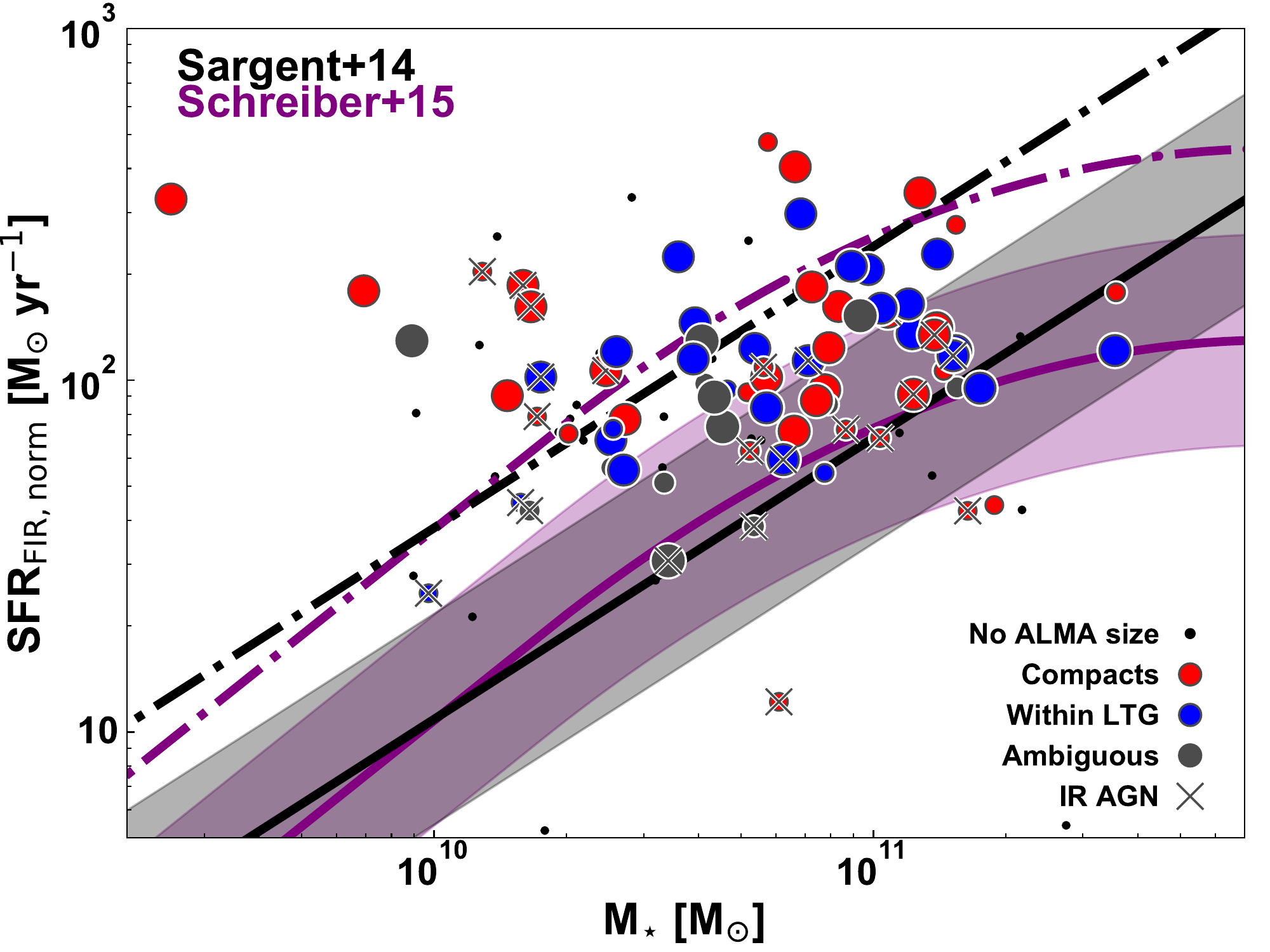}
\caption{Star formation rate as a function of stellar mass for our sample.
Solid lines indicate the main sequence locus and the shaded areas highlight the 0.3 {\it dex} main sequence scatter. 
The dash-dotted lines represent the 3.5$\times$ main sequence threshold above which we classify galaxies as starbursts. 
The black curves define these {\it loci} according to the parametrisation of \citet{Sargent14} at $z \sim 1.25$. 
The violet curves correspond to the \citet{Schreiber15} parametrisation at $z \sim 1.25$.
Blue and red circles highlight extended and compact galaxies respectively. Grey circles indicate ambiguous sources. 
{Larger symbols indicate galaxies with CO(5-4) or CO(2-1) observations that are considered for the analysis presented in Sect. \ref{sec:Results}. We highlight with crosses the AGN.
Black dots indicate galaxies without ALMA size measurements.}}
\label{fig:main_sequence}
\end{figure}

\section{Results}
\label{sec:Results}

In the following we study the molecular gas properties of our sample as a function of the compactness and the main sequence offset.
For consistency with our previous studies, we quantify the main sequence offset using the main sequence parametrization from \cite{Sargent14}.
{
We also checked that using the widely-adopted parametrisation from \citet{Schreiber15}, which accounts for the bending of the main sequence at high stellar masses, does not affect the main sequence position of our sample.}
For this analysis, unless stated otherwise, we consider only galaxies with reliable CO fluxes or robust CO flux upper limits. The latter corresponds to reliable upper limits on the line flux, given the presence of alternative sub-mm lines confirming the redshift obtained from the ALMA spectra (z$_{\rm spec, sub-mm}$).
We then compute the $L'_{\rm CO}$ luminosities using the flux and $z_{\rm spec, sub-mm}$ measurements from our public catalogue \citep{Valentino20}.

\subsection{Gas excitation}
\label{subsec:Excitation}

\subsubsection{The $L'_{\rm CO(5-4)}/L'_{\rm CO(2-1)}$ ratio}
\label{subsubsect:R52}

In Figure \ref{fig:CO_ratios_MS_comp} we show the $R_{52} = L'_{\rm CO(5-4)}/L'_{\rm CO(2-1)}$ ratio as a function of $\Delta$MS (left panel) and $C_{\rm gas}$ (right panel). 
The $R_{52}$ ratio is a proxy for the CO excitation and this plot allows us to infer the excitation properties of each galaxy as a function of their structural and star formation rate properties. 

The left panel of Figure \ref{fig:CO_ratios_MS_comp} shows that there is a substantial number of compact galaxies within a factor of $\pm 3.5$ around the main sequence. 
These compact galaxies within the main sequence have an enhanced $R_{52}$ ratio with respect to their extended counterparts.
The $R_{52}$ ratio of compact main-sequence galaxies is instead similar to that of galaxies above the main sequence.
{This plot suggests that compact and ambiguous galaxies within the main sequence contribute to the scatter observed in the $R_{52}$-$\Delta$MS relation reported in \citet{Valentino20} and obtained using the same sample considered in this work (green dotted line in the left panel of Figure \ref{fig:CO_ratios_MS_comp}).
To quantify the contribution of these galaxies to the scatter observed in the \citet{Valentino20} relation, we fit $R_{52}$ as a function of $\Delta$MS excluding compact and ambiguous galaxies with $\Delta$MS $\leq \pm 3.5$. We also exclude the strongest outlier above the main sequence, for consistency with our previous analysis.}
We apply a Bayesian regression analysis in the log-log space using the Python version of the \textsc{linmix$\_$err.pro} package \citep{Kelly07}. The results of this fit are listed in Table \ref{tab:linmix_results}. The best-fit line is shown as a dark red solid line in the left panel of Figure \ref{fig:CO_ratios_MS_comp}.
{The slope is $\beta = 0.45 \pm 0.14$ and this is $\sim 2 \times$ steeper at $\sim 1.3 \sigma$ than the trend reported in \cite{Valentino20} when considering all galaxies regardless of their compactness.}
The intrinsic scatter of the best-fit relation ($\sigma_{\rm int} = 0.11 \pm 0.03$) is slightly reduced. 
Finally, the correlation between $R_{52}$ and $\Delta$MS strengthens when excluding compact and ambiguous main sequence galaxies from the fit.
{This suggests that the presence of compact/ambiguous galaxies (i.e. non typical disks) within the main sequence blurs the correlation between $R_{52}$ and $\Delta$MS.}
We note however that the compactness classification criterion applied in this paper reduces the source statistics with respect to the analysis presented in \citet{Valentino20}. 

The right panel of Figure \ref{fig:CO_ratios_MS_comp} shows a trend of increasing $R_{52}$ as a function of the compactness.
To quantify the correlation between $R_{52}$ and $C_{\rm gas}$ we apply a Bayesian regression analysis in the log-log space similarly to that described above. The results of this fit are listed in Table \ref{tab:linmix_results}. We show the best-fit line as a red line in the right panel of Fig. \ref{fig:CO_ratios_MS_comp}.
The slope of the $C_{\rm gas}$-$R_{52}$ trend is $\beta = 0.40 \pm 0.30$ and this is similar to the slope of the $\Delta$MS-$R_{52}$ best-fit relation discussed above. However, the slope of the $C_{\rm gas}$-$R_{52}$ relation is poorly constrained.
The $C_{\rm gas}$-$R_{52}$ correlation index is smaller than the $R_{52}$-$\Delta$MS correlation index obtained above.
We note however that our data only allow us to probe a limited compactness range due to an average $\sim 1 "$ beam. 
Observations with a smaller beam would be required to probe a wider range of sizes and explore in more details the relation between $R_{52}$ and the compactness.

\begin{table*}
\begin{center}
\caption{Scaling relations between CO properties, main sequence offset and compactness. }
\label{tab:linmix_results}
\begin{tabular}{lccccc} 
\hline
\hline
		Relation & Slope  & Intercept & Intrinsic scatter & Correlation & {$N_{\rm det}$,$N_{\rm lim}^{\rm (a)}$} \\
		$x, y$ & $\beta$  & $\alpha$ & $\sigma_{\rm int}$ & $\rho$ &  \\
\hline
\multicolumn{6}{c}{\multirow{1}{*}{Distance from the main sequence}} \\
\hline
		$\Delta$MS, $R_{52}^{\dagger}$ & 0.45 $\pm$ 0.14 & -0.83 $\pm$ 0.09  & 0.11 $\pm$ 0.03 &  0.78 & {19,4}\\
		$\Delta$MS, $L'_{\rm CO(2-1)}/L_{\rm IR, SF}^{\dagger}$ &  -0.42 $\pm$ 0.14 & -1.73 $\pm$ 0.09  &  0.22 $\pm$ 0.04 & -0.58 & {21,4} \\ 
\hline
\multicolumn{6}{c}{\multirow{1}{*}{Compactness}} \\
\hline
		$C_{\rm gas}$, $R_{52}$ & 0.40 $\pm$ 0.30 & -0.63 $\pm$ 0.10  & 0.10 $\pm$ 0.04 &  0.57 & {25,5}\\
		$C_{\rm gas}$, $L'_{\rm CO(2-1)}/L_{\rm IR, SF}$ & -0.56 $\pm$ 0.31 & -1.85 $\pm$ 0.11 & 0.23 $\pm$ 0.05 & -0.5 & {29,4} \\
		$C_{\rm gas}$, $\mu_{\rm gas, dust}$ &  -1.44 $\pm$ 0.5 &  0.05 $\pm$ 0.17  &  0.44 $\pm$ 0.06 &  -0.53 & {69,-} \\	
		\hline
\end{tabular}
\end{center}
\begin{flushleft}
{\bf Notes.} In this table we quote the mean of the best-fit parameters from \textsc{linmix} and corresponding 1$\sigma$ errorbars. 
\\$^{\dagger}$Excluding compacts and ambiguous within the main sequence.
{\\$^{\rm (a)}$ Number of sources with CO line ratio measurements and upper/lower limits used to infer the relations presented in this table. 
We note that the \textsc{linmix$\_$err.pro} package does not allow us to account for the presence of upper and lower limits at the same time. 
Therefore, the best-fit parameters for the $\Delta$MS-$R_{52}$ and $C_{\rm gas}$-$R_{52}$ relation reported here are derived considering $R_{52}$ detections only. We have however tested that including $R_{52}$ detections and upper limits or $R_{52}$ detections and lower limits we obtain fully consistent results. This is analogous to what we have performed in our previous analysis \citep{Valentino20}. }
\end{flushleft}
\end{table*}

\begin{figure}
\includegraphics[width=\columnwidth]{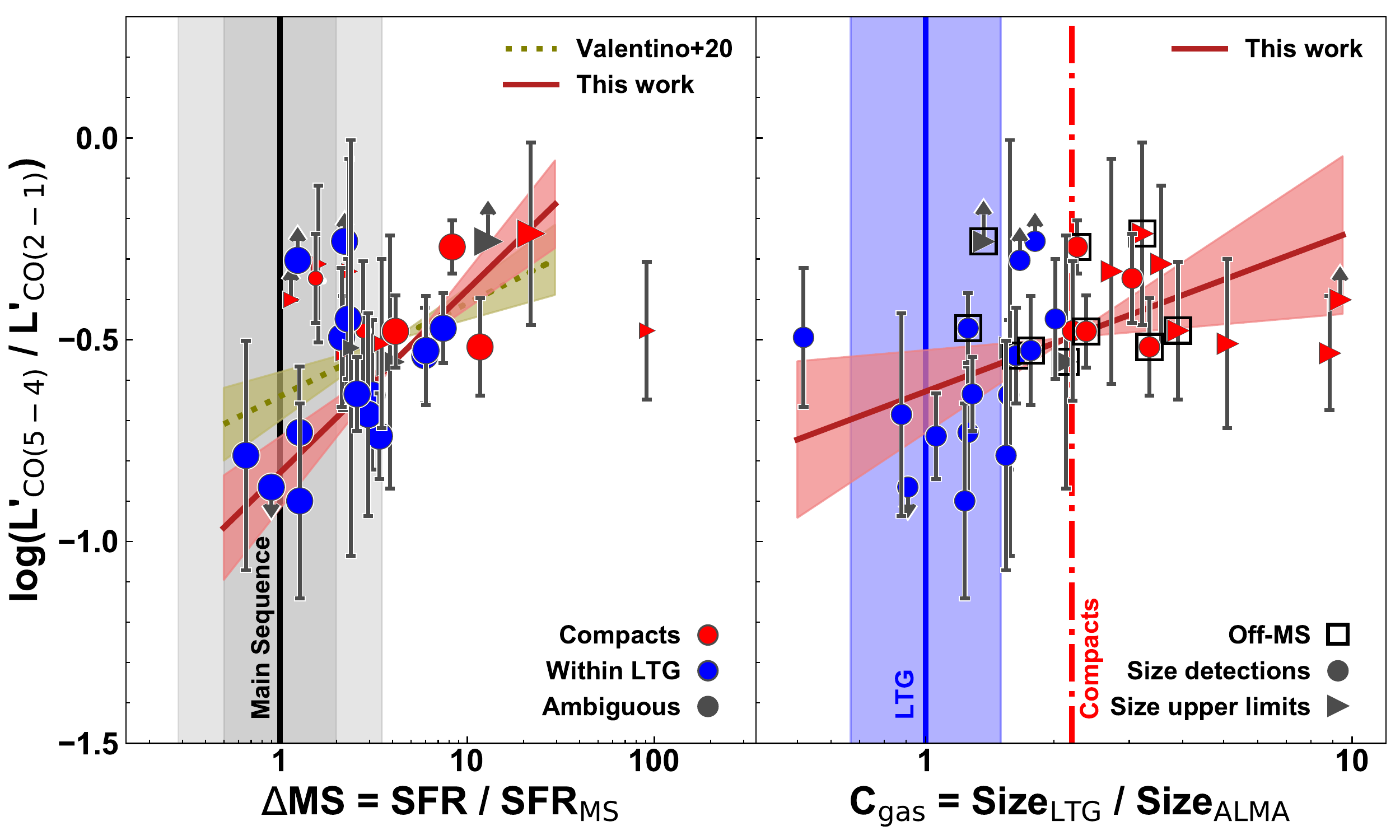}
\caption{ $R_{52}$ as a function of the main sequence offset ({\it left}) and the compactness ({\it right}). 
In the left panel, the black line, dark grey and grey shaded areas highlight the main sequence position, the $1 \sigma$ scatter and the $\pm 3.5 \times$ $\Delta$MS region respectively.
The green dotted line and shaded area indicate the $R_{52}$-$\Delta$MS trend and 1$\sigma$ confidence interval from\citet{Valentino20}. 
The dark red solid line and shaded area indicate the $R_{52}$-$\Delta$MS trend and 1$\sigma$ confidence interval obtained after excluding compact and ambiguous main-sequence galaxies from the fit.
{The data points considered for this fit are highlighted with larger symbols.}
In the right panel, the blue area shows the LTG relation and scatter at the average stellar mass of our sample.
The red dash-dotted line marks the threshold above which galaxies are classified as compacts. 
The dark red solid line and shaded area indicate the $R_{52}$-$C_{\rm gas}$ trend and 1$\sigma$ confidence interval.
Blue circles indicate extended ALMA galaxies. Red symbols highlight the compacts. Grey symbols represent ambiguous sources. 
{Right facing triangles indicate compactness upper limits.} 
In the right panel we highlight with open squares galaxies with $\Delta$MS $\geqslant 3.5$.
{Errorbars on $R_{52}$ are obtained by propagating the 1$\sigma$ uncertainty associated to the CO(5-4) and CO(2-1) flux measurements.
The typical 1$\sigma$ error on $\Delta$MS is 0.2 $\it dex$ accounting for observational uncertainties on SFR and $M_{\star}$. The 1$\sigma$ typical error on $C_{\rm gas}$ is 0.2 $\it dex$ considering observational uncertainties on $R_{\rm eff, ALMA}$.} }
\label{fig:CO_ratios_MS_comp}
\end{figure}

\subsubsection{{Average CO spectral line energy distributions as a function of the compactness}}

In the previous section we used the $R_{52}$ ratio to study the CO excitation as a function of the main sequence offset and the compactness in individual galaxies in our sample. 
{However, CO(4-3) and CO(7-6) observations available for a subset of our sources (see Tables \ref{tab:L_CO_comp} and \ref{tab:L_CO_comp_ms}) allow us to construct average CO spectral line energy distributions (SLEDs) for the three classes of sources identified in Sect. \ref{subsec:compactness}}. 
{We thus construct average CO SLEDs for the galaxies that have CO(2-1) and CO(5-4) emission line detections/upper limits, and a measurement of the compactness}.
We split those sources into different sub-samples according to their compactness and main sequence position and we compute average $L'_{\rm CO}$ luminosities using a survival analysis technique to account for the presence of upper limits \citep{KM}. 
We report in Tables \ref{tab:L_CO_comp} and \ref{tab:L_CO_comp_ms} the average CO luminosities of each sub-sample along with the detection statistics for each transition.
We then convert the average $L'_{\rm CO}$ luminosities into CO fluxes at $z \sim 1.25$, corresponding to the average redshift of the sample. 
{Errors on the average $L'_{\rm CO}$ luminosities and average $I_{\rm CO}$ fluxes quoted throughout this section correspond to the interquartile range from the distribution of individual measurements in each sub-sample.}

In Figure \ref{fig:CO_SLEDs}, we show the average SLEDs for extended, ambiguous and compact galaxies in our sample.
{Here we include as a reference the average CO SLEDs from \citet{Valentino20}, which uses the same sample considered in this work but classifies galaxies according to their main sequence offset.}
A comparison with other galaxy types and QSO SLEDs for this sample has already been performed in \citet{Valentino20} and we refer the reader to this paper for more details in this regard (see in particular their Sect. 4.4).
Here we find that the CO SLED of extended galaxies closely resembles that of main sequence galaxies at $z \sim 1.25$.
The CO excitation ladder of compact galaxies is consistent with that of the strongest main-sequence outliers, defined as galaxies with $\Delta$MS $\geqslant 7$.
Likewise, ambiguous galaxies have a CO SLED similar to that of strong starbursts, possibly suggesting that these sources are more compact than indicated by their loose ALMA size upper limits.
We note however that the CO lines statistics are significantly limited for this class of objects (see Table \ref{tab:L_CO_comp}) and future observations would be required to better understand their nature. Given the limited statistics, we do not investigate any further the CO SLED properties of the ambiguous galaxies sub-sample.

{One might expect that the different excitation properties of extended and compact galaxies might be due to a higher AGN contribution to the latter sample, since we have shown in Sect. \ref{subsec:compactness} that compact galaxies have a marginally higher AGN fraction.
However, we have explored the contribution of AGN to the observed CO line ratios and to the average CO SLEDs in our sample in a dedicated study \citep{Valentino21}. This analysis shows that there are no statistically significant differences between the $R_{52}$ ratios and the average CO SLEDs of galaxies with and without an AGN when considering galaxies with $f_{\rm AGN} < 80 \%$ as in this work. Furthermore, the authors find no differences in the CO(7-6)/CO(2-1) ratios of galaxies with and without an AGN. 
This suggests that the differences seen in the CO excitation of extended and compact galaxies are driven by the different star formation properties of the two sub-samples up to CO(7-6).
These findings are consistent with other studies showing that the $R_{52}$ ratios does not correlate with the AGN fraction \citep{Liu21}. This is also in agreement with other studies showing that there are not statistically-significant differences in the CO SLED of star-forming versus AGN-dominated galaxies, at least when considering low-to-mid-J CO transition \citep{Brusa18, Kirkpatrick19, Boogaard21}. This supports the idea that the low-to-mid J CO emission is dominated by processes associated to star formation. }

\begin{figure}
\includegraphics[width=\columnwidth]{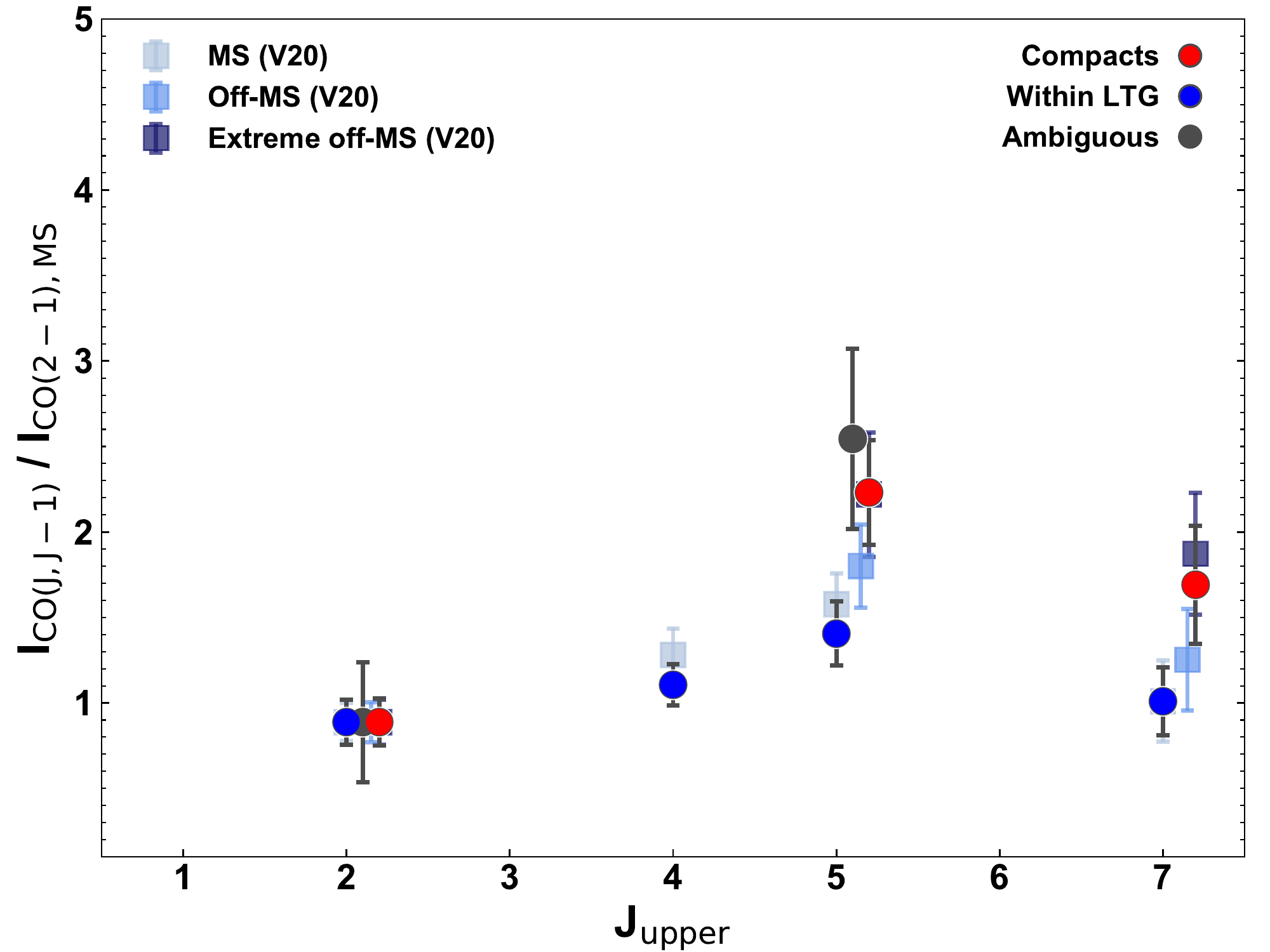}
\caption{Average CO SLED of extended, ambiguous and compact galaxies (blue, grey and red circles, respectively).
{As a reference we show the average SLEDs of main sequence ($\Delta$MS$\leqslant 3.5$), off main sequence ($\Delta$MS$\geqslant 3.5$) and extreme off main sequence ($\Delta$MS$\geqslant 7$) galaxies from \citet{Valentino20} with light blue, blue and dark blue shaded squares, respectively.
We normalise the data-points to the mean CO(2-1) flux + 1$\sigma$ flux error of $z \sim 1.25$ main sequence galaxies.} }
\label{fig:CO_SLEDs}
\end{figure}

To investigate additional dependences of the CO SLED shape on the main sequence offset, we split extended and compacts between on- and off- main sequence sources. 
For this exercise, we classify as on-main sequence galaxies with $\Delta$MS$< 3.5$ and off-main sequence galaxies with $\Delta$MS$\geqslant 3.5$.
We show in Figure \ref{fig:CO_SLEDs_onoffMS} the average CO SLEDs of these sources.
Extended galaxies within and above the main sequence have slightly different CO SLEDs. 
In particular, extended main-sequence galaxies seem slightly less excited than extended galaxies above it. 
We also note that the CO SLED of main sequence galaxies in the \cite{Valentino20} analysis (light blue squares in Figure \ref{fig:CO_SLEDs_onoffMS}) is intermediate between that of extended galaxies on and above the main sequence presented here. 
This suggests that main sequence galaxies may not represent a homogeneous population, similarly to that reported in Sect. \ref{subsubsect:R52}.
On the other hand, compact galaxies on and above the main sequence have remarkably similar CO SLEDs and these are consistent with that of the most extreme main-sequence outliers (dark blue squares in Figure \ref{fig:CO_SLEDs_onoffMS}). 

\begin{figure}
\includegraphics[width=\columnwidth]{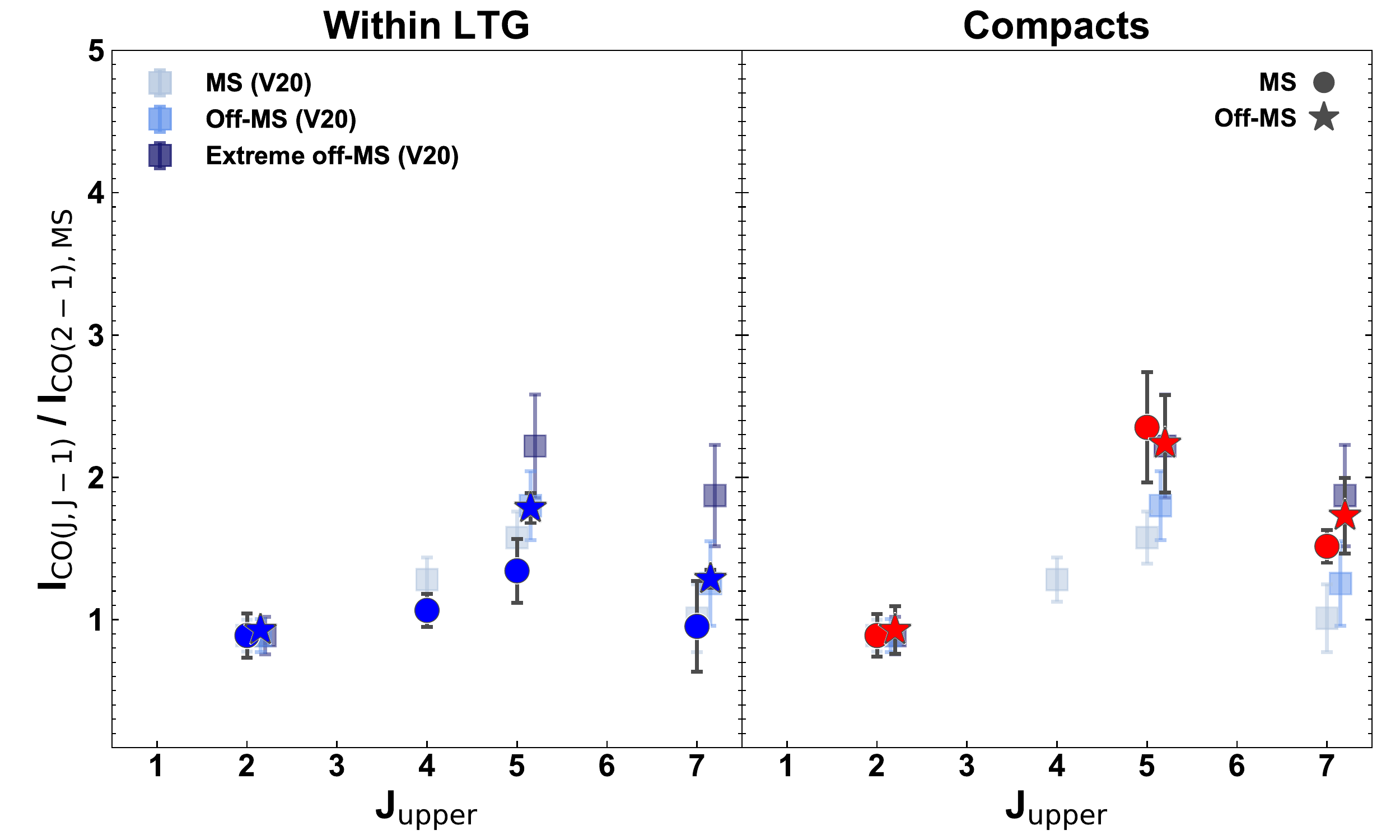}
\caption{ Average CO SLED of extended ({\it left}) and compact ({\it right}) galaxies in our sample, split according to their main sequence position. 
In both panels, circles indicate the average SLED of galaxies with $\Delta$MS \textless 3.5 and stars show the SLED for galaxies with $\Delta$MS $\geqslant 3.5$.
}
\label{fig:CO_SLEDs_onoffMS}
\end{figure}

Figures \ref{fig:CO_SLEDs} and \ref{fig:CO_SLEDs_onoffMS} suggest that the CO excitation increases as a function of the source compactness. This dependence seems more significant than the variations of the CO excitation with the main sequence offset.
To quantify this effect, we compute line ratios as a function of the compactness and the main sequence position (see Table \ref{tab:R_SLEDs}).
The average $R_{52}$ and $R_{72}$ ratios of compact galaxies are $1.6 \times$ and $1.8 \times$ higher than that of extended galaxies.
Instead, compact galaxies have $R_{52}$ and $R_{72}$ ratios that are $1.8 \times$ higher than in extended galaxies within the main sequence. 
Focusing on the off-main sequence population, compact galaxies are 1.3$\times$ and 1.4$\times$ more excited in CO(5-4) and CO(7-6) than their few extended counterparts.
The $R_{52}$ and $R_{72}$ excitation ratios of compact galaxies on and above the main sequence are consistent with each other within the errorbars,  suggesting homogeneous CO excitation properties.  
On the other hand, we find marginal evidence for variations in the excitation ratios of extended sources as a function of the MS position. 
In particular, extended galaxies above the main sequence have $R_{52}$ and $R_{72}$ ratios that are 1.3$\times$ and 1.4$\times$ higher than in extended main sequence galaxies.
This might indicate an evolutionary trend with extended galaxies above the main sequence being merging pairs in an early stage of the interaction. Alternatively, this might suggest that extended off-main sequence galaxies are gas-rich starbursting disks with enhanced star formation rates  
after anomalous gas accretion episodes \citep[e.g.][]{Scoville16}. 
{Observations with increased spatial resolution will allow us to identify any unresolved merger pairs among extended galaxies above the main sequence.}

{These results suggest that main sequence sources do not represent a homogeneous population in terms of CO excitation properties, as also anticipated in \cite{Valentino20}.
Thus, adopting a CO excitation correction that is based on the main sequence position of a galaxy would introduce significant uncertainties in, e.g., the derived CO fluxes due to the presence of a variety of CO SLEDs within the main sequence. This is indicated by the measured $R_{52}$ ratios within the main sequence ranging from a minimum value of $R_{\rm 52, MS, min} = 0.13 \pm 0.03$ to a maximum value of $R_{\rm 52, MS, max} \gtrsim 0.55$.} }
The sub-millimetre compactness classification, on the other hand, allows us to select galaxies with more homogeneous CO excitation properties, in agreement with the results from individual CO line ratios discussed in the previous section. 
{Given the relatively small number of sources, we caution the reader that the observed variations should be taken as indicative.
Future studies with larger statistics will allow us to better investigate variations of the CO SLED with the main sequence offset and the galaxy compactness.}

\begin{table}
\centering
\caption{ Average CO and [CI] luminosities and detection statistics for {galaxies classified as a function of their compactness}.
The $L'$ luminosities are expressed in 10$^{10}$ K km s$^{-1}$ pc$^{2}$.
The average I fluxes in Figure \ref{fig:CO_SLEDs} are expressed in Jy km s$^{-1}$ and are computed from $L'$ luminosities by adopting $z = 1.25$.
\\$\dagger$Formally biased mean value, as the first upper limit was turned into a detection for the calculation of the KM estimator \citep{KM}.}
\label{tab:L_CO_comp}
\begin{tabular}{cccc} 
\hline
\hline
\multicolumn{4}{c}{\multirow{1}{*}{{Within LTG}}} \\
		\hline
		Transition & N$_{\rm det}$, N$_{\rm up}$  & Mean & Median\\
		\hline
		$L'_{\rm CO(2-1)}$ & 13, 2 & 2.13 $\pm$ 0.32$\dagger$ & 1.70$^{+ 0.97}_{-0.11}$\\
		$L'_{\rm CO(4-3)}$ & 4, 0 & 0.66  $\pm$ 0.07&  0.59$^{+ 0.12}_{-0.12}$\\
		$L'_{\rm CO(5-4)}$ & 14, 1 & 0.54  $\pm$ 0.07$\dagger$ & 0.46$^{+ 0.21}_{-0.14}$\\
		$L'_{\rm CO(7-6)}$ & 5, 0 & 0.20  $\pm$ 0.04 & 0.18$^{+ 0.03}_{-0.09}$\\
		\hline
		$L'_{\rm CI[1-0]}$ & 6, 0 &  0.41 $\pm$  0.06  & 0.368$^{+ 0.07}_{-0.12}$\\
		$L'_{\rm CI[2-1]}$ & 5, 0 & 0.18 $\pm$  0.02  & 0.19$^{+ 0.01}_{-0.08}$ \\
		\hline		
\hline
\multicolumn{4}{c}{\multirow{1}{*}{Ambiguous}} \\
		\hline
		Transition & N$_{\rm det}$, N$_{\rm up}$  & Mean & Median\\
		\hline
		$L'_{\rm CO(2-1)}$ & 2, 2 &  0.66 $\pm$ 0.26$\dagger$ & - \\
		$L'_{\rm CO(4-3)}$ & 1, 0 &  0.72 & - \\
		$L'_{\rm CO(5-4)}$ & 3, 1 &  0.30 $\pm$ 0.06$\dagger$ & 0.22$^{+ 0.19}_{-0.08}$\\
		$L'_{\rm CO(7-6)}$ & - & - & - \\
		\hline
		$L'_{\rm CI[1-0]}$ &  0, 1 & 0.07 & - \\
		$L'_{\rm CI[2-1]}$ &  - & - & - \\
		\hline		
\hline
\multicolumn{4}{c}{\multirow{1}{*}{Compacts}} \\
		\hline
		Transition & N$_{\rm det}$, N$_{\rm up}$  & Mean & Median\\
		\hline
		$L'_{\rm CO(2-1)}$ & 11, 1 & 1.67$\pm$0.26$\dagger$ & 1.49$^{+ 0.61}_{-0.60}$\\
		$L'_{\rm CO(4-3)}$ & - & - & - \\
		$L'_{\rm CO(5-4)}$ & 12, 0 & 0.67 $\pm$ 0.09 & 0.54$^{+ 0.43}_{-0.12}$\\
		$L'_{\rm CO(7-6)}$ & 4, 0 & 0.26$\pm$0.05 & 0.19$^{+ 0.09}_{-0.04}$\\
		\hline
		$L'_{\rm CI[1-0]}$ & 2, 1 & 0.29 $\pm$ 0.10 $\dagger$ & - \\
		$L'_{\rm CI[2-1]}$ & 4, 0 & 0.26 $\pm$ 0.04 & 0.25$^{+ 0.001}_{-0.1}$ \\
		\hline		
\end{tabular}
\end{table}

\begin{table}
\centering
\caption{As in Table \ref{tab:L_CO_comp} but for extended {and compact} galaxies on and above the main sequence.}
\label{tab:L_CO_comp_ms}
\begin{tabular}{cccc} 
\hline
\hline
\multicolumn{4}{c}{\multirow{1}{*}{{Within LTG}, on MS ($\Delta$MS $\leqslant$ 3.5)}} \\
		\hline
		Transition & N$_{\rm det}$, N$_{\rm up}$  & Mean & Median\\
		\hline
		$L'_{\rm CO(2-1)}$ & 10, 2 &  2.22 $\pm$  0.39$\dagger$ & 1.70$^{+ 1.88}_{-0.22}$\\
		$L'_{\rm CO(4-3)}$ & 4, 0 &  0.66 $\pm$  0.07 & 0.59$^{+ 0.12}_{-0.12}$\\
		$L'_{\rm CO(5-4)}$ & 11, 1 & 0.54 $\pm$ 0.09$\dagger$ & 0.39$^{+ 0.44}_{-0.09}$\\
		$L'_{\rm CO(7-6)}$ & 3, 0 & 0.19 $\pm$ 0.06 & - \\
		\hline
		$L'_{\rm CI[1-0]}$ & 6, 0 & 0.41 $\pm$ 0.06  & 0.37$^{+ 0.07}_{-0.12}$\\
		$L'_{\rm CI[2-1]}$ & 3, 0 & 0.17 $\pm$  0.04 & -\\
		\hline		
		\hline		
\multicolumn{4}{c}{\multirow{1}{*}{{Within LTG}, off MS ($\Delta$MS \textgreater 3.5)}} \\
		\hline
		Transition & N$_{\rm det}$, N$_{\rm up}$  & Mean & Median\\
		\hline
		$L'_{\rm CO(2-1)}$ & 3, 0 &  1.80 $\pm$ 0.06 & - \\
		$L'_{\rm CO(4-3)}$ & - & - & - \\
		$L'_{\rm CO(5-4)}$ & 3, 0 & 0.56 $\pm$  0.03 & - \\
		$L'_{\rm CO(7-6)}$ & 2, 0 & 0.20 $\pm$  0.01 & - \\
		\hline
		$L'_{\rm CI[1-0]}$ & - & - & - \\
		$L'_{\rm CI[2-1]}$ & 2, 0 & 0.19 $\pm$ 0.01 & -\\
		\hline		
		\hline		
\multicolumn{4}{c}{\multirow{1}{*}{Compacts, on MS ($\Delta$MS $\leqslant$ 3.5)}} \\
		\hline
		Transition & N$_{\rm det}$, N$_{\rm up}$  & Mean & Median\\
		\hline
		$L'_{\rm CO(2-1)}$ & 6, 1 &  1.22 $\pm$ 0.21$\dagger$ & 1.04$^{+ 0.49}_{- 0.30}$\\
		$L'_{\rm CO(4-3)}$ & - & -  & -\\
		$L'_{\rm CO(5-4)}$ & 7, 0 & 0.52 $\pm$  0.08 & 0.43$^{+ 0.11}_{-0.08}$\\
		$L'_{\rm CO(7-6)}$ & 2, 0 & 0.17 $\pm$  0.01 & - \\
		\hline
		$L'_{\rm CI[1-0]}$ &  1, 1 & 0.19 $\pm$ 0.09 $\dagger$  & -\\
		$L'_{\rm CI[2-1]}$ &  2, 0 & 0.20 $\pm$ 0.03  & - \\
		\hline		
		\hline		
\multicolumn{4}{c}{\multirow{1}{*}{Compacts, off MS $\Delta$MS \textgreater 3.5)}} \\
		\hline
		Transition & N$_{\rm det}$, N$_{\rm up}$  & Mean & Median\\
		\hline
		$L'_{\rm CO(2-1)}$ & 5, 0 &  2.30 $\pm$ 0.41  & 2.09$^{+ 0.66}_{-0.98}$\\
		$L'_{\rm CO(4-3)}$ & - & - & - \\
		$L'_{\rm CO(5-4)}$ & 5, 0 & 0.89 $\pm$  0.14 & 0.77$^{+0.36 }_{-0.26}$\\
		$L'_{\rm CO(7-6)}$ & 2, 0 & 0.35 $\pm$  0.05 & - \\
		\hline
		$L'_{\rm CI[1-0]}$ & 1, 0 & 0.48 & - \\
		$L'_{\rm CI[2-1]}$ &  2, 0& 0.31 $\pm$  0.05 & - \\
		\hline		
\end{tabular}
\end{table}

\begin{table*}
\centering
\caption{ Average line luminosities ratios for galaxies at $z \sim 1.25$.
The ratios and their {1$\sigma$} uncertainties are computed analytically based on the mean $L'$ luminosities in Tables \ref{tab:L_CO_comp} and \ref{tab:L_CO_comp_ms}.}
\label{tab:R_SLEDs}
\begin{tabular}{ccccccc} 
\hline
\hline
		Transition & Within LTG & Within LTG, on MS  & Within LTG, above MS & Compacts & Compacts, on MS & Compacts, above MS \\
		\hline
		$R_{42}$ & 0.31 $\pm$ 0.06 &  0.30 $\pm$ 0.03 & -  & - & - & - \\
		$R_{52}$ & 0.25 $\pm$ 0.05 & 0.24 $\pm$ 0.06  & 0.31 $\pm$ 0.02 & 0.40 $\pm$ 0.08 & 0.43 $\pm$ 0.10 & 0.39 $\pm$ 0.09 \\
		$R_{72}$ & 0.09 $\pm$ 0.02 & 0.08 $\pm$ 0.03  & 0.11 $\pm$ 0.01 & 0.16 $\pm$ 0.04 & 0.14 $\pm$ 0.02 & 0.15 $\pm$ 0.04 \\
\hline
\end{tabular}
\end{table*}

\subsubsection{Large Velocity Gradient modelling}

To better understand the CO excitation properties of our galaxies, we apply a Large Velocity Gradient (LVG) modelling to the observed average SLEDs of our sample. We report few details about the modelling below and we refer the reader to \cite{Daddi15} and \citet{Liu15} for a detailed description of the approach.
{We create a grid of LVG models using the RADEX tool \citep{vanderTak07}. As RADEX requires the input of the [CO/H$_2$] abundance ratio, turbulence Doppler line width ($\delta\mathrm{V}$) and a H$_2$ column density ($N_{\mathrm{H_2}}$) separately, we set them to [CO/H$_2$]~$=5\times10^{-5}$, $\delta\mathrm{V}=50\,\mathrm{km s^{-1}}$ and $N_{\mathrm{H_2}}/n_{\mathrm{H_2}}=10$~pc. This implies a velocity gradient of 5~$\mathrm{km s^{-1} pc^{-1}}$, consistent with or in the range of the findings for Galactic center or relatively dense and warm clouds 
\citep{GoldreichKwan, Dahmen98, Ao13} and extragalactic molecular gas in local actively star-forming galaxies \citep{Curran01, Weiss01, Weiss05, Zhang14} and high-redshift galaxies \citep{Weiss07, Dannerbauer09}. We note that these quantities are degenerate. Given that the SLEDs of our galaxies are not fully sampled, leaving the velocity gradient or the abundance free to vary would result in much larger uncertainties. Due to the degeneracy of the models, a different set of assumed values of velocity gradients or abundances will lead to no difference in determining the $n_{\mathrm{H_2}}$ and $T_{\mathrm{kin}}$, but could systematically bias the optical depth, the filling factor and hence the total mass from LVG. Thus in this work we only use the LVG fitting to infer $n_{\mathrm{H_2}}$ and $T_{\mathrm{kin}}$. }
{To better constrain the fit, we include the [CI] transitions available for a subset of the sample (see Tables \ref{tab:L_CO_comp} and \ref{tab:L_CO_comp_ms}). This allows us to provide additional information to the total gas content, since this transition correlates with the total infrared luminosity similarly to low-J CO transitions \citep{Valentino18}. }
Here we assume that the neutral atomic carbon is co-spatial with CO and a fixed [C$^0$/H$_2$] = 3 $\times$ 10$^{-5}$ abundance \citep{Weiss03, Papadopoulos04}. 

We compute the grid at the median redshift of the sample spanning a density and temperature range of $n({\rm H_2}) = 10^2 - 10^6$ cm$^{-3}$ and $T_{\rm kin} = 5 - 300$ K, including the appropriate value of the temperature of the cosmic microwave background. 
We derive the best-fit model, the best-fit parameters and their 1$\sigma$ uncertainties using a customized $\chi^2$ minimization algorithm, optimized for the exploration of highly multi-dimensional spaces \citep[MICHI2\footnote{\url{https://ascl.net/code/v/2533}}][]{Liu21}. 
We iteratively sampled the $\chi^2$ distribution 15000 and 10000 times for the single and two components modelling, respectively (see below), randomizing the parameters within normal distributions centred on the minimal $\chi^2$ derived from the previous iteration, artificially inflating their widths.

\begin{figure}
\includegraphics[width=\columnwidth]{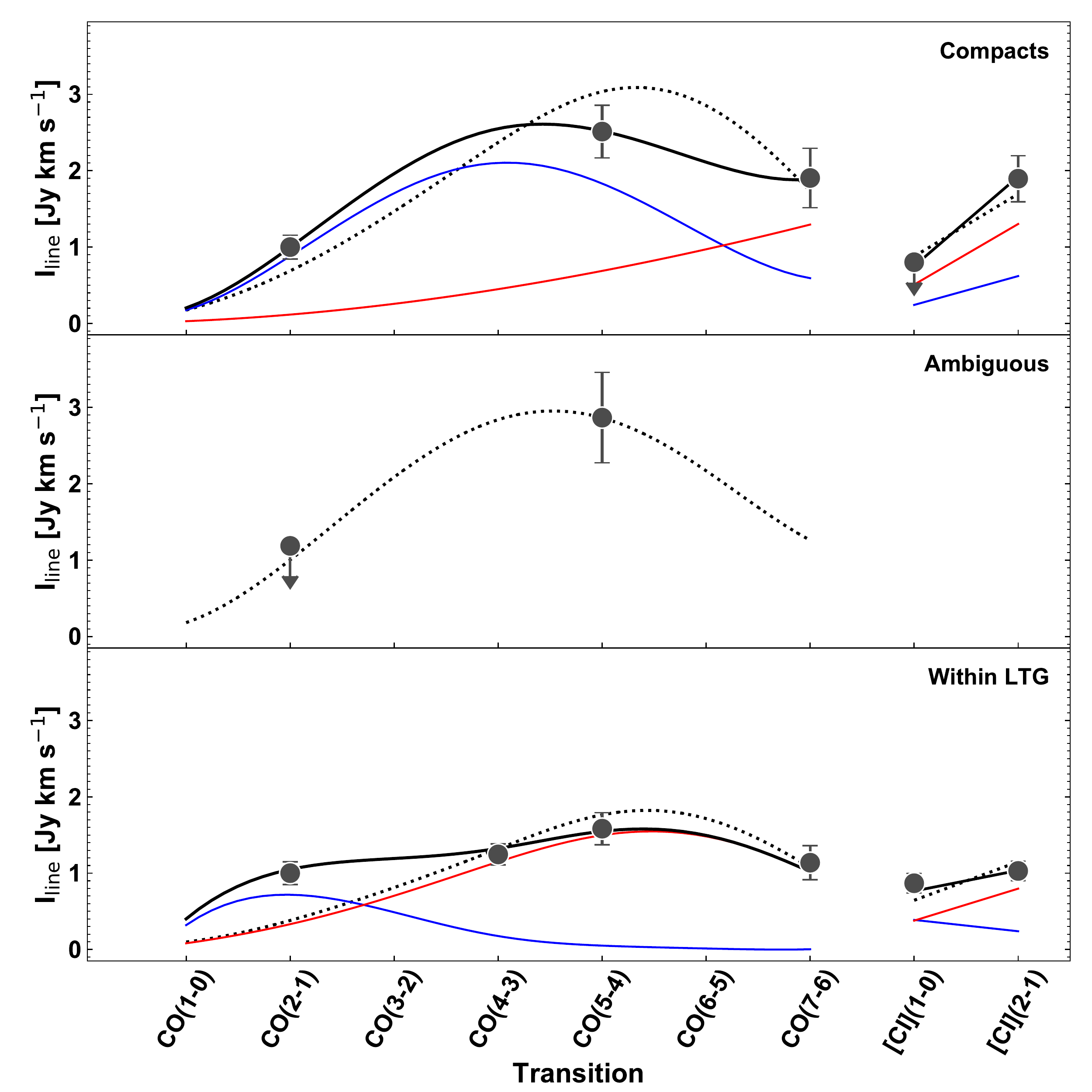}
\caption{ LVG modelling of the observed CO+[CI] SLEDs for galaxies within the LTG relation ({\it bottom panel}), galaxies with a size upper limit within the LTG relation (ambiguous, {\it central panel}), and compact galaxies ({\it top panel}). 
The filled symbols show the mean fluxes while arrows indicate 3$\sigma$ upper limits.
The dotted black line shows the best-fit model with a single component. The blue and red lines show the low- and high-excitation components of the two components LVG modelling, with the black solid line indicating their sum. }
\label{fig:LVG_compactness}
\end{figure}

\begin{figure}
\includegraphics[width=\columnwidth]{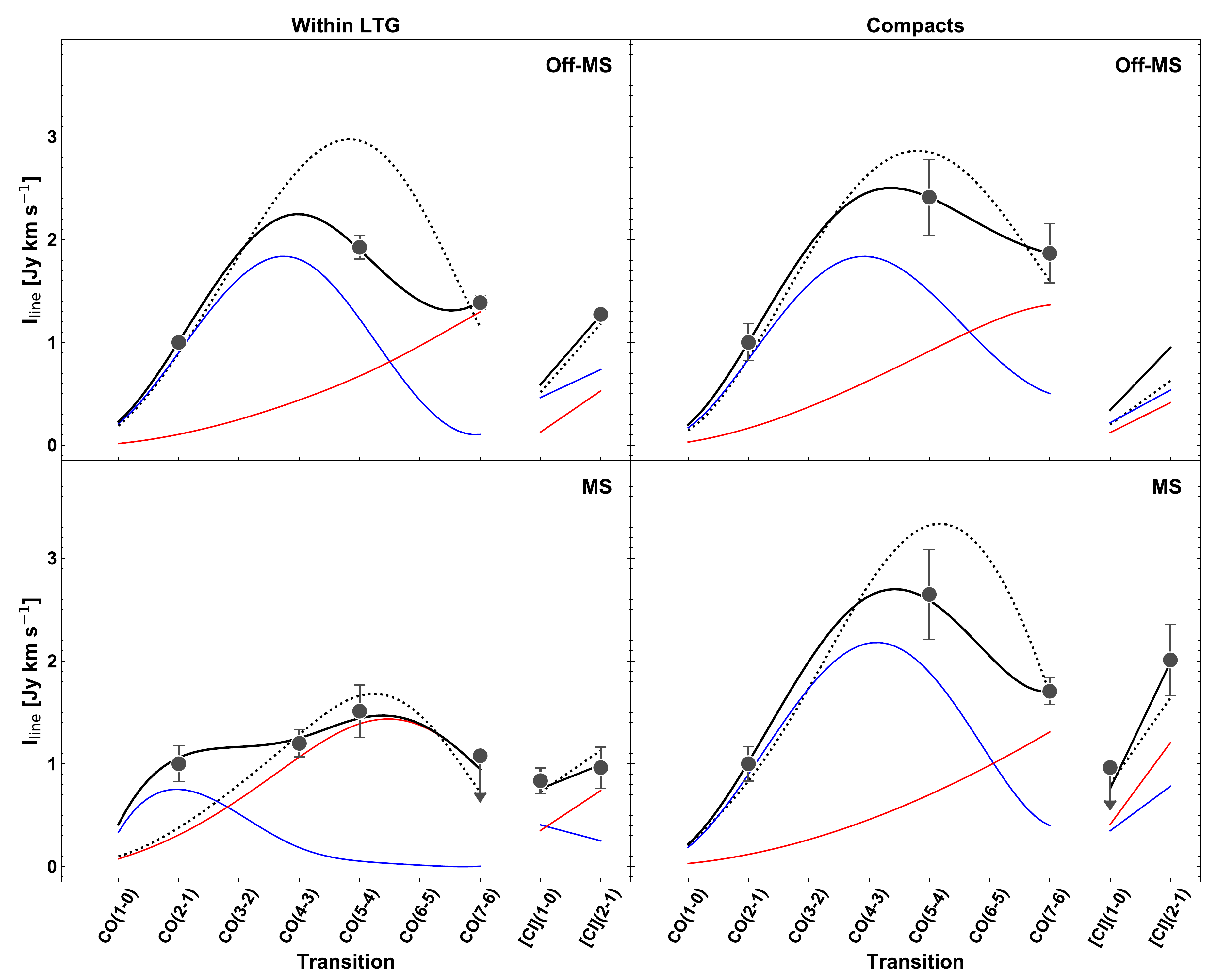}
\caption{ LVG fit to the observed CO+[CI] SLEDs for extended ({\it left}) and compact ({\it right}) galaxies in our sample, split according to their main sequence position as in Figure \ref{fig:CO_SLEDs_onoffMS}. Colour code is analogous to Figure \ref{fig:LVG_compactness}.}
\label{fig:LVG_compactness_MS}
\end{figure}

A single component model can only reproduce the CO SLED of ambiguous galaxies (see central panel in Figure \ref{fig:LVG_compactness}), likely because of the limited amount of information available. Both the density and temperature are poorly constrained for this sub-sample.
Instead, Figure \ref{fig:LVG_compactness} shows that a single component model does not properly fit the observed CO SLEDs of the rest of the sub-samples. In particular, it significantly underestimates the CO(2-1) emission of both extended and compact galaxies (bottom and top panels in Figure \ref{fig:LVG_compactness}) while overestimating their CO(5-4) emission. Similar trends are observed when we further distinguish between galaxies based on their main sequence position (see Figure \ref{fig:LVG_compactness_MS}).
This is consistent with previous studies \citep[e.g.][]{Riechers11, Hodge13, Zhang14, Kamenetzky14, Kamenetzky17, Liu15, Daddi15}.

We thus perform a two components LVG modelling, by assuming the presence of a diffuse and dense gas phase with $n({\rm H_2, low}) < n({\rm H_2, high})$. This allows us to better constrain $n({\rm H_2})$, but does not allow us to place robust constraints on $T_{\rm kin}$. 
We report the best-fit parameters from the two components modelling in Table \ref{tab:LVG_2component}.
We find that all galaxies are characterized by low- and high- density gas components, with compacts having marginally higher gas densities than extended sources {($n_{\rm H_{2}, low} \sim 10^{2} - 10^{4}$ cm$^{-3}$ and $n_{\rm H_{2}, high} \sim 10^{4} - 10^{6}$ for the compacts, while $n_{\rm H_{2}, low} \sim 10^{2} - 10^{3}$ cm$^{-3}$ and $n_{\rm H_{2}, high} \sim 10^{3} - 10^{4}$ cm$^{-3}$ for the extended sample, see also Table \ref{tab:LVG_2component})}.
We also find that $\sim 40 \%$ of the total molecular gas mass of extended galaxies is in the dense component. 
{The best-fit high-density component for the compacts is rising at high J (see top panel in Figure \ref{fig:LVG_compactness} and right panels in Figure \ref{fig:LVG_compactness_MS}). While we caution that we have no constraints on the CO SLEDs of our sources beyond J=7 and the LVG best-fit models are highly extrapolated for higher-J CO observations, this might suggest that the average CO SLED of this population is dominated by an excited and dense gas component, similarly to what is observed in the starburst-dominated local ULIRG Arp220 \citep{Rangwala11} and high-redshift SMGs \citep{Yang17, Canameras18, Birkin21}. On the other hand, it is unlikely that the CO SLED of our sources up to J=7 is dominated by the AGN component. In fact, even in local QSOs such as Mrk 231, the CO excitation up to J=8 can be explained by heating from star formation \citep{vanderWerf10}.
We also note that our best-fit LVG models seem to suggest a somewhat different behaviour for compact galaxies on and above the main sequence (see top and bottom right panels in Figure \ref{fig:LVG_compactness_MS}). However, this is likely due to the lack of constraints on the [CI] transitions for the off main-sequence compact population, since the shape of the observed CO SLEDs is nearly identical for the two classes of galaxies up to J=7 (see right panel in Figure \ref{fig:CO_SLEDs_onoffMS}). The best-fit parameters derived from the LVG analysis of the two sub-samples are consistent within the uncertainties (see Table \ref{tab:LVG_2component}). }

\begin{table*}
\centering
\caption{ Best-fit parameters of a double-component LVG modelling of the {average} CO+[CI] SLEDs of $z \sim 1.25$ galaxies {presented in Figures \ref{fig:CO_SLEDs} and \ref{fig:CO_SLEDs_onoffMS}, and Tables \ref{tab:L_CO_comp} and \ref{tab:L_CO_comp_ms}.}
\\The average values and their uncertainties are the best-fit estimates and their statistical 1$\sigma$ errors, where we impose that $n_{\rm H_{2}, low} < n_{\rm H_{2}, high}$.
\\$\dagger$ { We note that absolute values of the gas mass from LVG modelling depend on the adopted CO abundance, constant for the various populations analysed here. Therefore, they are subject to the uncertainties already described in Section \ref{subsec:gas}. Relative comparisons between the two phases for each population still hold, under the assumption that dense and diffuse gas reservoirs share the same metallicity.}
}
\label{tab:LVG_2component}
\begin{tabular}{ccccccc} 
\hline
\hline
		Parameter & Within LTG & Within LTG, on MS  & Within LTG, above MS & Compacts & Compacts, on MS & Compacts, above MS \\
		\hline
		log($n_{\rm H_{2}, low}$/[cm$^{-3}$])  & 2.2 $\pm$ 0.2 &  2.2 $\pm$ 0.3  & 3.3 $\pm$ 0.6 & 3.1 $\pm$ 0.8 & 3.3 $\pm$ 0.8 & 3.0 $\pm$ 0.8 \\
		log($n_{\rm H_{2}, high}$/[cm$^{-3}$]) & 3.9 $\pm$ 0.1 & 3.9  $\pm$ 0.2 & 4.8 $\pm$ 1.2 & 6 $\pm$ 1.3 & 5.2 $\pm$ 1.3 & 4.1 $\pm$ 1.6 \\
		$T_{\rm kin, low}$/[K] & 45 $\pm$ 98 &  45 $\pm$ 113 & 45 $\pm$  130 & 180 $\pm$ 135 & 80 $\pm$ 135 & 215 $\pm$ 138 \\
		$T_{\rm kin, high}$/[K]& 45 $\pm$  10 & 45  $\pm$ 23 & 250 $\pm$ 130 & 135 $\pm$ 138 & 125 $\pm$ 138 & 105 $\pm$ 138 \\
		log($M_{\rm H_{2}, low}$/[M$_{\odot}$])$\dagger$   & 10.3 $\pm$  0.3 & 10.3  $\pm$ 0.3 & 10.1 $\pm$ 0.2 & 9.7 $\pm$ 0.4 & 9.7 $\pm$ 0.4 & 9.8 $\pm$ 0.6 \\
		log($M_{\rm H_{2}, high}$/[M$_{\odot}$])$\dagger$  & 10.1 $\pm$  0.2 & 10.1  $\pm$ 0.3 & 9.6 $\pm$ 1.1 & 11.1 $\pm$ 1.5 & 10.2 $\pm$ 1.5 & 9.6 $\pm$ 1.8 \\
\hline
\end{tabular}
\end{table*}

\subsection{Star formation efficiency and depletion time}
\label{subsect:SFE}

\begin{figure*}
    \centering
    \begin{subfigure}[t]{0.45\textwidth}
        \centering
        \includegraphics[width=\columnwidth]{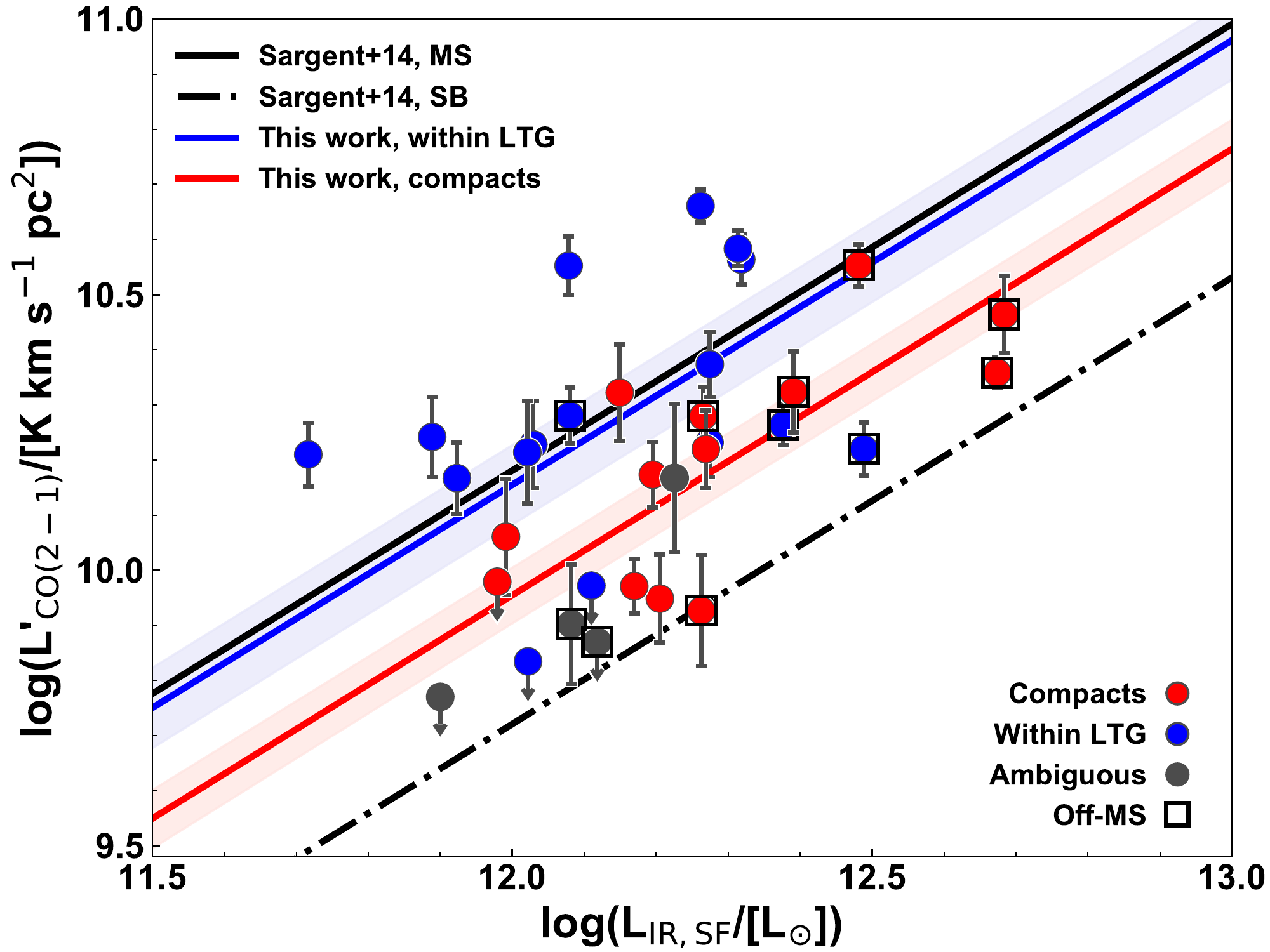}
    \end{subfigure}
    \begin{subfigure}[t]{0.45\textwidth}
        \centering
        \includegraphics[width=\columnwidth]{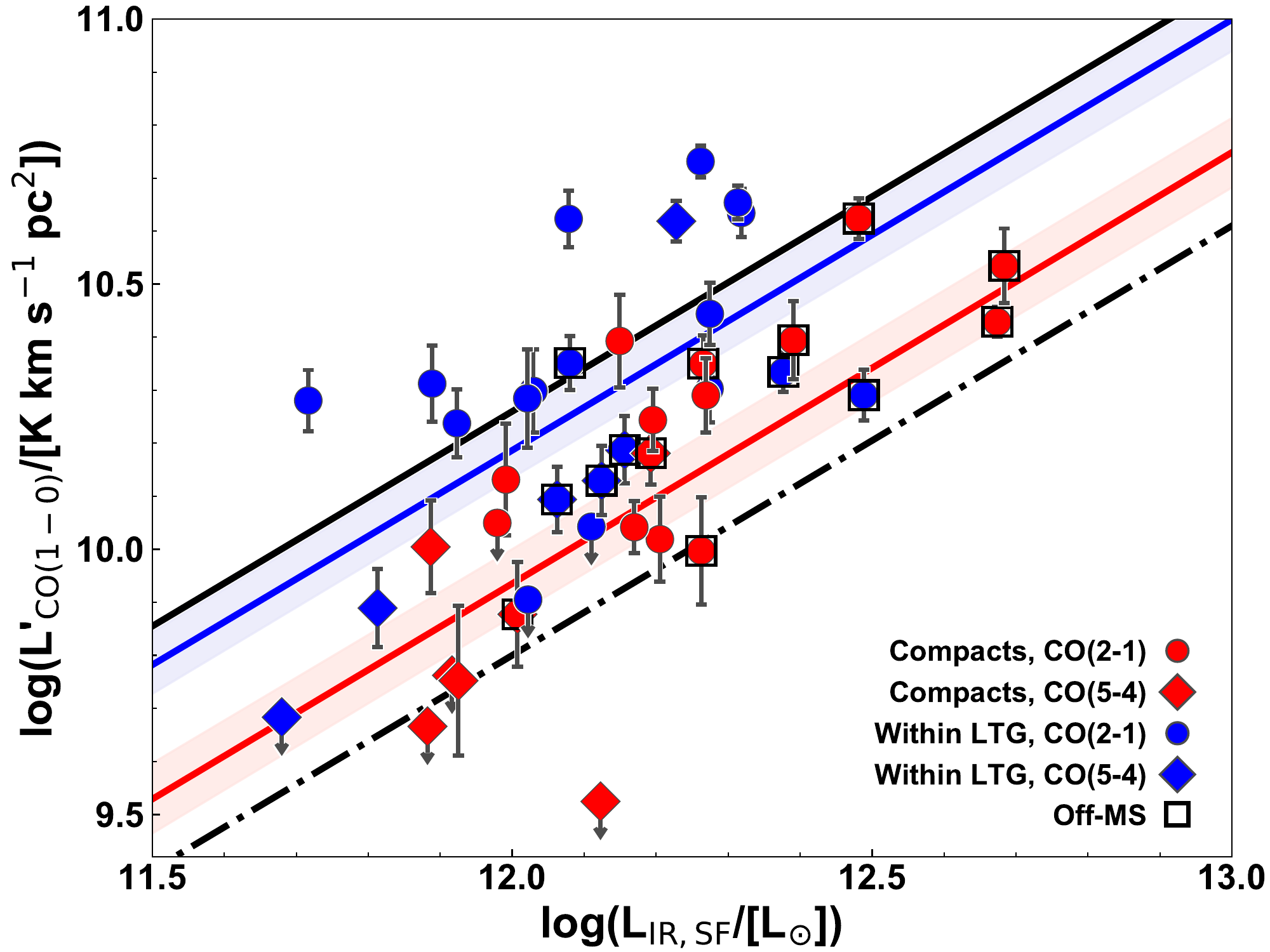}
    \end{subfigure}
    \caption{ $L'_{\rm CO(2-1)}$ [K km s$^{-1}$ pc$^2$] ({\it left}) and $L'_{\rm CO(1-0)}$ [K km s$^{-1}$ pc$^2$] ({\it right}) luminosity as a function of the infrared luminosity from star formation $L_{\rm IR, SF}$ [$L_{\odot}$]. 
The solid and dashed black lines represent the models for main sequence and starburst galaxies from \protect\cite{Sargent14}.
In the left panel, we convert this equation to $L'_{\rm CO(2-1)}$ using $R_{21}$ = 0.85.
In both panels, the blue and red solid lines are the best-fit lines with a slope $\beta = 0.81$ obtained by fitting separately the extended and compact population, respectively. The shaded areas represent the 1$\sigma$ confidence interval on the best-fit normalisation. 
{The symbols color-code is analogous to Figure \ref{fig:CO_ratios_MS_comp}.
In the right panel we show with filled diamonds measurements of $L'_{\rm CO(1-0)}$ extrapolated from the CO(5-4) observed flux. 
The errorbars in this figure correspond to the 1$\sigma$ uncertainty on the CO(2-1) or CO(5-4) flux measurements.}
}
\label{fig:LCO_LIR}
\end{figure*}

In the left panel of Figure \ref{fig:LCO_LIR} we show the $L'_{\rm CO(2-1)}$ luminosity as a function of $L_{\rm IR, SF}$ for {the galaxies with CO(2-1) detections or upper limits in our sample.}
This is the so-called integrated Schmidt-Kennicutt plane and it allows us to study the relation between the molecular gas content and star formation rate,  hence the nature of star formation in our sample \citep[e.g.][]{Daddi10letter, Sargent14}. 
Extended galaxies are on average close to the region of disks \citep[][solid black line in Figure \ref{fig:LCO_LIR}]{Sargent14}. Compacts seem to be globally shifted towards the dash-dotted line in Figure \ref{fig:LCO_LIR}, defining the ``strong starbursts'' ($\Delta$MS $\gtrsim$ 15) locus in the $L'_{\rm CO(2-1)}$ versus $L'_{\rm IR, SF}$ plane according to the \cite{Sargent14} model \citep[see also][]{Solomon05, Greve05}.
In the right panel of Figure \ref{fig:LCO_LIR} we show the $L'_{\rm CO(1-0)}$ luminosity as a function of $L_{\rm IR, SF}$. 
{To construct this plot, we use CO(2-1) fluxes, where available.
To increase the statistics, we also include galaxies without CO(2-1) observations by converting their CO(5-4) flux to CO(2-1) using a $R_{52}$ that is appropriate for their average SLED shape (see Sect. \ref{subsec:Excitation} and Table \ref{tab:R_SLEDs}). 
We do not include galaxies in the ambiguous sample because their average excitation properties are poorly constrained.}
We stress however that mid/high-J CO transitions sample the denser phase of the molecular gas associated with star formation \citep{Liu15, Daddi15, Valentino20} and should not be used as a proxy of the molecular gas mass {when no direct constraints on the excitation correction are available}.
{We finally convert the CO(2-1) flux to $L'_{\rm CO(1-0)}$ using $R_{21} = 0.85$, which is the average CO(2-1)-to-CO(1-0) ratio measured in star-forming galaxies at high redshift \citep{Bothwell13, Daddi15, Boogaard21}. 
Using an homogeneous excitation correction for the CO(2-1) luminosity is a conservative choice since galaxies with an enhanced $R_{52}$ ratio are expected to also show an enhanced $R_{21}$ ratio \citep[see e.g.][]{Daddi15}. Implementing a differential excitation correction would thus exacerbate the tension between the extended and compact populations.}
The right panel of Figure \ref{fig:LCO_LIR} shows that the offset between compacts and extended galaxies increases when accounting for differences in the excitation properties of the two populations. 

To quantify the separation between extended and compact galaxies in the $L'_{\rm CO(1-0)}$ versus $L_{\rm IR, SF}$ plane, we fit the two populations independently using a linear function in the log-log space and accounting for upper limits as described in Sect. \ref{subsec:Excitation}. 
We then consider only the best-fit solutions yielding a slope $\beta = 0.81 \pm 0.03$, i.e. consistent with the value reported in \cite{Sargent14}. 
We find that extended galaxies have a best-fit normalization value of $0.46 \pm 0.22$ which is consistent with the \cite{Sargent14} value for disks. 
We find a best-fit normalization of $0.18 \pm 0.22$ for the compacts. 
This corresponds to a $\sim 2 \times$ offset between compact and extended galaxies. This value is intermediate between the factor of $3\times$ offset from \cite{Sargent14} and the $1.7\times$ offset reported by \cite{Silverman15} for $z \sim 1.5$ starbursts. We obtain consistent results when considering the $L'_{\rm CO(2-1)}$ luminosity in the left panel of Figure \ref{fig:LCO_LIR}.
To quantify the significance of the offset between compact and extended galaxies, we compare the distributions of the two sub-samples after subtracting the $\beta \times L_{\rm IR, SF}$ trend from the observed $L'_{\rm CO(1-0)}$ luminosity. We perform a log-rank test accounting for the presence of upper limits as described in the previous section.
We find a 99.4 \% probability that the distributions of compacts and extended galaxies in the right panel of Figure \ref{fig:LCO_LIR} are different $(p - value = 0.0058)$. 

The low-J CO transitions are proxies for the molecular gas content \citep{Carilli13,Tacconi20} while $L_{\rm IR, SF}$ quantifies the star-forming activity of a galaxy. Therefore, Figure \ref{fig:LCO_LIR} suggest that compact galaxies in our sample have an enhanced star formation efficiency (SFE = $M_{\rm gas}$/SFR) and a lower depletion time ($\tau_{\rm depl} =$ SFR/$M_{\rm gas} = 1/$SFE) than extended galaxies.
It also appears that extended and compacts off-main sequence galaxies (open black squares in Fig. \ref{fig:LCO_LIR}) preferentially occupy the region with high $L_{\rm IR, SF}$ and $L'_{\rm CO(2-1)}$ or $L'_{\rm CO(1-0)}$. This suggests that compact galaxies above the main sequence have enhanced star formation rates and gas masses, as expected from literature scaling relations \citep{Magdis12, Sargent14, Genzel15,Tacconi18,Tacconi20,Scoville14,Scoville16,Silverman15,Silverman18, Elbaz18, Franco20, Feldmann20}. 
These galaxies also lie closer to the "strong starburst" locus, which also implies a higher star formation efficiency.
{On the other hand, compact galaxies within the main sequence occupy the region with low $L_{\rm IR, SF}$ and $L'_{\rm CO(2-1)}$ or $L'_{\rm CO(1-0)}$ along the red solid line in Figure \ref{fig:LCO_LIR}. This suggests that these sources have similar star formation efficiency (or $\tau_{\rm depl}$) to compact off-main sequence sources, but lower star formation rates and gas masses.}
Finally, we highlight that ambiguous objects have low CO(2-1) fluxes suggesting high star formation efficiency. Hence, these sources are likely adding to the compacts population (see also Sect. \ref{subsec:Excitation}).

\begin{figure}
\includegraphics[width=\columnwidth]{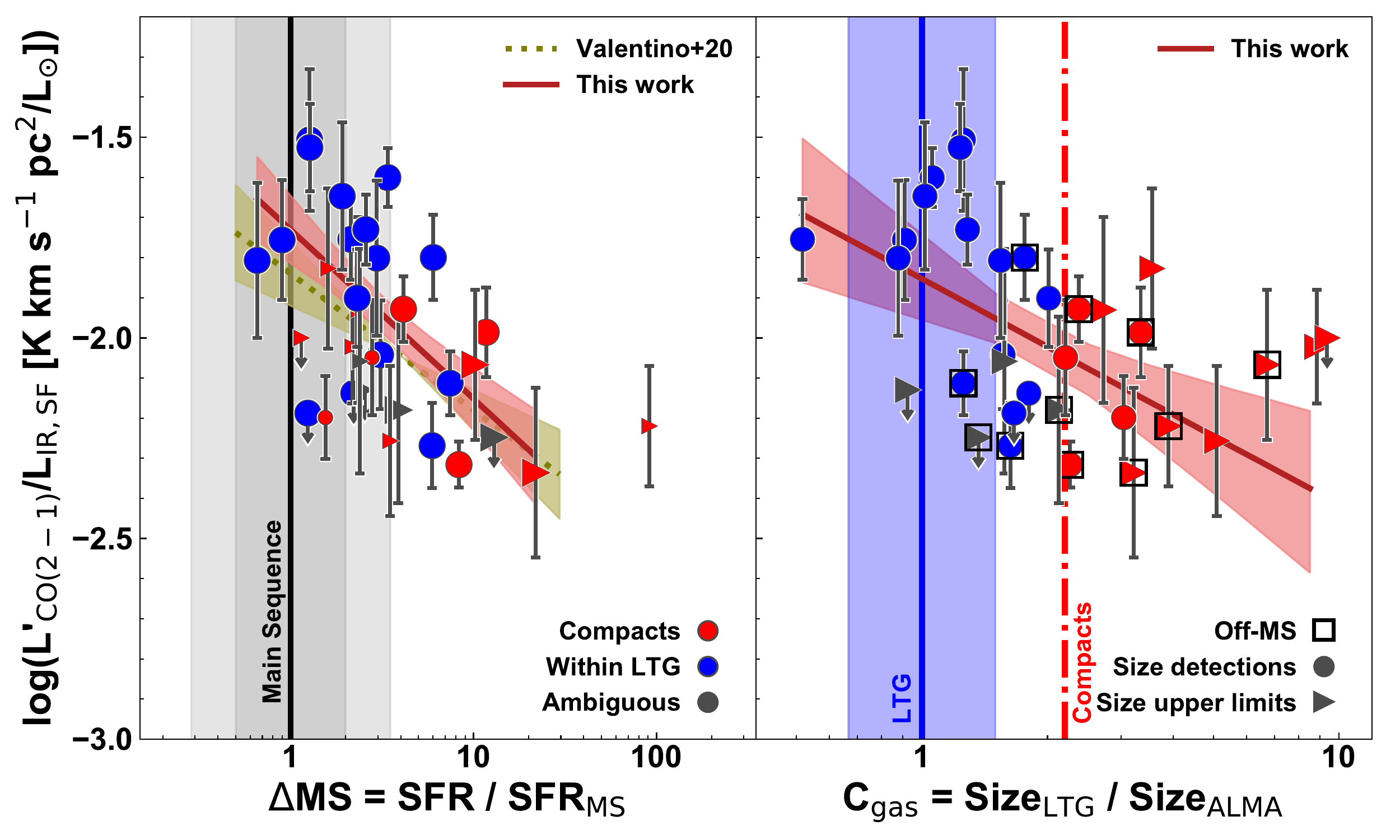}
\caption{ $L'_{\rm CO(2-1)}$ [K km s$^{-1}$ pc$^2$] / $L_{\rm IR, SF}$ [$L_{\odot}$] as a function of the main sequence offset ({\it left}) and the compactness ({\it right}). The green dotted line and shaded area in the left panel represent the best-fit model and 1$\sigma$ confidence interval from \protect\cite{Valentino20}. 
The dark red solid line and shaded area in the left panel indicate the $L'_{\rm CO(2-1)}$ [K km s$^{-1}$ pc$^2$] / $L_{\rm IR, SF}$-$\Delta$MS trend and 1$\sigma$ confidence interval obtained after excluding compact and ambiguous main sequence galaxies from the fit.
The dark red solid line and shaded area in the right panel mark the best fit model and 1$\sigma$ confidence interval of the $L'_{\rm CO(2-1)}$ [K km s$^{-1}$ pc$^2$] / $L_{\rm IR, SF}$ [$L_{\odot}$]-$C_{\rm gas}$ trend. 
The colour-code and symbols are analogous to Fig. \ref{fig:CO_ratios_MS_comp}.
{We measure the uncertainty on the $L'_{\rm CO(2-1)}$/$L_{\rm IR, SF}$ ratio by propagating the 1$\sigma$ errors on the CO(2-1) flux measurements and on $L_{\rm IR, SF}$.}
}
\label{fig:LCO_LIR_MS}
\end{figure}

{To explore the relation between the star formation efficiency, main sequence offset and compactness in our sample, we plot the $L'_{\rm CO(2-1)}/L_{\rm IR, SF}$ ratio as a function of the main sequence position and compactness in Figure \ref{fig:LCO_LIR_MS}.}
{As already observed in Sect. \ref{subsec:Excitation} for the molecular gas excitation, we find that compact galaxies within the main sequence have a $L'_{\rm CO(2-1)}/L_{\rm IR, SF}$ ratio smaller than their extended counterparts, and similar to that measured in galaxies above the main sequence.
Also in this case, compact main-sequence galaxies seem to contribute to the scatter observed in the $L'_{\rm CO(2-1)}/L_{\rm IR, SF}$-$\Delta$MS relation from \citet{Valentino20}.
Similarly to the approach described in Sect. \ref{subsec:Excitation}, we quantify the contribution of compact main-sequence galaxies to the scatter of the $L'_{\rm CO(2-1)}/L_{\rm IR, SF}$-$\Delta$MS trend by fitting the $L'_{\rm CO(2-1)}/L_{\rm IR, SF}$ ratio as a function of the main sequence offset excluding the compact and ambiguous galaxies within the main sequence scatter and the strongest outlier.} As for the previous case, excluding these sources from the fit improves the correlation between $L'_{\rm CO(2-1)}/L_{\rm IR, SF}$ and $\Delta$MS (see Table \ref{tab:linmix_results}). 
We also find a slightly steeper slope and smaller intrinsic scatter than the trend reported in \cite{Valentino20}.
The right panel of Figure \ref{fig:LCO_LIR_MS} shows that the $L'_{\rm CO(2-1)}/L_{\rm IR, SF}$ ratio decreases as a function of the compactness (red line in Figure \ref{fig:LCO_LIR_MS}, see also Table \ref{tab:linmix_results}). 
We note that the slope of this correlation ($\beta = -0.56$) seems steeper than the slope of the $L'_{\rm CO(2-1)}/L_{\rm IR, SF}$ versus $\Delta$MS correlation ($\beta = -0.42$, red line in the left panel of Figure \ref{fig:LCO_LIR_MS}). This is only a suggestion at this stage, however, due to the large uncertainties associated with this parameter.

\subsection{Star formation rate surface density, intensity of the radiation field and dust temperature}
\label{subsect:Sigma_SFR}

\begin{figure}
\includegraphics[width=\columnwidth]{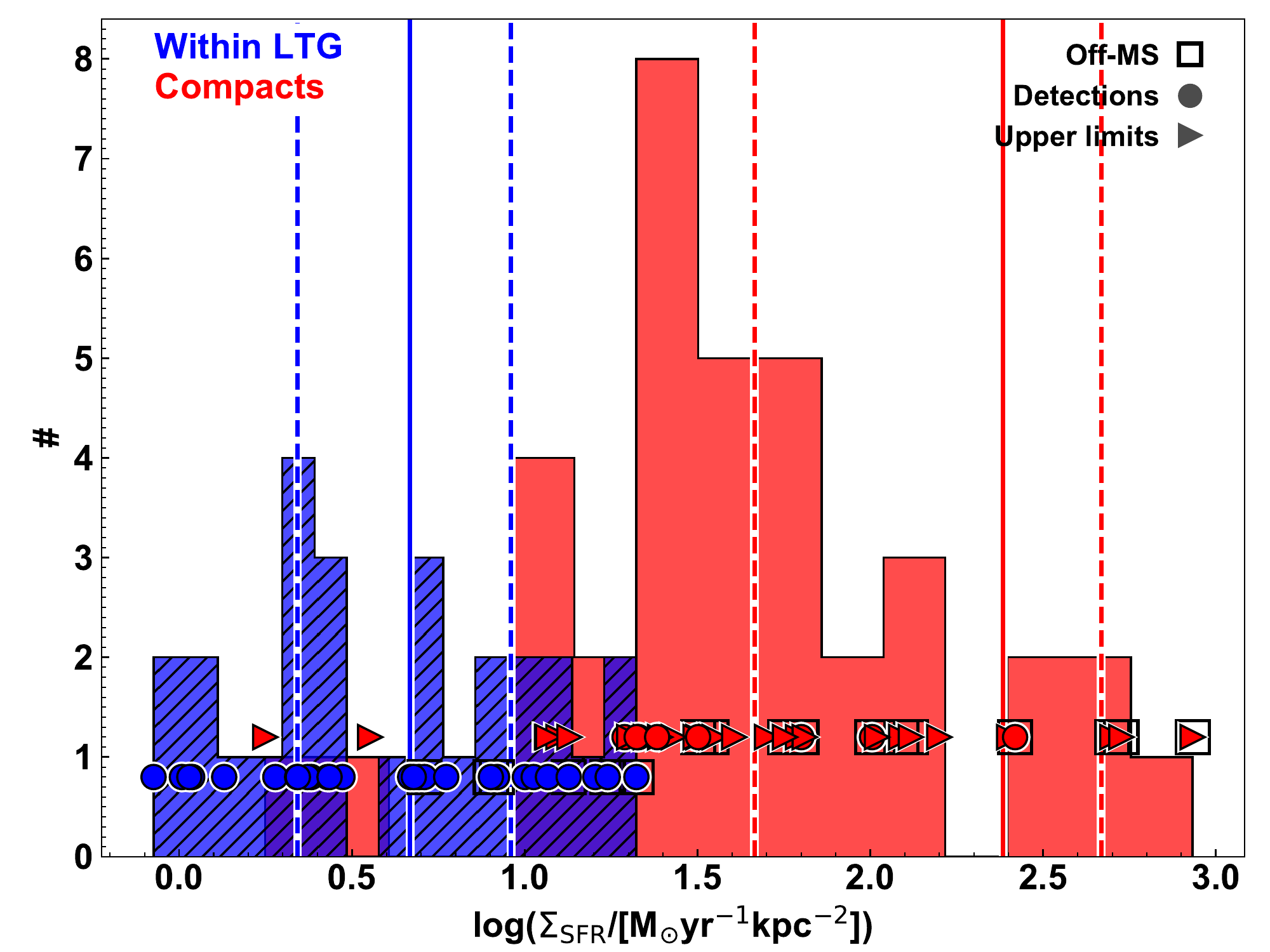}
\caption{ {Star formation rate surface density distribution for compact (red) and extended (hatched blue) galaxies in our sample. Solid and dashed lines indicate the median and interquartile range of each distribution, respectively.
Measurements and upper limits are highlighted with circles and triangles, respectively.
Open black squares indicate galaxies above the main sequence, with $\Delta$MS$\geqslant3.5$.}}
\label{fig:SigmaSFR_distribution}
\end{figure}

{We measure the star formation rate surface density by dividing the far-infrared star formation rate by the area $\Sigma_{\rm SFR} =$ SFR$/(2\pi R_{\rm eff}^{2})$, where $R_{\rm eff}$ is the ALMA effective radius. In Figure \ref{fig:SigmaSFR_distribution} we show the star formation rate surface density distribution for our sample.  
The median star formation rate surface density is $\Sigma_{\rm SFR} = 242^{+223}_{-196}$ M$_{\odot}$yr$^{-1}$kpc$^{-2}$ for compact galaxies and $\Sigma_{\rm SFR} = 5^{+5}_{-3}$ M$_{\odot}$yr$^{-1}$kpc$^{-2}$ for the extended population, with errorbars indicating the interquartile range of the distribution. A log-rank test to the two distributions, accounting for the presence of upper limits on $\Sigma_{\rm SFR}$ shows that the two distributions are significantly different ($p-value \textless 0.001$). 
This shows that compact galaxies have enhanced star formation rate surface density with respect to extended galaxies and similar to local (U)LIRGs \citep[e.g.][]{Liu15b} and galaxies above the main sequence at high redshift \citep{JimenezAndrade19}. This is somewhat expected, since compact and extended galaxies have similar star formation rates (Figure \ref{fig:main_sequence}) while very different sub-millimetre sizes (Figure \ref{fig:mass_size_selection}). This fits the idea that the compactness allows us to identify starburst galaxies, as discussed in Sect. \ref{sec:discussion}. }

\begin{figure}
\includegraphics[width=\columnwidth]{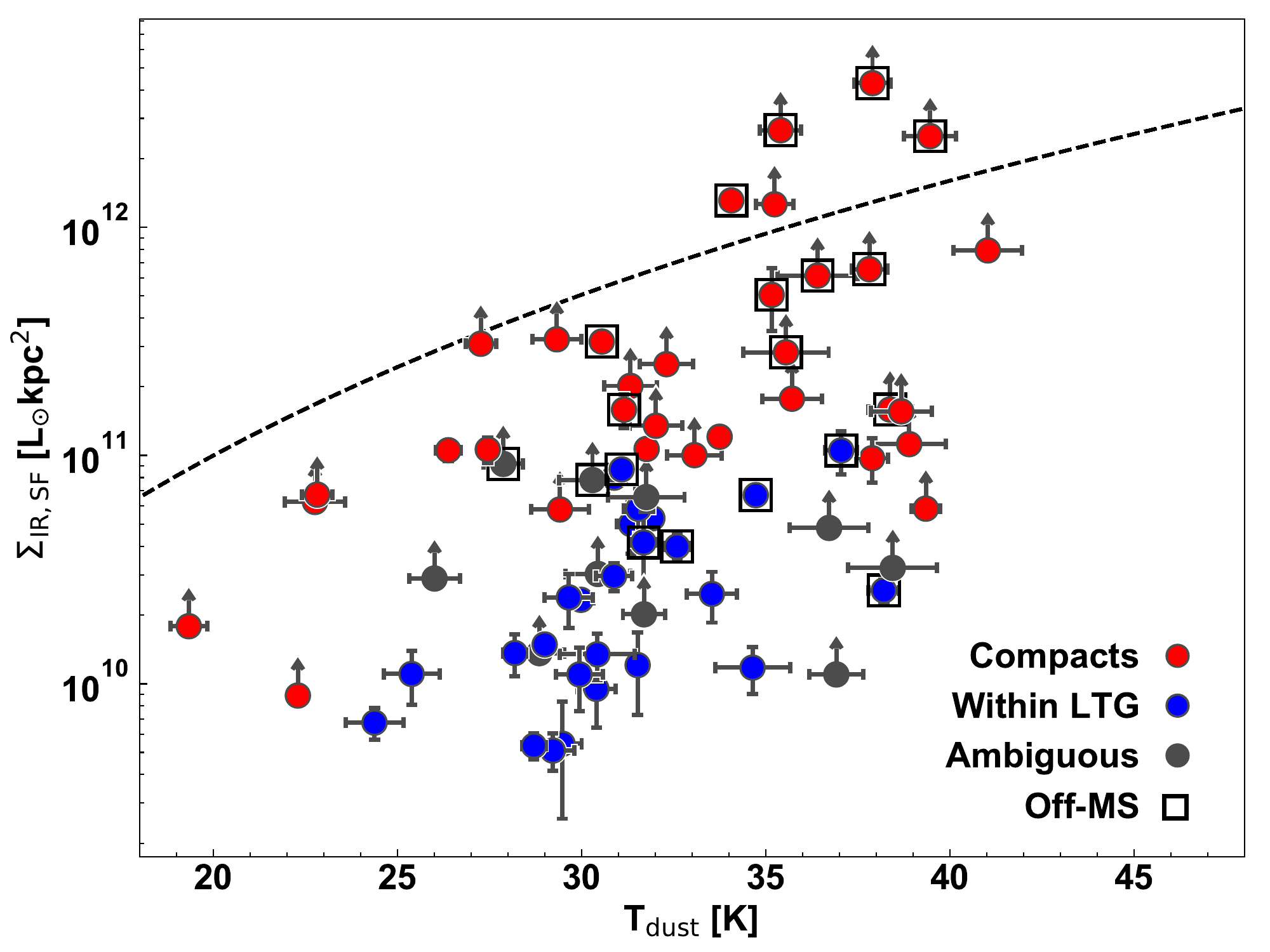}
\caption{ { IR luminosity surface density as a function of the dust temperature for our sample.  The colour-code is analogous to Fig. \ref{fig:CO_ratios_MS_comp}. The dashed line represents the Stefan-Boltzman law for optically-thick dust clouds.}}
\label{fig:SigmaIR_Tdust}
\end{figure}

{In Figure \ref{fig:SigmaIR_Tdust} we show the far-infrared surface density as a function of the dust temperature. Here, $T_{\rm dust}$ is derived from the intensity of the radiation field $\langle U \rangle = (T_{\rm dust}/18.9 \ {\rm K})^{6.04}$, following \citet{Magdis17}. 
Figure \ref{fig:SigmaIR_Tdust} shows that compact galaxies have higher infrared surface density than extended galaxies. The infrared surface density of compact galaxies is instead similar to that measured in starbursts and sub-millimetre galaxies at high redshift \citep{Ikarashi15, Simpson17, Jin19, Hodge19}.  
We note that some compact galaxies in Figure \ref{fig:SigmaIR_Tdust} lie above the limit for optically-thick dust clouds. This might suggests that these galaxies are optically thick and their dust temperature is even warmer than what inferred from the peak of the far-infrared SED.
As a result, the dust mass in those object might be overestimated \citep{Jin19, Cortzen20}.
The overall distribution of compact galaxies is somewhat skewed towards high values of dust temperatures. For the compact galaxies we find a median $T_{\rm dust} = 34 \pm 4$ K, while for the extended population $T_{\rm dust} = 31 \pm 1$ K with the uncertainties corresponding to the interquartile range of the dust temperature distributions of each population.}
 
\begin{figure}
\includegraphics[width=\columnwidth]{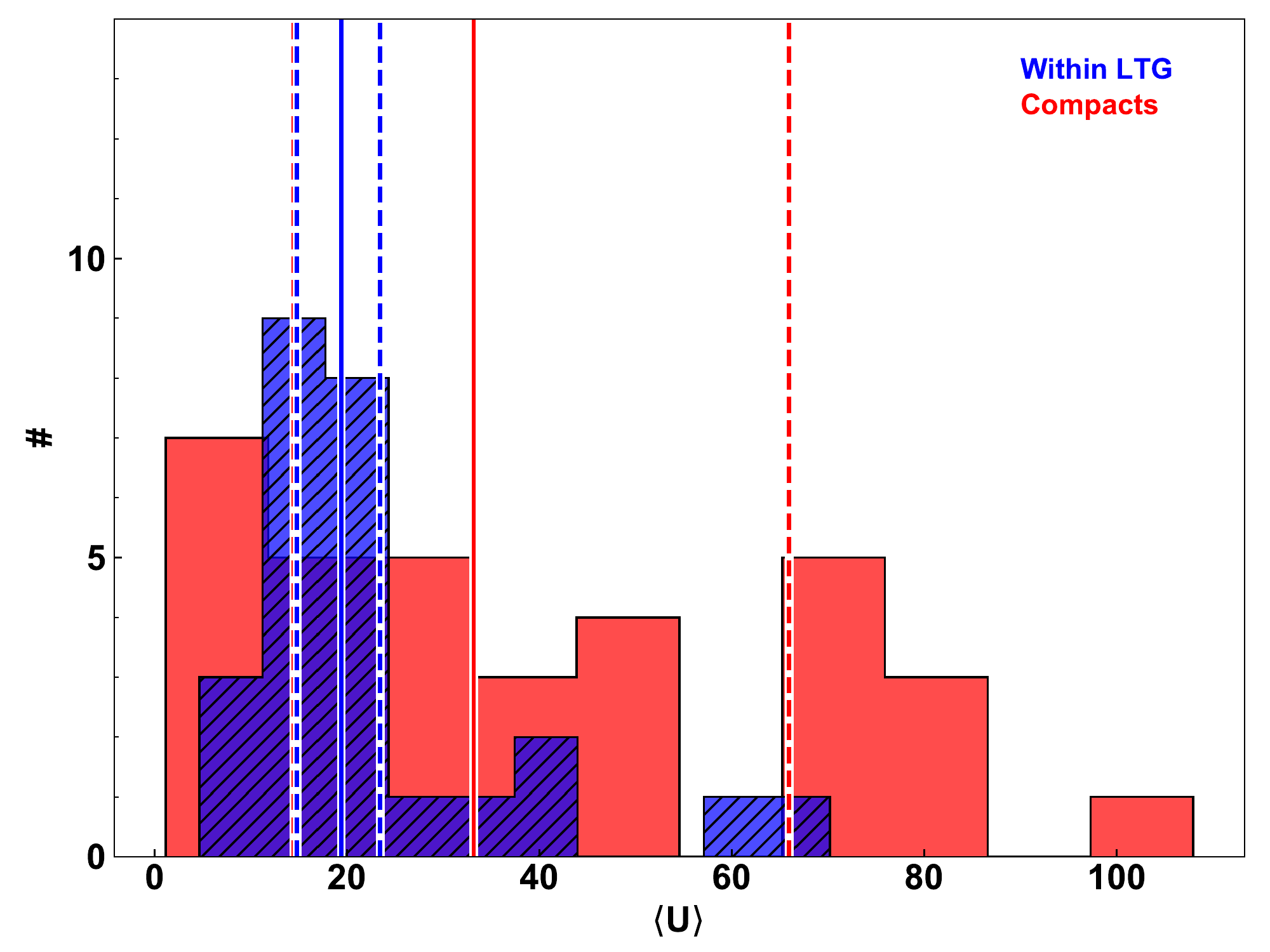}
\caption{ {Distribution of the intensity of the radiation field for compact (red) and extended (hatched blue) galaxies in our sample. 
Solid lines indicate the median of each distribution. Dashed lines represent the interquartile ranges.}}
\label{fig:U_distribution}
\end{figure}

{Figure \ref{fig:U_distribution} shows that compact galaxies have higher intensity of the radiation field than extended galaxies. 
For the compact galaxies we find a median $\langle U \rangle = 33^{+33}_{-19}$, while for the extended population $\langle U \rangle = 19^{+4}_{-5}$. The uncertainties indicate the interquartile range of the distributions.
The intensity of the radiation field is a metallicity-weighted measurement of the star formation efficiency \citep{Magdis12}. Therefore, Figure \ref{fig:U_distribution} provides additional indications of the fact that compact galaxies have enhanced star formation efficiency with respect to the extended population.}

\subsection{Gas content}
\label{subsec:gas}

The mass of the molecular gas reservoir is a crucial quantity to understand the future evolution of galaxies. 
However, measuring the amount of molecular gas in galaxies is notoriously a difficult task \citep[see e.g.][]{Birkin21}. 
This stems from the fact that measuring the molecular gas mass requires choosing an observable-to-gas conversion factor depending on the molecular gas tracer adopted.
When inferring gas masses from the CO luminosity or the dust mass, for example, one needs to assume a CO-to-H$_2$ or a gas-to-dust ratio ($\alpha_{\rm CO}$ and $\delta_{\rm GDR}$, respectively). Both conversion factors have complex dependences on e.g. the state of the ISM, the gas-phase metallicity or galaxy type \citep{Leroy11,Bolatto13,Narayanan12,Magdis12,Genzel15,Silverman18_Pacs787} that are not yet fully understood.
For this reason, we first explore the molecular gas properties of our sample by considering observed $L'_{\rm CO(2-1)}$ luminosities as a proxy. 
We also use $M_{\rm dust}$ as an independent tracer of the molecular gas reservoir.

\subsubsection{Observed quantities}

\begin{figure}
\includegraphics[width=\columnwidth]{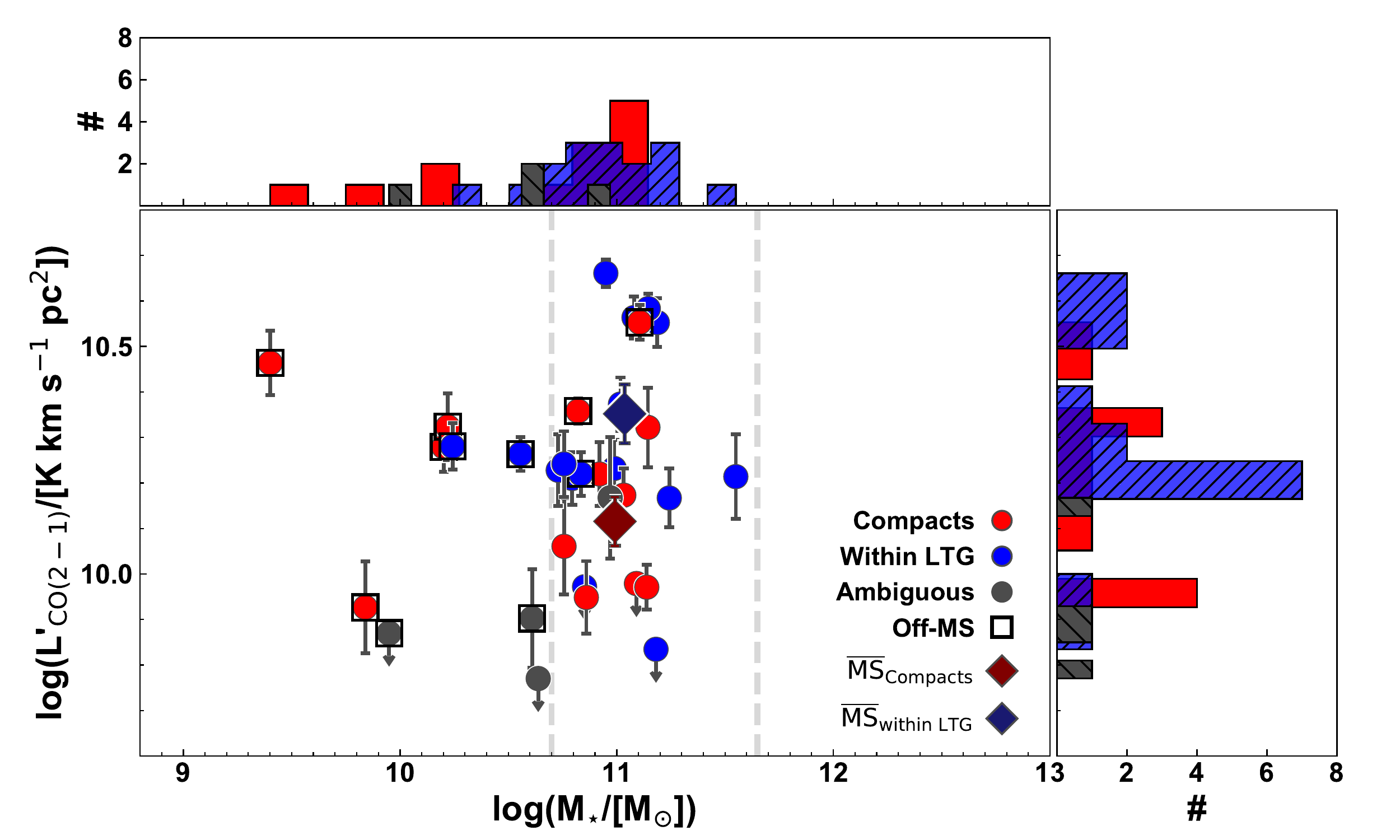}
\caption{$L'_{\rm CO(2-1)}$ [K km s$^{-1}$ pc$^2$] as a function of the stellar mass.
Large filled diamonds are the average $L'_{\rm CO(2-1)}$ luminosities for compact and extended main sequence galaxies with $M_{\star} \geqslant 5 \times 10^{10} \ M_{\odot}$. Vertical dashed grey lines highlight the $M_{\star}$ range considered for computing average $L'_{\rm CO(2-1)}$ values.
Blue circles highlight extended galaxies. Red circles indicate the compacts. Grey large circles represent ambiguous sources. The open black squares highlight galaxies with $\Delta$MS$\geqslant 3.5$. 
{The errorbars represent the 1$\sigma$ uncertainty on the CO(2-1) flux.}
The upper histogram indicate the $M_{\star}$ distribution of galaxies in our sample, split according to their compactness. The histogram on the right indicate their $L'_{\rm CO(2-1)}$ distribution.}
\label{fig:LCO_Mstar}
\end{figure}

The analysis performed in the previous sections suggests that compacts are characterized by a highly excited ISM, enhanced star formation efficiency and shorter depletion time. 
Also, compact galaxies within the main sequence seem under-luminous in CO(2-1), possibly suggesting that their gas content is reduced. 
However, the gas fraction is predicted to decrease with stellar mass \citep[e.g.][and references therein]{Magdis12,Tacconi18,Tacconi20, Liu19}.
To understand whether the trends observed in the previous sections are driven by the stellar mass, we plot in Figure \ref{fig:LCO_Mstar} the $L'_{\rm CO(2-1)}$ luminosity as a function of the stellar mass for our sample.
Despite the scatter, this plot confirms that compact main sequence galaxies are less luminous in $L'_{\rm CO(2-1)}$ than extended and compact off-main sequence galaxies with similar stellar mass, on average. 
To quantify this effect, we compute the average $L'_{\rm CO(2-1)}$ luminosity of compact and extended galaxies within the main sequence. For this computation we consider only galaxies with $M_{\star} \geqslant 5 \times 10^{10} \ M_{\odot}$ {to account for the fact that the gas fraction (hence $L'_{\rm CO(2-1)}$) decreases as a function of the stellar mass. This results in 7 compact and 13 extended galaxies within the main sequence with an average stellar mass of $M_{\star} = 10^{11} \ M_{\odot}$.}
We find $L'_{\rm CO(2-1), compacts, MS} = 1.3 \pm 0.2 \times 10^{10}$ [K km s$^{-1}$ pc$^2$] and $L'_{\rm CO(2-1), extended, MS} = 2.2 \pm 0.3 \times 10^{10}$ [K km s$^{-1}$ pc$^2$]\footnote{Formally biased estimators as we turned the first upper limit into a detection to compute the mean value of the distribution.}. That is, compact main sequence galaxies are $\sim 1.7 \times$ less luminous in CO(2-1) at $\sim 5 \sigma$ significance.

\begin{figure*}
\includegraphics[scale=0.55]{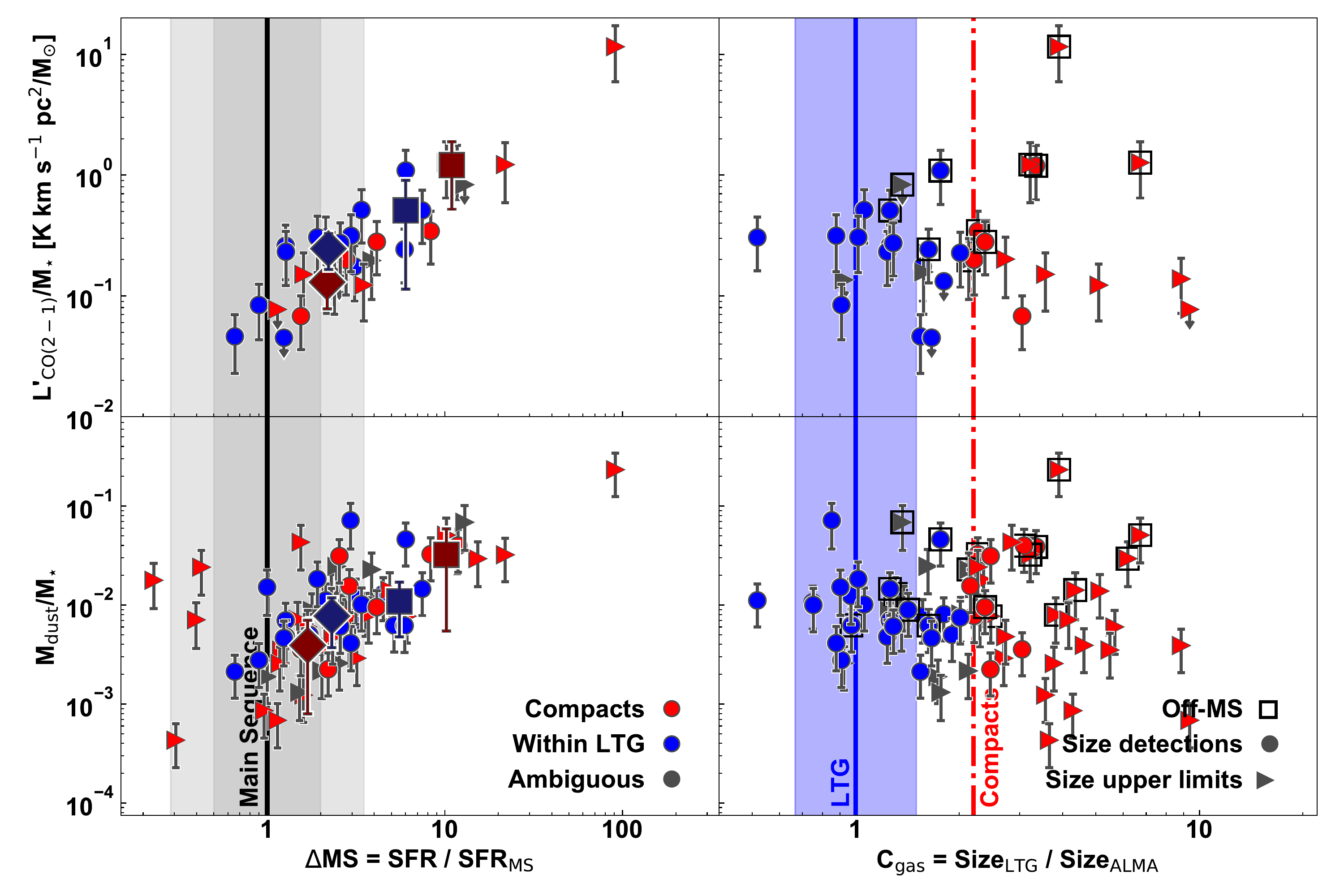}
\caption{ {\it Top row:} $L'_{\rm CO(2-1)}$/$M_{\star}$ ratio as a function of the main sequence offset ({\it left}) and the compactness ({\it right}). 
{\it Bottom row:} $M_{\rm dust}/M_{\star}$ ratio as a function of the main sequence offset ({\it left}) and the compactness ({\it right}). 
In the left panels, large filled diamonds indicate median $L'_{\rm CO(2-1)}$/$M_{\star}$  and $M_{\rm dust}/M_{\star}$ ratios for extended and compact galaxies with $1 \leq \Delta$MS$ \leq 3.5$ while large filled squares indicate median gas fractions for extended and compact sources with $\Delta$MS$\geq 3.5$ {(dark blue and dark red symbols, respectively)}.
The colour code and symbols for individual data-points are analogous to Figure \ref{fig:CO_ratios_MS_comp}.
{The errorbars associated to the individual data-points are obtained by propagating the 1$\sigma$ uncertainty on the CO(2-1) flux measurements and a 0.2 {\it dex} uncertainty on the stellar mass.}
}
\label{fig:fgas_CO_Mdust_MS_compactness}
\end{figure*}

Finally, in Figure \ref{fig:fgas_CO_Mdust_MS_compactness} we show the ratio between the $L'_{\rm CO(2-1)}$ luminosity and stellar mass as a function of the main sequence offset and the compactness. 
In addition, we show the ratio between the dust and stellar mass as a function of the same quantities.
The top-left panel of Figure \ref{fig:fgas_CO_Mdust_MS_compactness} shows that the $L'_{\rm CO(2-1)}$/$M_{\star}$ ratio increases as a function of the main sequence offset.
Compact galaxies on the main sequence, on the other hand, seem to be under-luminous in CO(2-1) with respect to extended sources. 
We find a similar trend for the $M_{\rm dust}$/$M_{\star}$ ratio in the bottom-left panel of Fig.  \ref{fig:fgas_CO_Mdust_MS_compactness}. 
In the right panels of this figure we see that both the  $L'_{\rm CO(2-1)}$/$M_{\star}$ and $M_{\rm dust}$/$M_{\star}$ ratios decrease as a function of the compactness, albeit with a large scatter.

\subsubsection{Derived quantities}
\label{subsub:gas_21}

Building on the results presented in the previous sections, we now convert CO luminosities and dust masses into gas masses by adopting conversion factors that are appropriate for the excitation conditions and star formation efficiency properties of each galaxy.
Clearly our approach will still suffer from the classical uncertainties related to the choice of the observable-to-gas conversion factors. 
{However, the detailed knowledge of the molecular gas properties of both individual galaxies in our sample (see Table \ref{tab:Summary}) and the availability of average CO SLEDs for each of the galaxy sub-populations analysed in this work} allow us to physically motivate the choice of $\alpha_{\rm CO}$/$\delta_{\rm GDR}$ for each galaxy class. 
A detailed analysis of the $\alpha_{\rm CO}$/$\delta_{\rm GDR}$ conversion factors in individual sources will be presented in future papers. 
We highlight here that a common approach in literature is to use the main sequence position of a source to define the preferred observable-to-gas conversion factor \citep[e.g.][]{Magnelli12, Sargent14, Accurso17, Aravena19, Cassata20}. While this might appear a reasonable assumption lacking additional constraints on the galaxy ISM conditions, we urge caution against this approach considering that galaxies within the main sequence display a wide range of compactnesses, CO excitations and star formation efficiencies.

When inferring CO-based gas masses, we only consider galaxies with CO(2-1) observations because this low-J transition is directly tracing the total molecular gas reservoir. We convert $L'_{\rm CO(2-1)}$ to $L'_{\rm CO(1-0)}$ using $R_{21} = L'_{\rm CO(2-1)}/L'_{\rm CO(1-0)} = 0.85$. We then compute the gas mass as $M_{\rm gas} = \alpha_{\rm CO} \times L'_{\rm CO(1-0)}$. 
We adopt $\alpha_{\rm CO} = 3.6$ M$_{\odot}$(K km s$^{-1}$pc$^{2}$)$^{-1}$ for extended galaxies since this value has been suggested to be appropriate for high-redshift, highly star-forming disks \citep[e.g.][]{Daddi15}.
Ambiguous and compact galaxies are instead characterized by a compact CO emission, enhanced CO excitation and high $L'_{\rm CO(2-1)}/L_{\rm IR, SF}$ ratios, resembling the ISM conditions of starbursting objects in the local Universe for which $\alpha_{\rm CO} = 0.8$ M$_{\odot}$(K km s$^{-1}$pc$^{2}$)$^{-1}$ \citep{Solomon87} . Indeed, a starburst-like $\alpha_{\rm CO}$ has been shown to be appropriate for compact galaxies at high redshift \citep{Tadaki17b}. 

When computing gas masses from the dust mass, we consider only galaxies with a a reliable dust mass measurement ($M_{\rm dust}$/ $M_{\rm dust, err} \geqslant 5$). This corresponds to 69 galaxies, significantly improving the statistics with respect to CO-based gas mass estimates.
We compute the gas mass as $M_{\rm gas} = \delta_{\rm GDR} \times M_{\rm dust}$, following \cite{Magdis12}. 
We adopt $\delta_{\rm GDR} = 85$ for the extended population, corresponding to a metallicity dependent gas-to-dust ratio at Z=12 + log(O/H) = Z$_{\sun}$ \citep{Magdis12}. 
On the other hand, we use $\delta_{\rm GDR} = 30$ for compact and ambiguous galaxies since we show that these have "starburst-like" ISM conditions {such as an enhanced CO excitation (as seen from both individual $R_{52}$ ratios and average CO SLEDs, see Sect. \ref{subsec:Excitation}), enhanced efficiency (see Sect. \ref{subsect:SFE}), high SFR and infrared surface density, dust temperatures and intensity of the radiation field (see Sect. \ref{subsect:Sigma_SFR})}, and starbursts at high redshift are reported to have super-solar metallicities \citep{Puglisi17}. 

\begin{figure}
\includegraphics[width=\columnwidth]{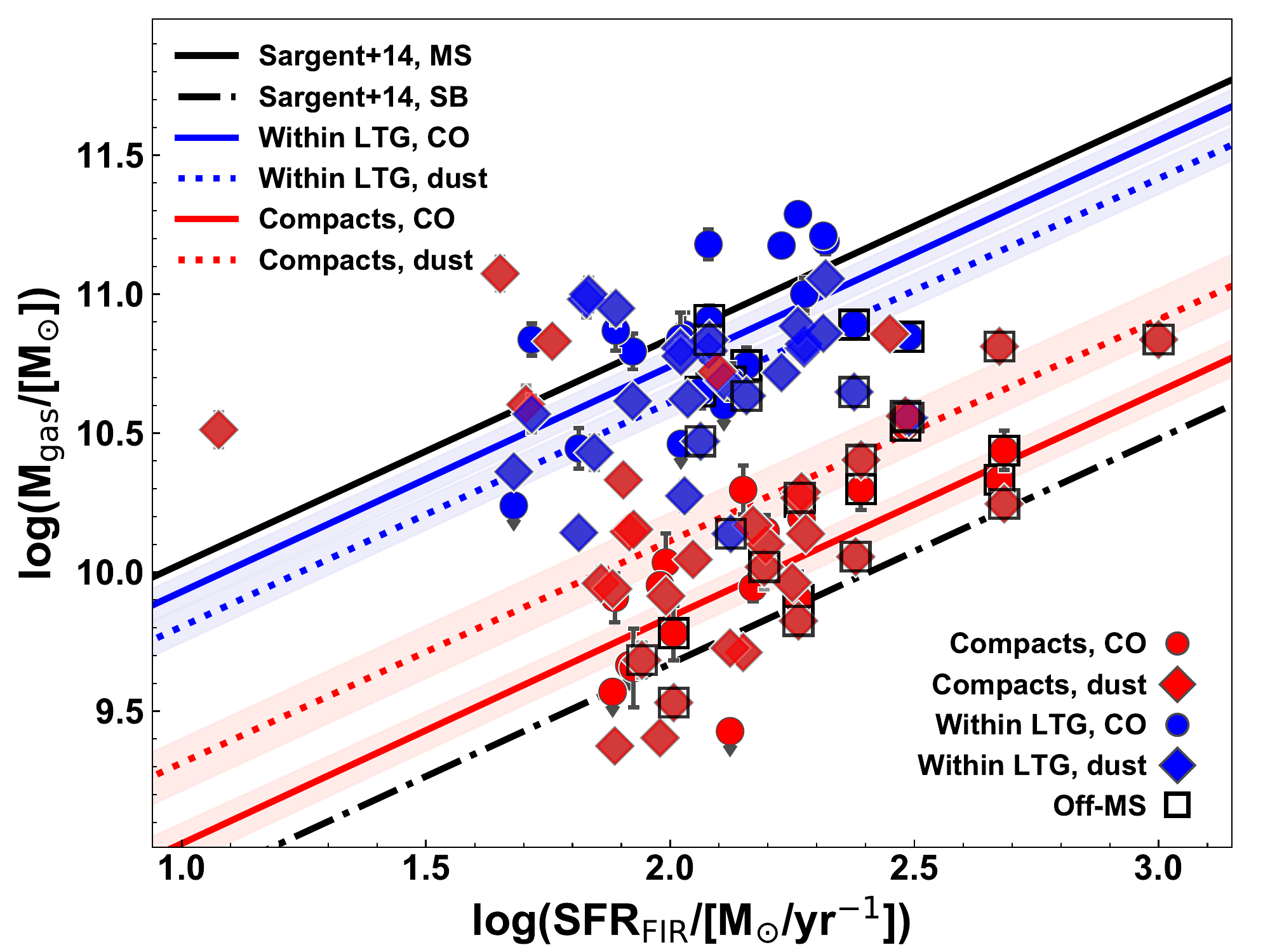}
\caption{Inverse, integrated Schmidt-Kennicutt relation between the star formation rate and the molecular gas mass.
{Coloured filled circles show gas masses derived from CO lines. 
Coloured filled diamonds indicate dust-based gas masses. We report both CO-based and dust-based gas masses when available.}
Blue and red solid (dotted) lines represent the best-fit lines with a slope $\beta$ = 0.81 obtained by fitting the CO- (dust-) based gas mass measurements for the extended and compact population, respectively. The shaded areas represent the 1$\sigma$ confidence interval on the best-fit normalisations.
}
\label{fig:Integrated_SK}
\end{figure}

\begin{figure*}
\includegraphics[scale=0.55]{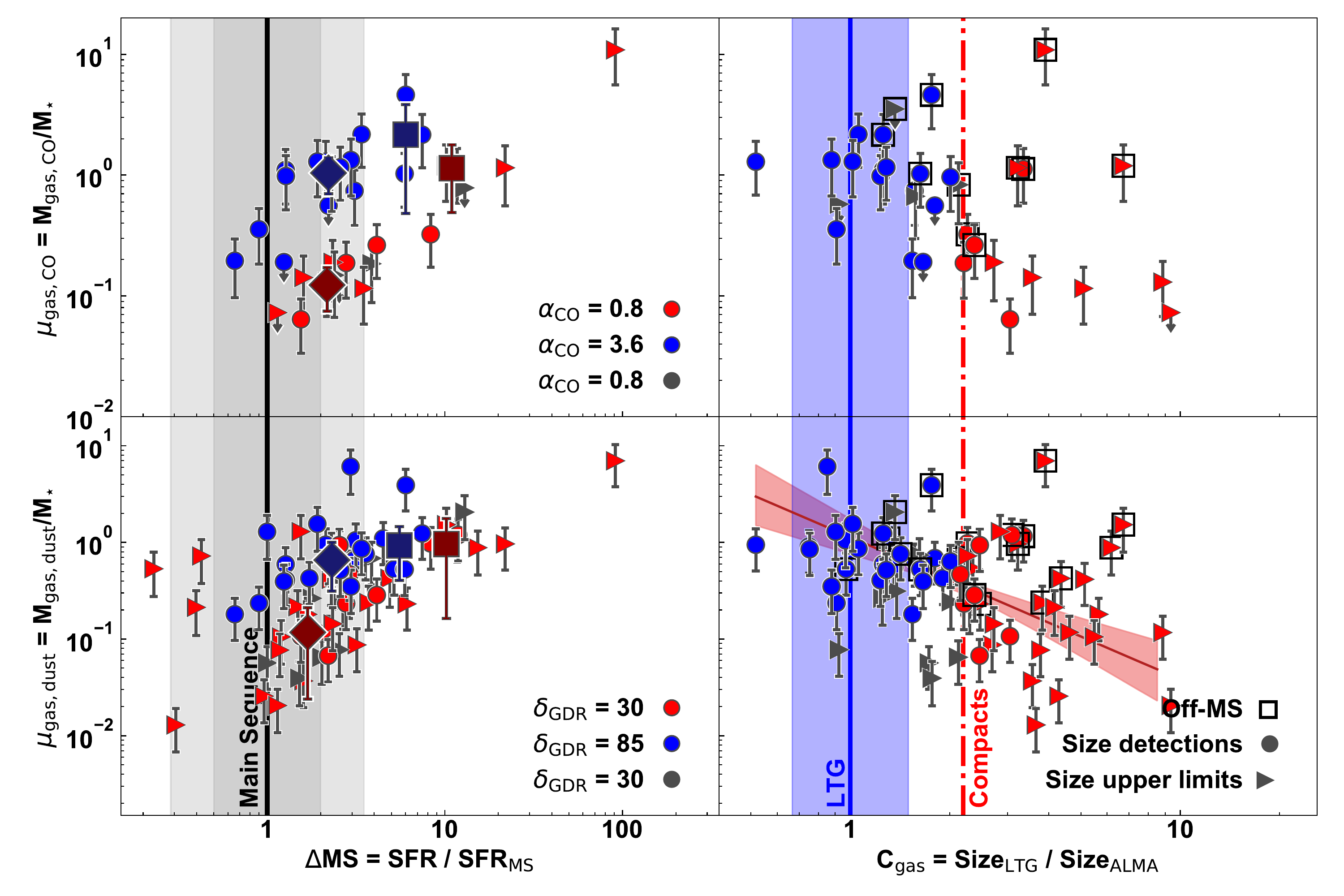}
\caption{ Gas fractions from $L'_{\rm CO(2-1)}$ [K km s$^{-1}$ pc$^2$] ({\it top row}) and $M_{\rm dust}$ [M$_{\odot}$] ({\it bottom row}), as a function of the main sequence offset ({\it left}) and the compactness ({\it right}). 
We convert $L'_{\rm CO(2-1)}$ or $M_{\rm dust}$ to a gas mass by assuming $\alpha_{\rm CO}$ or $\delta_{\rm GDR}$ conversion factors tailored to the compactness properties of each source, as reported in the legend.
Symbols and colour-code are analogous to Figure \ref{fig:fgas_CO_Mdust_MS_compactness}.
{Similarly to Figure \ref{fig:fgas_CO_Mdust_MS_compactness}, error-bars in this plot are obtained by propagating the 1$\sigma$ error on the CO flux and a typical $0.2 dex$ uncertainty on the stellar mass. Therefore, the error-bars do not account for the systematic uncertainty associated to the $\alpha_{\rm CO}$ or $\delta_{\rm GDR}$ conversion factors.}}
\label{fig:fgas_MS_compactness}
\end{figure*}

{In Figure \ref{fig:Integrated_SK}, we show the correlation between the molecular gas mass and the star formation rate for our sample, that, is the integrated Schmidt-Kennicutt relation considering derived quantities rather than pure observables (as instead shown in Figure \ref{fig:LCO_LIR}). 
Here we measure gas masses from both CO lines and dust masses, when available. 
The inclusion of dust-based molecular gas masses allows us to study the relation between the gas mass and SFR with increased statistics with respect to gas mass measurements from the CO(2-1).
We note that Figure \ref{fig:Integrated_SK} would be equivalent to show the ``resolved'' Schmidt-Kennicutt relation considering the molecular gas and star formation rate surface densities for our sample. This is because we find that molecular gas sizes (as sampled by the dust continuum) and SFR sizes (traced by the CO(5-4) emission) are nearly equivalent for our galaxies (see Sec. \ref{subsub:ALMA_biases}), and therefore both axes would be rescaled by the same quantity.}
This plot confirms the results of Figure \ref{fig:LCO_LIR} that extended and compact galaxies have different star formation efficiency properties. 
These galaxies occupy distinct regions of the integrated Schmidt-Kennicutt plane and the offset increases when translating observables into physical quantities, accounting for the ISM properties of each source. 
In particular, if we consider CO-based molecular gas masses, we obtain that the normalisation of the log$(M_{\rm gas})$-$\beta \times$log(SFR$_{\rm FIR}$) relation is $8.21 \pm 0.08$ for compact galaxies and $9.12 \pm 0.07$ for extended sources (red and blue solid lines in Fig. \ref{fig:Integrated_SK}, respectively).
The average offset of compact galaxies with respect to the locus for extended sources is 0.91 {\it dex}. 
This corresponds roughly to a factor of 8 enhancement in star formation efficiency, further stressing the starbursting nature of the ISM in these sources.
When considering dust-based molecular gas masses, the offset of compact galaxies reduces to 0.5 {\it dex} or a factor of 3.2 star formation efficiency enhancement, on average (see red and blue dotted lines in Figure \ref{fig:Integrated_SK}). 
However, the fit seems to be driven by a small number of compact galaxies with a high gas mass in this case. 
{These sources might have been misclassified as compacts since they lie only $\sim 1 \sigma$ below the Mass-Size relation (see also Figure \ref{fig:fgas_MS_compactness}). }
Alternatively, this might reflect the fact that the separation between compact and extended galaxies is not purely bimodal.
However, we still observe that most compact galaxies are shifted towards the starburst locus in the \citet{Sargent14} model. 

We define the gas fraction as $\mu_{\rm gas} = \frac{M_{\rm gas}}{M_{\star}}$ and we show this quantity as a function of the main sequence offset and the compactness in Figure \ref{fig:fgas_MS_compactness}.
To quantify the difference in the gas fraction of compact and extended galaxies on and above the main sequence, we split our sample in two $\Delta$MS bins and we compute the average gas fraction of compact and extended sources within each bin. 
When considering CO-based gas fractions, we find $\mu_{\rm gas, Compacts, MS} = 0.12 \pm 0.05$ and $\mu_{\rm gas, Extended, MS} = 1.04 \pm 0.34$ for compact and extended galaxies within the main sequence. 
If we consider dust-based gas masses, we find $\mu_{\rm gas, Compacts, MS} = 0.12 \pm 0.09$ and $\mu_{\rm gas, Extended, MS} = 0.66 \pm 0.34$. 
{That is, compact galaxies have $\sim 6 - 9 \times$ reduced gas fractions with respect to extended sources, in agreement with previous results for small samples of one or two compact galaxies \citep[][for CO-based gas fractions]{Tadaki17b, Popping17, Brusa18}. }
The median gas fractions are different at $1.2 - 2.7 \sigma$ significance.
On the other hand, galaxies above the main sequence ($\Delta$MS$\geqslant 3.5$) have similar gas fractions as we obtain $\mu_{\rm gas, Compacts, off-MS} = 1.14 \pm 0.70$ and $\mu_{\rm gas, Extended, off-MS} = 2.2 \pm 1.7$ when considering the CO(2-1) luminosity as a molecular gas tracer.
This is confirmed when considering dust masses as proxies for the molecular gas ($\mu_{\rm gas, Compacts, off-MS} = 0.97 \pm 0.80$ and $\mu_{\rm gas, Extended, off-MS} = 0.93 \pm 0.52$).
We also find a tentative trend of decreasing gas fraction as a function of the compactness when considering both CO(2-1) luminosities and dust masses.
We explore the dependence of $\mu_{\rm gas}$ on the compactness by applying a linear regression analysis to the bottom panel of Fig. \ref{fig:fgas_MS_compactness}, owing to the larger statistics available when considering the dust mass as a gas mass tracer.
Indeed, we find a correlation between the gas fraction and the compactness. The results are reported in Table \ref{tab:linmix_results} and the best-fit correlation is shown as a red solid line in the right panel of Figure \ref{fig:fgas_MS_compactness}. 

A small number of compact galaxies within the main sequence show {``main sequence like''} gas fractions when considering dust-based measurements. 
{These sources lie only 1$\sigma$ below the mass-size relation and are close to the compactness limit that we use to discriminate between compact and extended galaxies (red dash-dotted line in the right panels of Figure \ref{fig:fgas_MS_compactness}, e.g.). Hence, these could have been misclassified (due to e.g. noise in the ALMA size measurements) and might rather belong to the extended sample.}
Alternatively our data might suggest that galaxies do not follow a bimodal distribution, but are gradually distributed in the $\mu_{\rm gas}$-$C_{\rm gas}$ plane similarly to our previous findings \citep[see Figure \ref{fig:Integrated_SK} and e.g. Figure 3 in][]{Valentino20}. This seems to be suggested by the right panel of Figure \ref{fig:fgas_MS_compactness} where we see that $\mu_{\rm gas}$ broadly decreases as a function of the compactness. 
However, we argue that Figure \ref{fig:fgas_MS_compactness} overall suggests a more complex dependence of the gas fraction on galaxy properties, possibly as a result of evolutionary trends. We will explore this aspect in Section \ref{sec:discussion}.

We finally note that the difference in $\mu_{\rm gas}$ between compact and extended galaxies on the main sequence depends on the choice of the $\delta_{\rm GDR}$ or $\alpha_{\rm CO}$ conversion factors. {In particular, the tension between compact and extended galaxies within the main sequence is reduced to a factor of $2$ when considering metallicity-dependent conversion factors (see Appendix \ref{Appendix:A2}).}
However, the extensive analysis of the ISM conditions presented in the previous sections, the results obtained from the observables as well as the agreement between molecular gas tracers when using observables-to-gas conversion factors tailored to the compactness properties of each source 
{(see Figures \ref{figA:Mgas_CO_dust} and \ref{figA:LcoMdust_dms_cgas})} corroborate our choice of the conversion factors for each galaxy sub-sample. 
Finally, the use of different conversion factors for compacts and extended galaxies is also corroborated by the results presented in Figure \ref{fig:LCO_LIR} and \ref{fig:Integrated_SK}, suggesting that these sources have different star formation efficiency properties.

\section{Discussion}
\label{sec:discussion}

\subsection{The role of compactness in galaxy evolution}

While the ISM properties of galaxies at $z \sim 1.3$ are weakly correlated with the offset from the main sequence \citep{Valentino20}, a diversity of ISM properties has been observed within the main sequence scatter itself \citep{Elbaz18, Puglisi19}. 
In agreement with these results, here we find that $\sim 46 \%$\footnote{This number would actually be even larger if one would consider galaxies with loose size upper limits as sub-millimetre compact sources, see also Sec. \ref{subsec:Excitation}} of galaxies in our sample have a compact molecular gas reservoir. 
These galaxies have excited CO line ratios, enhanced star formation efficiencies, and are spread on and above the main sequence blurring the $R_{52}$ and $L'_{\rm CO(2-1)}/L_{\rm IR, SF}$ vs $\Delta$MS correlations (see Figures \ref{fig:CO_ratios_MS_comp} and \ref{fig:LCO_LIR_MS}).
These results suggest that galaxies within the main sequence scatter are not all largely unperturbed gas-rich disks.
However, this does not seem to be simply due to a large scatter in the properties of star-forming galaxies.
Our results suggest instead that the diversity of gas excitation conditions and efficiency observed in MS galaxies is associated with the compactness of the molecular gas reservoir.
Indeed, excluding sub-millimetre compact galaxies from the fits in Figures \ref{fig:CO_ratios_MS_comp} and \ref{fig:LCO_LIR_MS} improves the correlation between 
galaxy ISM properties and the main sequence offset.
Furthermore, we find correlations between galaxy ISM properties and the sub-millimetre compactness (see Table \ref{tab:linmix_results}).
Distinguishing galaxies for their sub-millimetre compactness also allows us to select objects with significantly different CO SLEDs (see Figures \ref{fig:CO_SLEDs} and \ref{fig:CO_SLEDs_onoffMS}). 
These results suggest that the compactness of the molecular gas reservoir traces the ISM state of a source.
We thus suggest here that using a sub-millimetre compactness threshold:
\begin{center}
\begin{equation}
C_{\rm gas} = \frac{R_{\rm eff, vdW+14}(z)}{R_{\rm eff, ALMA}} = 
	\frac{A(z) \times (M_{\star}/7 \times 10^{10} M_{\odot})^{\alpha(z)}}{R_{\rm eff, ALMA}} \geqslant 2.2
\label{eqn:Comp}
\end{equation}
\end{center}
would allow us to better distinguish between high redshift, gas-rich disks and galaxies harbouring a highly excited, starbursting ISM.
Here $A(z)$ and $\alpha(z)$ are the best-fit coefficients for the Mass-Size relation of \cite{vanDerWel14}. 
The value $C_{\rm gas}=2.2$ corresponds to $\sim 1 \sigma$ below the optical Mass-Size relation of disks.
We note that this proposed criterion is qualitatively similar to using the star formation rate surface density ($\Sigma_{\rm SFR}$) to select starbursting sources \citep[e.g.][and references therein]{JimenezAndrade19, Valentino20}.
{In fact, compact galaxies have higher star formation rate surface density than extended sources (see e.g. Figure \ref{fig:SigmaSFR_distribution}).}
However, considering the sub-millimetre compactness as in Equation \ref{eqn:Comp} allows us to account for the stellar mass dependence of the star formation rate and size thus rescaling to the structural properties of each object.
In line with our results, various studies \citep[e.g.][]{DownesSolomon98, Combes13, Narayanan14, Bournaud15} have highlighted the importance of the star formation rate surface density as a proxy for the star formation properties of galaxies since this parameter depends on the gas density, temperature and optical depth \citep{Narayanan14}. 

One caveat here is that, as a result of the far-infrared selection, our observations sample the upper stripe of the main sequence scatter at $M_{\star} \sim 5 \times 10^{10} \ M_{\odot}$. Our observations fully probe the 1$\sigma$ scatter of the main sequence ($\pm 0.3$ {\it dex}) only at $M_{\star} \geqslant 10^{11} \ M_{\odot}$. 
Similarly to previous ALMA studies at high redshift \citep[e.g.,][]{Elbaz18, Tadaki20}, our sample thus appears to be biased towards highly star-forming massive galaxies and it might be not trivial to extrapolate our results to lower stellar mass regimes. This is because the high-mass end of the main sequence is the locus where galaxies are expected to quench soon \citep{DekelBirnboim06, Zolotov15}, and/or where the more numerous population of passive galaxies might be temporarily boosted by rejuvenation \citep{Mancini19}. 
Deeper observations of mass-selected samples at $M_{\star} \leqslant 10^{10} \ M_{\odot}$ will be required to understand if our results apply to the main sequence population at lower stellar mass.

\subsection{On the nature of {sub-millimetre} compact galaxies within the MS}
\label{subsec:compacts}

As discussed in the introduction, the discovery of a significant number of sub-millimetre compact galaxies within the scatter of the main sequence conflicts with the idea that galaxies along this sequence are mostly secularly evolving disks.
The lack of a clear trend between galaxy ISM properties and the main sequence offset seems to disfavour alternative scenarios according which galaxies oscillate around the main sequence as a result of compaction episodes (e.g. \citealt{Tacchella16a}).
It has also been proposed that sub-millimetre compact main sequence galaxies are the result of major mergers with a moderate star formation rate enhancement because of the enhanced gas fractions \citep{JimenezAndrade19}. While some of the sub-millimetre compact main sequence galaxies in our analysis might be consistent with being "failed burst" mergers, this would fail to explain why most of these sources have a reduced gas fraction (see Figure \ref{fig:fgas_MS_compactness}).

The investigation of the ISM conditions presented in this paper allows us to shed light on possible formation mechanisms of such objects. 
Sub-millimetre compact galaxies on and above the main sequence have remarkably similar excitation properties (see in particular Figure \ref{fig:CO_SLEDs_onoffMS} and Table \ref{tab:R_SLEDs}) and star formation efficiency (see Sect. \ref{subsect:SFE}) and these properties are enhanced with respect to those of extended galaxies.
This hints at a common origin between {sub-millimetre} compact galaxies on and above the main sequence, likely associated to a merger event.
Mergers are in fact capable of inducing strong inflows to the nuclear regions reducing significantly the size of the molecular gas reservoir and enhancing the  efficiency of star formation \citep{MihosHernquist96}. Merger-driven starbursts are also predicted to have larger gas excitations than highly star-forming disks due to the prevalence of compressive tides enhancing the density of the gas \citep{Bournaud15}. 
On the other hand, sub-millimetre compact galaxies within the main sequence with $M_{\star} \geqslant 10^{10.7} M_{\odot}$ are under-luminous in $L'_{\rm CO(2-1)}$ for their stellar mass and SFR (see Figures \ref{fig:LCO_Mstar} and \ref{fig:fgas_CO_Mdust_MS_compactness}) translating into reduced gas fractions (see Figure \ref{fig:fgas_MS_compactness}).
The reduced gas fractions of sub-millimetre compact galaxies within the main sequence would naturally result from efficient gas consumption during the preceding starburst phase.
Furthermore, when fitting a linear function in the log($\mu_{\rm gas}$)-log($\Delta$MS) plane, we find that sub-millimetre compact galaxies have a larger intrinsic scatter than extended sources ($\sigma_{\rm int, Compacts} = 0.49 \pm 0.08$ versus $\sigma_{\rm int, Extended} = 0.22 \pm 0.09$ in Figure \ref{fig:fgas_MS_compactness}). 
This fits the idea that sub-millimetre compacts within the main sequence are the relic of a previous starburst episode, as in this case we expect to detect sources in different stages of the post-starburst phase.

\begin{figure}
\includegraphics[width=\columnwidth]{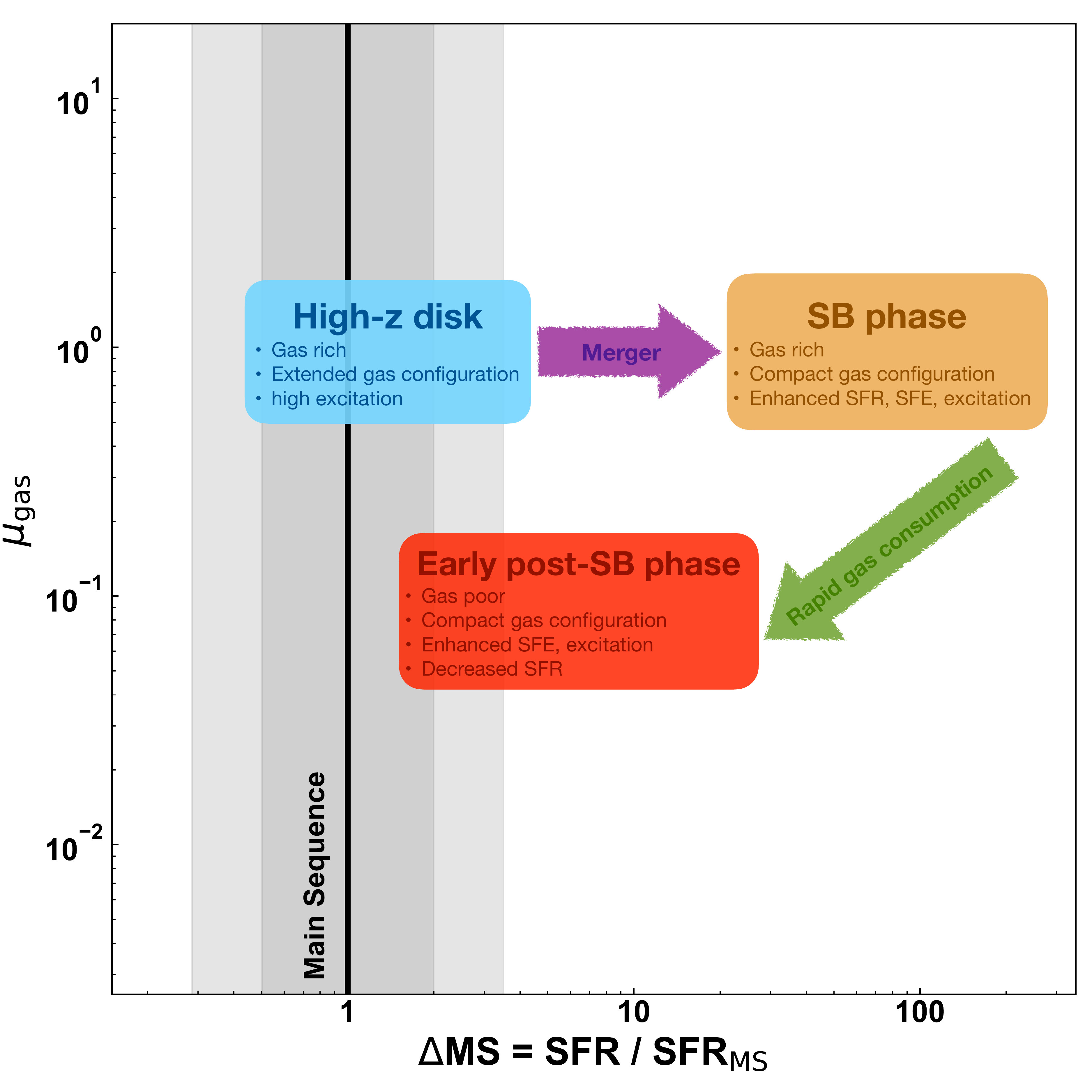}
\caption{A schematic figure summarising how the gas fraction evolves as a function of the main sequence offset according to our proposed {evolutionary scheme}. 
Strong gravitational perturbations (i.e. a major merger) induce an enhancement of the ISM excitation, star formation efficiency and of the star formation rate, and push the galaxy above the main sequence. {The galaxy is gas-rich, has a compact molecular gas reservoir and} moves roughly horizontally acros the plane.
The galaxy rapidly consumes the gas and reduces its star formation rate while retaining a compact {molecular gas} configuration, enhanced star formation efficiency and ISM conditions. 
The galaxy goes back to the main sequence, moving to the bottom left corner of the plane.}
\label{fig:evo_trend_cartoon}
\end{figure}

Therefore, the results of this paper support the scenario proposed in \cite{Puglisi19} and suggest that sub-millimetre compact massive galaxies within the main sequence represent transient objects in an ``early post-starburst phase'' following a merger-driven starburst episode (see also \citealt{Elbaz18} and \citealt{Franco20}). 
We schematically summarize this proposed evolutionary trend in Figure \ref{fig:evo_trend_cartoon}.
Our results and proposed scenario are consistent with cosmological simulations suggesting that compact galaxies form by repeated major mergers of small progenitors \citep{Chabanier20}. These simulations predict gas fractions of $\sim 20 \%$ in compact star-forming galaxies versus $\sim 50 \%$ for extended main-sequence galaxies at $z \sim 2$, as a result of gas consumption by star formation, as well as gas exhaustion in major mergers which can efficiently consume and/or expel gas and in good agreement with our results.

\begin{figure}
\includegraphics[width=\columnwidth]{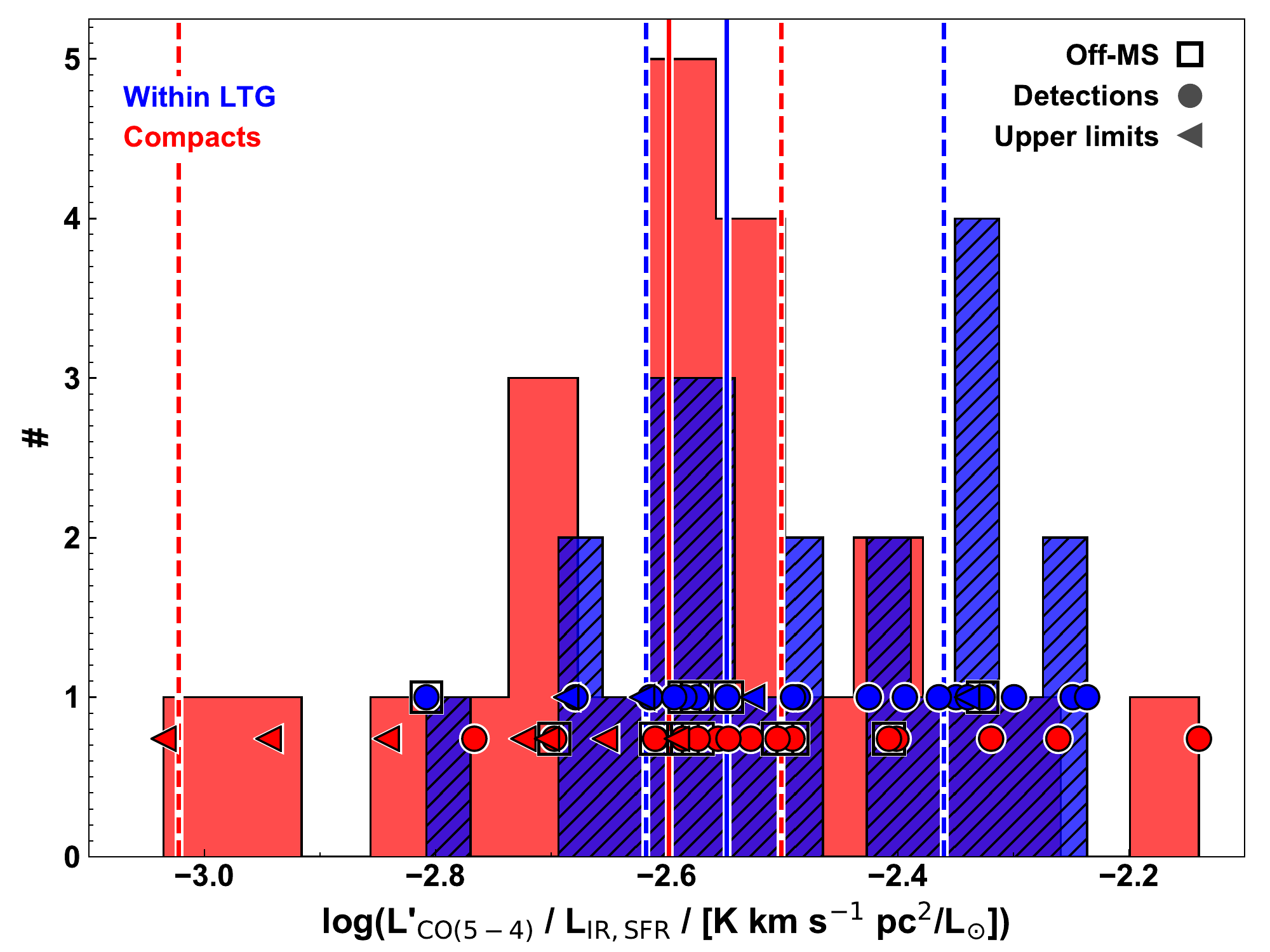}
\caption{ {Distribution of the $L'_{\rm CO(5-4)}/L_{\rm IR, SF}$ logarithmic ratio for sub-millimetre compact (red) and extended (hatched blue) galaxies in our sample. Solid and dashed lines indicate the median and interquartile range of each distribution, respectively. 
The log($L'_{\rm CO(5-4)}/L_{\rm IR, SF}$) measurements and upper limits are highlighted with circles and triangles, respectively.
Open black squares indicate galaxies above the main sequence, with $\Delta$MS$\geqslant3.5$.}}
\label{fig:L54_IR_compactness}
\end{figure}

{A visual inspection of the HST imaging available for this sample does not reveal any clear evidence of an enhanced merger fraction in the sub-millimetre compact population within the main sequence. However, this does not necessarily contradict the idea that sub-millimetre compact galaxies are associated to mergers. In fact, only F814W HST imaging is homogeneously available for the full sample, sampling the rest-frame UV emission at $z \sim 1.3$. Therefore, strong dust attenuation effects might hamper an accurate merger classification \citep[e.g.]{Cibinel19}. Furthermore, in our proposed interpretation, sub-millimetre compact galaxies within the main sequence are ``early post-starburst'' galaxies, hence likely observed at or somewhat past the coalescence phase.
Therefore, we would not necessarily classify these sources as mergers. In fact, morphological classification criteria are able to identify mergers up to the coalescence phase, hence before disturbances or asymmetries in the imaging fade away \citep[see discussion in][and references therein]{Puglisi19}.}
{On the other hand, the $L'_{\rm CO(5-4)}/L_{\rm IR, SF}$ ratio might provide further indications on the evolutionary stage of our sources.
Figure \ref{fig:L54_IR_compactness} shows the distribution of the $L'_{\rm CO(5-4)}/L_{\rm IR, SF}$ logarithmic ratio for galaxies in our sample, distinguished for their sub-millimetre compactness. We measure log($L'_{\rm CO(5-4)}/L_{\rm IR, SF}/ {\rm [K \ km s^{-1} pc^2] / L_{\odot}]})= -2.6^{+ 0.4}_{- 0.1}$ in sub-millimetre compact galaxies and 
log($L'_{\rm CO(5-4)}/L_{\rm IR, SF}/{\rm [K \ km s^{-1} pc^2] / L_{\odot}} = -2.5^{+ 0.10}_{- 0.2}$) in sub-millimetre extended sources. Furthermore, we apply a log-rank test to the two distributions, accounting for the presence of upper limits on the log($L'_{\rm CO(5-4)}/L_{\rm IR, SF}$) ratio and we find a $\sim 87\%$ probability that the two distributions are different ($p-value = 0.13$). 
That is, we find marginal evidence that sub-millimetre compact galaxies have lower $L'_{\rm CO(5-4)}/L_{\rm IR, SF}$ ratio compared to sub-millimetre extended sources.
The CO(5-4) luminosity correlates linearly with the far-infrared luminosity from star formation and this has been interpreted as an evidence that $L'_{\rm CO(5-4)}$ traces dense, star-forming molecular gas \citep[][]{Bayet09,Greve14, Liu15, Daddi15, Valentino20, Cassata20}. Therefore, this might suggest that the sub-millimetre compact population is caught in a declining phase of the starburst since the "instantaneous star formation rate", as traced by the CO(5-4) luminosity \citep{Daddi15}, is lower than the star formation rate averaged on a $\sim 100$ Myr time-scale, as traced by the far-infrared luminosity \citep{Kenni98}.
However, we measure a difference only at the $\sim 2 \sigma$ level. Future studies with larger statistics will allow us to confirm this result.}

We note that compact star-forming galaxies at high redshift have already been proposed as a key population for our understanding of massive, quenched galaxies formation \citep{Barro13, Barro14, Barro16, Nelson14, vanDokkum15}. 
These so-called ``blue nuggets'' are however selected for their compact size in the optical.
This selection might be therefore biased to a later phase of the quenching process where the compact stellar core has been already formed. 
Conversely, our selection seems to identify an early phase of the compact core build-up (see Figure \ref{fig:mass_size_selection}). 
This might explain why extended SFGs and ``blue nuggets'' present similar ISM conditions except than in the radio regime which samples the starburst activity on longer time-scales \citep[0-400 Myr, see][and references therein]{GomezGuijarro19}. The radio properties of ``blue nuggets'' suggest that these galaxies are old starbursts \citep{GomezGuijarro19} and might indicate the existence of an evolutionary link between sub-millimetre compacts, optically-compact galaxies and passive ellipticals.

Selecting compact galaxies by means of the molecular gas size likely allows us to identify galaxies in an early phase of the transition to a passive, bulge dominated galaxy retaining the imprints of their formation mechanism. 
Future studies of this population will provide a new perspective for our understanding of quenching processes and passive galaxies formation which typically display post-starburst features \citep{Belli19, Deugenio20}.
A possible connection with the quenched population is also suggested by the tentative evidence that {sub-millimetre} compact galaxies have an enhanced AGN fraction (see Figure \ref{fig:mass_size_selection}), in agreement with recent literature results \citep[][]{Elbaz18, Scholtz20}.
Consistently with this result, major mergers can in fact enhance the accretion activity onto the central black hole \citep{SpringelHernquist05} and expel large quantities of gas via, e.g., tidal tails \citep{Puglisi21} affecting the future star formation processes in the remnant. 
{Finally, the existence of an evolutionary link between ``sub-millimetre compact'' and quenched galaxies might also be suggested by the extremely compact size of high-redshift post-starburst galaxies that are likely to be formed through dissipative collapse of gas and rapid star formation prior to quenching \citep{Almaini17, Maltby18}, consistently with our findings. }

\section{Summary}
\label{sec:summary}

In this work we presented a characterisation of the molecular gas properties and molecular gas content of galaxies as a function of the compactness of the molecular gas reservoir by using ALMA observations of several CO and [CI] transitions for a sample of 82 far-infrared selected galaxies at $z \sim 1.3$ in COSMOS. We measured the compactness of the molecular gas reservoir by comparing measurements of the molecular gas size to the optical $M_{\star}$-Size relation for disks at $z \sim 1.25$. We then investigated the relation between the molecular gas properties, the offset from the main sequence and the sub-millimetre compactness. We further measured molecular gas masses from multiple molecular gas tracers to gain insights on the origin of sub-millimetre compact galaxies within the main sequence. 
The main findings of this paper are as follows:

\begin{itemize}

\item  The $\gtrsim$46\% of galaxies in our sample have a molecular gas reservoir more compact than the stellar size of typical massive star-forming galaxies at $z \sim 1.3$. The effective radius of the molecular gas reservoir in these sources is on average $\geqslant 3.3 \times$ smaller than their $K_{\rm s}$-band effective radius. 
The compactness of the molecular gas reservoir shows no significant correlation with the main sequence position

\item Sub-millimetre compact galaxies have enhanced CO(5-4)/CO(2-1) and reduced CO(2-1)/$L_{\rm IR, SF}$ luminosity ratios with respect to sub-millimetre extended galaxies, implying enhanced CO excitation and star formation efficiencies.
A significant number of these sources are located within the scatter of the main sequence, blurring the correlation between these ratios and the main sequence offset reported in our previous analysis. Both the CO(5-4)/CO(2-1) and CO(2-1)/$L_{\rm IR, SF}$ luminosity ratios correlate with the sub-millimetre compactness.

\item The average CO SLED of sub-millimetre compact galaxies up to CO(7-6) is consistent with that of the most extreme main sequence outliers at $z \sim 1.3$.  On the other hand, galaxies with an extended molecular gas reservoir have a less excited average CO SLED.
This mirrors results from individual line ratios and suggest that the sub-millimetre compactness provides a good indicator for the CO excitation conditions of a galaxy.
Furthermore, this stresses the fact that high-J CO transitions should not be used to derive total molecular gas masses without prior knowledge of the excitation conditions of a galaxy since large variations of the CO excitation are observed within the main sequence itself.

\item We find that sub-millimetre extended and compact galaxies occupy distinct regions in the integrated Schmidt-Kennicutt plane and, in particular, sub-millimetre compact galaxies have enhanced $L_{\rm IR, SF}$ for a given CO(2-1) luminosity (Figure \ref{fig:LCO_LIR}) indicating enhanced star formation efficiencies similarly to starbursts (Figure \ref{fig:Integrated_SK}). {A higher star formation efficiency for the sub-millimetre compact population is also indicated by the intensity of the radiation field, inferred from the far-infrared spectral energy distribution.}

\item {We find that sub-millimetre compact galaxies have higher SFR surface density than extended sources, similarly to local starburst galaxies and high-redshift SMGs. We find indications that sub-millimetre extended galaxies have higher dust temperatures than extended sources, althought the average dust temperatures are consistent within the errorbars.}

\item {We find that sub-millimetre compact galaxies within the main sequence are under-luminous in CO(2-1) with respect to sub-millimetre extended main sequence galaxies and off-main sequence sources with similar stellar mass (Fig. \ref{fig:LCO_Mstar}) and SFR (Fig. \ref{fig:fgas_CO_Mdust_MS_compactness}). 
Using both the CO(2-1) luminosity and the dust mass as molecular gas tracers, and using $\alpha_{\rm CO}$ and $\delta_{\rm GDR}$ conversion factors tailored to the ISM conditions of our sources, we find that the gas fraction mildly increases as a function of the main sequence offset, in qualitative agreement with published scaling relations. However, we find that sub-millimetre compact galaxies within the main sequence have reduced gas fractions on average, but with a large scatter. 
While the magnitude of this offset depends  on the $\alpha_{\rm CO}$ and $\delta_{\rm GDR}$ prescriptions, these results suggest that sub-millimetre compact galaxies within the main sequence have lower gas fractions with respect to their extended main-sequence counterparts.}

\end{itemize}

Overall, this study shows that the structural properties of galaxies at long wavelengths are a crucial ingredient for interpreting the main sequence of star-forming galaxies at $z \sim 1$. 
While currently limited to $M_{\star} \geqslant 10^{11} \ M_{\odot}$ where our selection allows us to fully probe the main sequence scatter, our analysis suggests that the compactness of the molecular gas reservoir allows to identify sources with a highly excited, starbursting ISM. Similarly to the star formation rate surface density, this parameter provides a good proxy for the ISM conditions of a galaxy while also allowing to rescale for its individual structural properties showing a dependence on the stellar mass.
Future crucial steps for our understanding of star formation in distant galaxies will include to perform studies of the molecular gas properties in $M_{\star}$-selected samples of galaxies with lower stellar masses, to understand if these results also apply to the full main sequence population. 
Another critical aspect to understand the properties of star-forming galaxies on and above the main sequence at high-$z$ will be to study large galaxy samples at high spatial resolution in the far-infrared/sub-millimetre regime, where spatially resolved studies have not yet reached the statistics of spatially resolved optical surveys \citep[e.g.][]{FS09, Stott16}. 
This study has also allowed us to shed light on the origin of sub-millimetre compact galaxies within the main sequence which are now commonly detected in sub-millimetre/radio surveys at $z \geqslant 1$. We argue that their high CO excitation and star formation efficiency, and reduced gas fractions suggest that these sources are associated with an evolutionary phase of the merger. In particular, these properties are consistent with sources in an ``early post-starburst'' phase in which the star formation rate has declined and the gas fraction has been reduced as a result of efficient gas consumption while the galaxy retains enhanced excitation and efficiency. 
Future studies of this population will be crucial for our understanding of passive galaxies formation. 

\section*{Acknowledgements}

We would like to thank the referee for their constructive report which significantly improved the content and clarity of the paper.
AP acknowledges funding from Region Île-de-France and Incoming CEA fellowship from the CEA-Enhanced Eurotalents program, co-funded by FP7 Marie-Skłodowska-Curie COFUND program (Grant Agreement 600382). AP also gratefully acknowledges financial support from STFC through grants ST/T000244/1 and ST/P000541/1.
{MTS acknowledges support from a Scientific Exchanges visitor fellowship (IZSEZO 202357) from the Swiss National Science Foundation.}
AP thanks Ivan Delvecchio and Ian Smail for helpful discussions.

\section*{Data availability}
The data used in this paper are available through the ALMA data archive.


\bibliographystyle{mnras}
\bibliography{bibliography}


\appendix

\section{Exploring the dependence of $\mu_{\rm gas}$ of main-sequence galaxies on $\alpha_{CO}$ and $\delta_{\rm GDR}$}
\label{Appendix:A2}

{In this appendix we explore the dependence of the results presented in Sect. \ref{subsec:gas} and Figure \ref{fig:fgas_MS_compactness} on the choice of the observable-to-gas conversion factor. The main goal of this test is to explore the difference in gas fraction between extended and compact galaxies on the main sequence, since this aspect is critical for our interpretation of the sub-millimetre compact population within the main sequence (see Sect. \ref{sec:discussion}).
To quantify the dependence of our results on the observable-to-gas conversion factors, we compute gas fractions for galaxies on the main sequence by adopting metallicity dependent $\alpha_{\rm CO}$ and $\delta_{\rm GDR}$ conversion factors, following the prescription of \citet{Magdis12}.}
The metallicity of our galaxies is $8.43 \leqslant Z \leqslant 8.78 $ with a median value of $Z = 8.71$, as derived from the fundamental mass metallicity relation of \citet{Mannucci10} {(FMR)} considering a \citet{PP04} metallicity scale.

\begin{figure*}
\includegraphics[scale=0.55]{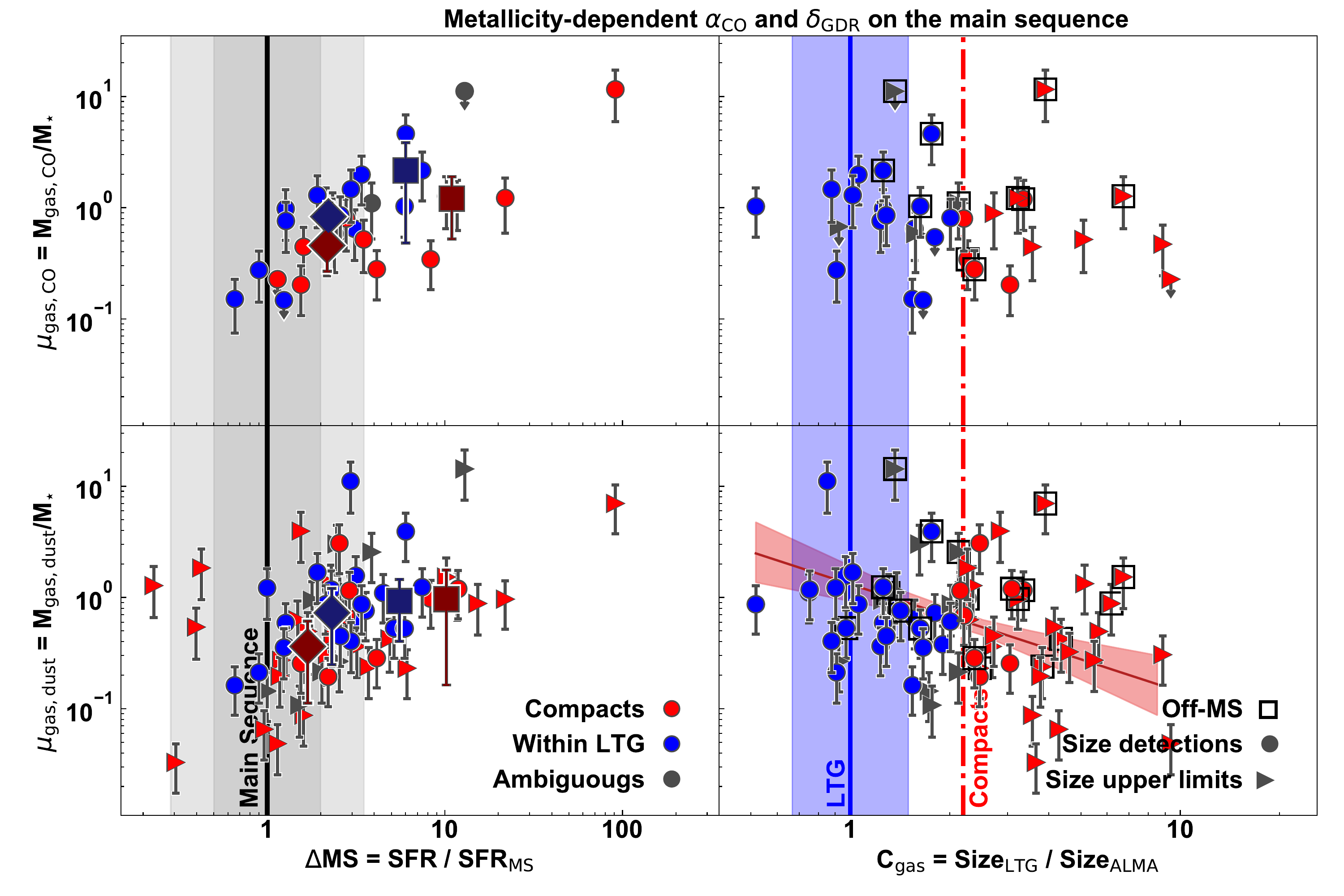}
\caption{ {As Figure \ref{fig:fgas_MS_compactness} but using metallicity-dependent $\alpha_{\rm CO}$ and $\delta_{\rm GDR}$ conversion factors for galaxies on the main sequence. We consider constant conversion factors for galaxies above the main sequence, as in the main text.}}
\label{figA:fgas_MS_compactness_Z}
\end{figure*}

{In the top row of Figure \ref{figA:fgas_MS_compactness_Z}}, we show the gas fraction from $L'_{\rm CO(2-1)}$ when adopting a metallicity-dependent $\alpha_{\rm CO}$ for galaxies within the main sequence. We find median gas fractions of $\mu_{\rm gas, Compacts, MS} = 0.46 \pm 0.20$ and $\mu_{\rm gas, Extended, MS} = 0.83 \pm 0.27$. 
{In the bottom row of Figure \ref{figA:fgas_MS_compactness_Z}, we show} the gas fractions from $M_{\rm dust}$ by adopting a $\delta_{\rm GDR}$ that varies as a function of the metallicit for galaxies within the main sequence. We find median gas fractions of $\mu_{\rm gas, Compacts, MS} = 0.13 \pm 0.10$ and $\mu_{\rm gas, Extended, MS} = 0.26 \pm 0.16$. 
Finally, using a metallicity-dependent $\delta_{\rm GDR}$ provides a slope, normalization and intrinsic scatter for the $C_{\rm gas}$-$\mu_{\rm gas, dust}$ relation that is consistent within the errorbars to that reported in Table \ref{tab:linmix_results} while providing a lower correlation index ($\rho = -0.3$).
{This figure} shows that compact main sequence galaxies show a reduced gas fraction, on average, although considering the metallicity dependence of the $\alpha_{\rm CO}$ and $\delta_{\rm GDR}$ reduces the difference with respect to the extended population.
We summarise our results for the average gas fractions measured for galaxies on and above the main sequence with different tracers and observable-to-gas conversion factors prescriptions in Table \ref{tabA:fgas_avg}.

\begin{table*}
\centering
\caption{Average gas fractions for extended and compact galaxies across the main sequence using different observable-to-gas conversion factors.
\\$^{\rm (a)}$ $1 \leqslant \Delta$MS$\textless 3.5$
\\$^{\rm (b)}$ $\Delta$MS$\geqslant 3.5$ }
\label{tabA:fgas_avg}
\begin{tabular}{ccccc} 
\hline
\hline
		Method & 
		$\mu_{\rm gas, Extended, MS}^{\rm (a)}$ & $\mu_{\rm gas, Compacts, MS}^{\rm (a)}$ & $\mu_{\rm gas, Extended, off-MS}^{\rm (b)}$  & $\mu_{\rm gas, Compacts, off-MS}^{\rm (b)}$  \\
		\hline
		Bimodal $\alpha_{\rm CO}$ - Compactness &  1.04 $\pm$ 0.34 & 0.12 $\pm$ 0.05 & 2.2 $\pm$ 1.7 & 1.14 $\pm$  0.7 \\
		$\alpha_{\rm CO}$(Z) &  0.83 $\pm$ 0.27 & 0.46 $\pm$ 0.20 & - & - \\
		Bimodal $\delta_{\rm GDR}$ - Compactness & 0.66 $\pm$ 0.34 & 0.12 $\pm$ 0.09 & 0.93 $\pm$ 0.52 &  0.97 $\pm$ 0.80 \\
		$\delta_{\rm GRD}$(Z) &  0.73 $\pm$ 0.48 &  0.36 $\pm$  0.25 &  - &  - \\
\hline
\end{tabular}
\end{table*}

\begin{figure}
\includegraphics[width=\columnwidth]{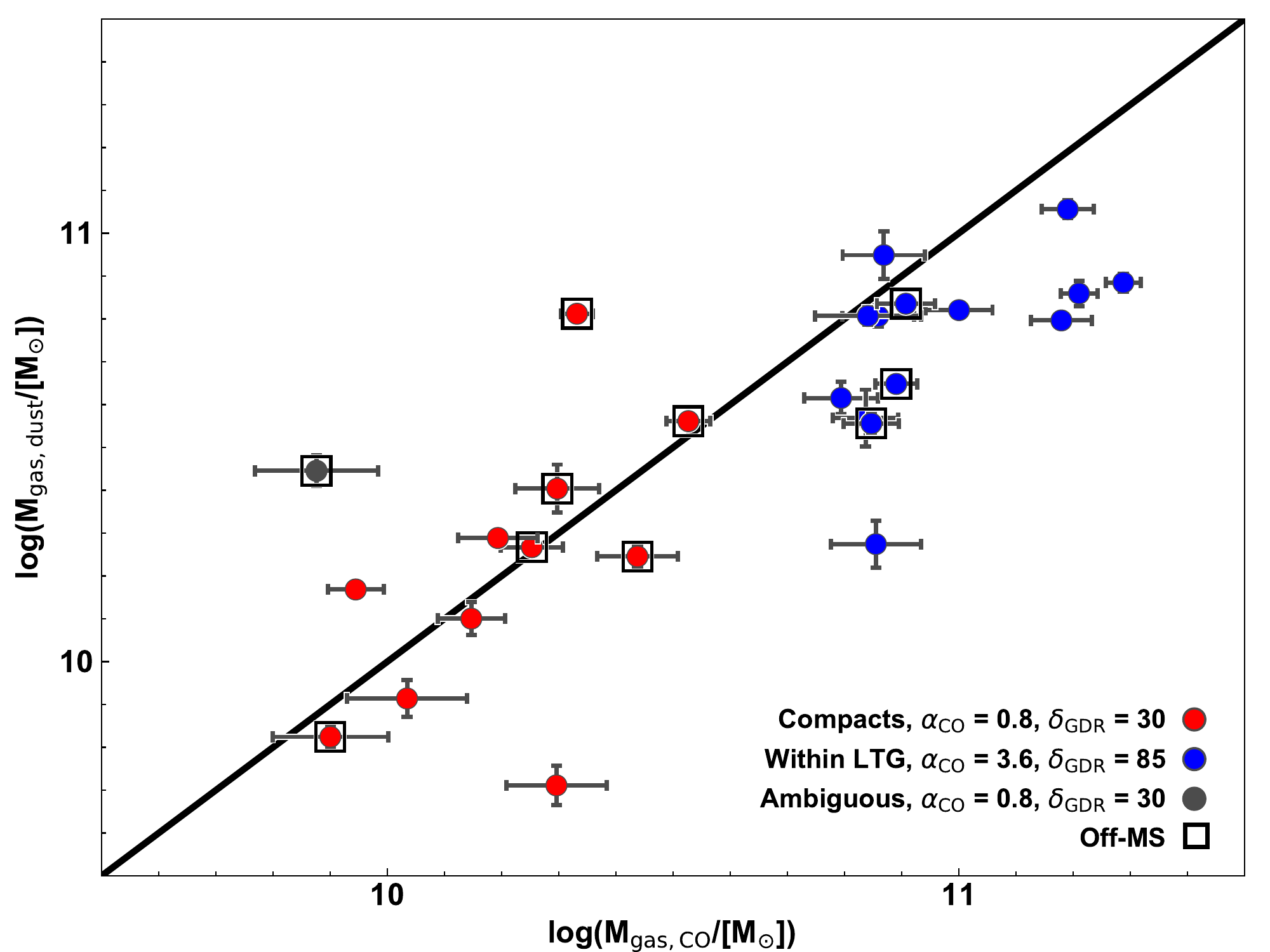}
\caption{ Comparison of the gas mass from $M_{\rm dust}$ [M$_{\odot}$] and from $L'_{\rm CO(2-1)}$ [K km s$^{-1}$ pc$^2$] for the galaxies with a robust dust mass estimate and a secure CO(2-1) detection. The solid line is the 1:1 relation.}
\label{figA:Mgas_CO_dust}
\end{figure}

Finally, in Figure \ref{figA:Mgas_CO_dust} we show the comparison between $M_{\rm gas}$ computed from $L'_{\rm CO(2-1)}$ and $M_{\rm dust}$, using the $\alpha_{\rm CO}$ and $\delta_{\rm GDR}$ specified in the caption. 
This figure shows a very good agreement between CO-based and dust-based gas masses for the compact galaxies, supporting our choice of starburst-like conversion factors for this class of objects. 
At the same time, this figure suggests that the gas mass of a subset of extended galaxies might be overestimated (underestimated) by a factor of $\sim 0.25$ {\it dex} when adopting $\alpha_{\rm CO} = 3.6$ ($\delta_{\rm GDR}$ =85).
This does not substantially affect our conclusions.

\begin{figure}
\includegraphics[width=\columnwidth]{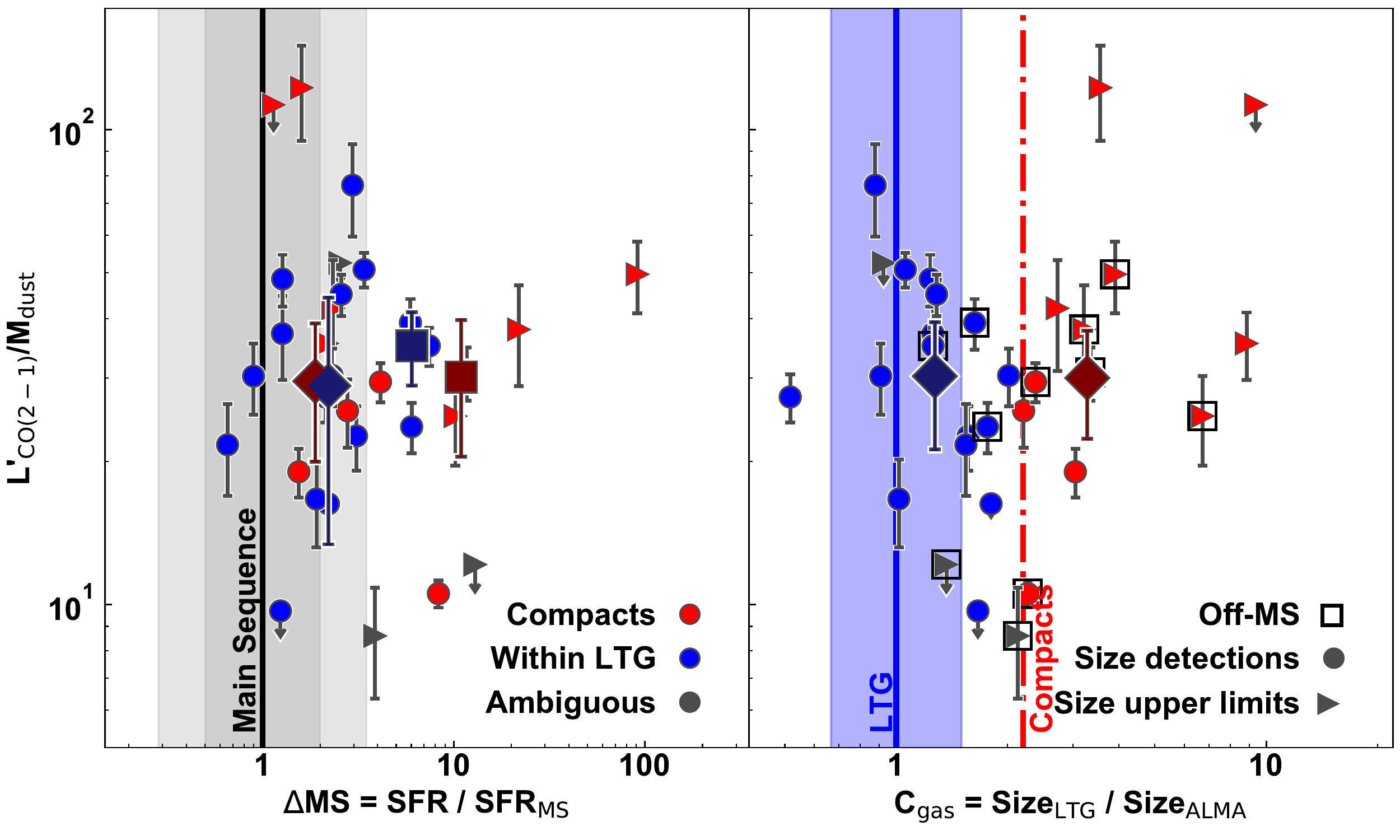}
\caption{ {The $L'_{\rm CO(2-1)}$/$M_{\rm dust}$ ratio as a function of the main sequence offset ({\it left}) and of the compactness ({\it right}) for galaxies with detections or reliable $L'_{\rm CO(2-1)}$ upper limits and robust measurements of $M_{\rm dust}$. In the left panel, large filled diamonds and large filled squares show the median $L'_{\rm CO(2-1)}$/$M_{\rm dust}$ ratio for compact and extended galaxies on and above the main sequence respectively. Large filled diamonds in the right panel show the median $L'_{\rm CO(2-1)}$/$M_{\rm dust}$ ratio for extended and compact galaxies. The errors on the average measurements are the interquartile range of the $L'_{\rm CO(2-1)}$/$M_{\rm dust}$ distribution. }}
\label{figA:LcoMdust_dms_cgas}
\end{figure}

{The small discrepancy in the gas masses measured from $L'_{\rm CO(2-1)}$ and $M_{\rm dust}$ might suggests variation in the $L'_{\rm CO(2-1)}$/$M_{\rm dust}$ ratio as a function of the main sequence offset and/or compactness. 
In Figure \ref{figA:LcoMdust_dms_cgas}, we thus investigate for the presence of variations in the $L'_{\rm CO(2-1)}$/$M_{\rm dust}$ ratio as a function of the main sequence offset and of the compactness for galaxies with detections or reliable $L'_{\rm CO(2-1)}$ upper limits and robust measurements of $M_{\rm dust}$. 
This plot shows that there is no clear dependence of the $L'_{\rm CO(2-1)}$/$M_{\rm dust}$ ratio on the main sequence offset nor the compactness. 
The median $L'_{\rm CO(2-1)}$/$M_{\rm dust}$ ratio for compact and extended galaxies on and above the main sequence is consistent within the errorbars. 
This suggests that CO and dust give equivalent results in terms of the molecular gas mass of compact and extended galaxies, without introducing biases.}

\bsp	
\label{lastpage}
\end{document}